\documentclass[useAMS,usenatbib,referee]{mn2e}
\input psfig.tex
\usepackage{graphicx}
\def\lsim{\lower.5ex\hbox{$\; \buildrel < \over \sim \;$}}
\def\gsim{\lower.5ex\hbox{$\; \buildrel > \over \sim \;$}}
\def \simeq{\lower.3ex\hbox{$\; \buildrel \sim \over - \;$}}
\def\ch{\lower-0.55ex\hbox{--}\kern-0.55em{\lower0.15ex\hbox{$h$}}}
\def\lh{\lower-0.55ex\hbox{--}\kern-0.55em{\lower0.15ex\hbox{$\lambda$}}}
\def\eg{{\it e.g.,} }
\def\etal{{\em et al.} }
\def\ie{{\em i.e.,} }

\title[Radiatively driven rotating pair-plasma jets from two component
accretion flows]{Radiatively driven rotating pair-plasma jets from two component
accretion flows}
\author[Indranil Chattopadhyay]{Indranil Chattopadhyay\thanks{E-mail:
Indranil.Chattopadhyay@wis.kuleuven.ac.be} \\
Centre for Plasma Astrophysics, Department of Mathematics, K. U. Leuven,
Celestijnenlaan 200B, Leuven 3001, Belgium}

\begin{document}

\date{Accepted .
      Received ;
      in original form }

\pagerange{\pageref{firstpage}--\pageref{lastpage}} \pubyear{}

\maketitle

\label{firstpage}

\begin{abstract}
Centrifugal pressure of matter spiralling onto black holes,
have long been known to produce standing or oscillating shocks.
The post-shock disc puffs up in the form
of a torus,
which intercepts soft photons from the outer Keplerian disc and inverse
Comptonizes them
to produce hard photons. The post-shock region also produces jets.
We study the interaction of both hard photons and soft photons,
with rotating electron-positron jets. We show that hard photons from the post-shock torus are instrumental in acceleration of jets, while soft
photons from the Keplerian disc is a better collimating agent.
We also show that if the jets are launched closer to the black hole,
relativistic and collimated jets are produced; if they are launched
at larger distances both collimation and acceleration are less.
We also show that if the shock location is relatively at larger distances from the black hole, collimation is better.
\end{abstract}

\begin{keywords}
Accretion, accretion discs - black hole physics - radiation mechanism:
general - radiative transfer - ISM: jets and outflows
\end{keywords}

\section{Introduction}

Jets in microquasars as well as in quasars shows relativistic terminal speed
[\eg GRS 1915+105, \citet{b28}; 3C 273, 3C 345, \citet{b44}; M87,
\citet{b1}],
%Mirabel&Rodriguez(1994), Zensus \etal(1995), Biretta \etal(1999)
though the actual acceleration process is an enigma.
It is well accepted in the scientific community that jets around compact objects
originate from the accretion disc accompanying such compact objects.
The study of interaction of radiation from the disc with outflowing jets,
is quite extensive.
The radiation field produced by a disc, depends on the geometry as well
as the dominant physical processes of the disc model. And hence the 
study of interaction of jets with radiation field produced by different
disc models, will, in general, draw different conclusions.

%Icke (1980){b23}
\citet{b23} studied the effect of radiative acceleration of gas flow above
a Keplerian disc, but did not consider the effect of radiation drag.
%Piran (1982){b34}
\citet{b34}, while calculating the radiative acceleration of outflows
about the rotation axis of thick accretion discs, found out that in order to
accelerate outflows to ${\gamma}>1.5$ (where ${\gamma}$ is the
bulk Lorentz factor), the funnels must be short and steep, but such funnels
are found to be unstable.
In a very important paper,
\citet{b24} considered blobby jets about the axis of symmetry of
%Icke (1989){b24}
thin discs (Shakura \& Sunyaev 1973; Novikov \& Thorne 1973, hereafter \citet{b31})
% shakura-sunyaev{b35}(citet{b31}; Novikov \& Thorne 1972, hereafter NT73)
and he obtained the `magic speed'
of $v_{m}=0.451c$ ($c$ --- the velocity of light), $v_m$being the
upper limit of terminal speed.
\citep{b19} extended this study for rotating flow above a thin
%Fukue (1996) {b19}
disc and drew similar conclusions, although for rotating flow, away
from axis of symmetry the terminal speed was found out to be a little
less than the magic speed of Icke.
It was shown that rotating winds above thin disc, generally
spreads as the radiation force along the azimuthal
direction (generated because of disc rotation) increases
the angular momentum of the flow, and are difficult to accelerate 
%\cite{b39,b40} tajima fukue 1996,1998
because of the presence of radiation drag term in the vertical direction
\citep{b39,b40}. Radiatively driven winds were also studied for
radiation coming from a slim disc \citep{b43}.
%{Watarai \& Fukue}{1999}]{b43}
Inspite of different disc temperature profile from
the thin disc, and also the inclusion
of advection term, radiatively driven outflows from slim
disc concluded, that these radiation fields will spread the outflows
and will suppress the vertical motion.
Later Fukue and his collaborators, achieved relativistic terminal speed
and collimation for
pair-plasma jets from a disc model which contains inner ADAF region (non-luminous) and outer slim disc (luminous), for non-rotating black hole
\citep{b20}, and repeated the same scheme for rotating black holes \citep{b32}.
%{Fukue \etal}{2001}]{b20}; {Orihara \& Fukue}{2003}]{b32}
We are working in a different regime. We consider
the TCAF (Two Component Accretion Flow) disc model
(see, Chakrabarti \& Titarchuk 1995, hereafter \citet{b4}), which consists
of less luminous outer Keplerian disc and more luminous post-shock torus in the
hard spectral-state of the disc.
\citet{b10} calculated the radiative force on optically thin jets
%{Chattopadhyay \& Chakrabarti}{2000}]{b10}
in and around the axis of symmetry for radiations coming from the post-shock 
torus.
\citet{b11} investigated the issue of radiative
%{Chattopadhyay \& Chakrabarti}{2002a}]{b11}
acceleration of normal plasma jets, by radiations only from the
post-shock torus of TCAF disc, and concluded
that normal-plasma jets are indeed accelerated to mildly relativistic
terminal speed. \citet{b12} also reported that hard
%{Chattopadhyay \& Chakrabarti}{2002b}]{b12}
radiations from
the post-shock torus do not impose {\it any} upper limit for terminal speed.
Chattopadhyay \etal (2004, hereafter \citet{b14}), solved the
%paper-1{b14}
equations of photo-hydrodynamics for on-axis pair dominated jets for
radiations coming from the whole of TCAF disc model. \citet{b14}
concluded that terminal speed ${\vartheta}{\sim}0.9c$ is easily achieved
if the shock-location in accretion $x_s{\sim}$ few ${\times}10r_g$
($r_g=$Schwarzschild radius), and the luminosity of the post-shock torus (${\equiv}
L_{\rm C}$) is
${\sim}$ few ${\times}10$\% the Eddington luminosity. \citet{b14} also showed that
%Paper-I
the special geometry of the post-shock torus ensures that, its contribution
to various radiative moments dominates over that due to Keplerian disc.
Thus radiations from Keplerian disc (of luminosity $L_{\rm K}$) has a very marginal effect
on determining the terminal speed of jets. Having shown this, it is only
natural to study how rotating jets behave in the presence of the radiation
field of the disc. The
azimuthal component of the radiative force generated by the disc rotation, will
try to spin up the jet, while radiative drag force in the same direction
will try to carry away the angular momentum from the jet. So it is particularly
interesting to study which of these two phenomena wins.
\vskip -0.3cm
In this paper we study the interaction of radiations from the whole of
TCAF disc and rotating jets. We show that the radiation from the CENBOL
(CENtrifugal pressure-dominated BOundary Layer; see Chakrabarti \etal 1996,
\citet{b6})
is the chief accelerating agent, while radiation from Keplerian disc
has greater degree of collimation property. Both collimation and acceleration
are much higher when the jet materials are launched closer
to the axis of symmetry. We show that, if radiative processes is the 
dominant accelerating agent then, we can get highly relativistic and
collimated jets only in moderate hard states (\ie $L_{\rm C}{\sim}L_{\rm K}$)
and not in extreme hard states ($L_{\rm C}{\gg}L_{\rm K}$).

In the next Section, we present a brief account of the TCAF disc model. 
In \S 3, we present the model assumptions and the equations of motion.
We give an account of the streamline coordinates
which we use to solve the governing equations.
We also compute ten independent moments of radiation field from such an disc.
In \S 4, we present our solutions and finally in \S 5,
we draw our conclusions.

\section[]{TCAF disc}

\begin {figure}
\vbox{
\vskip -0.0cm
\hskip 0.0cm
\centerline{
\psfig{figure=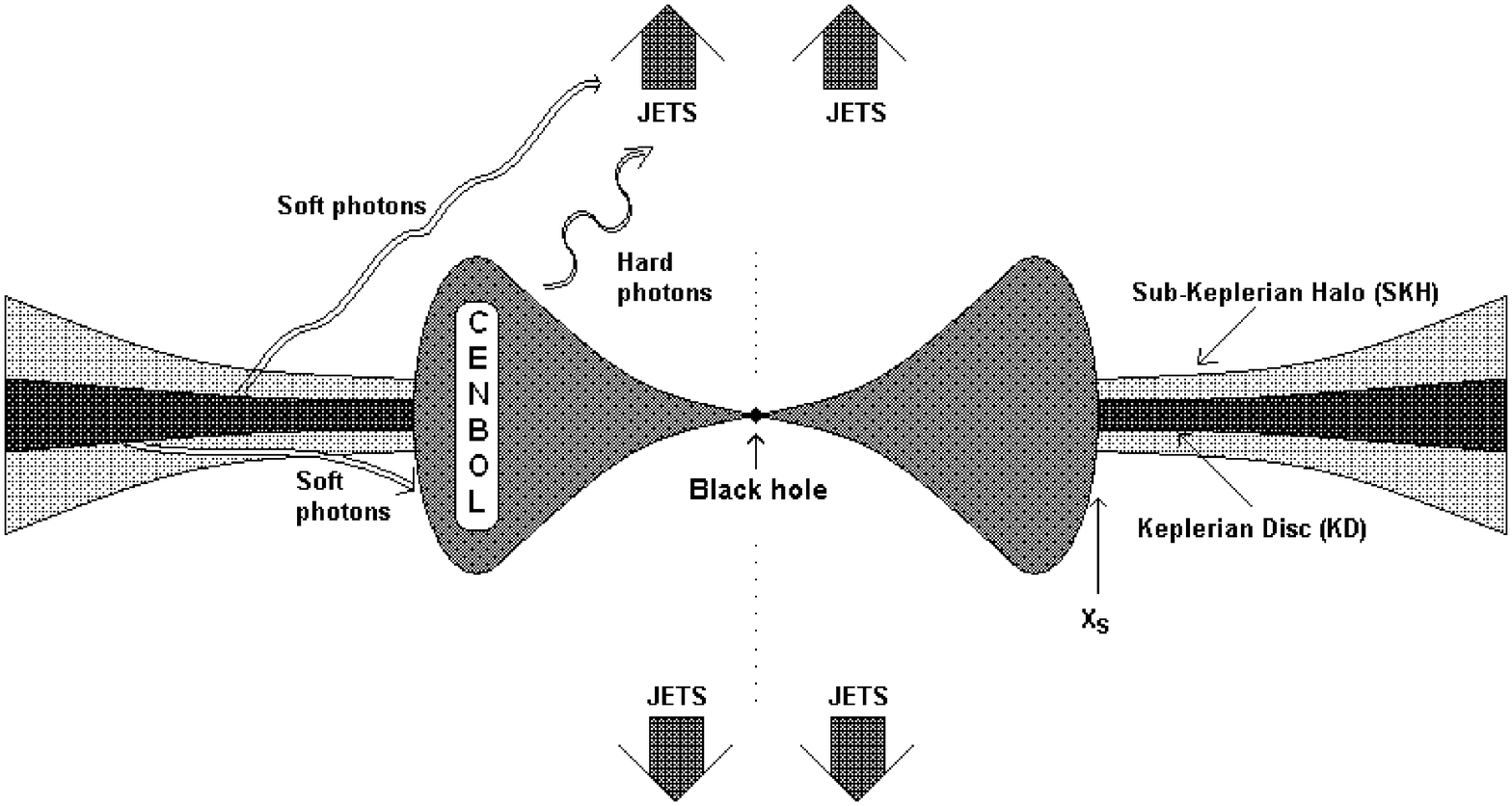,height=9truecm,width=16truecm}}}
\end{figure}
\begin{figure}
\vspace{0.0cm}
\caption[] {Cross-sectional view of Two Component Accretion Disc Model.}
\end{figure}

A detailed account of TCAF disc is given in Paper-I [also see,
\citet{b4}; \citet{b6}; Chakrabarti 1997,
hereafter \citet{b7}, \citet{b13}].
For the sake of completeness, let us now give a brief account of the TCAF disc.

Inner boundary condition for matter accreting onto black holes
are (i) supersonic and (ii) sub-Keplerian. Accretion topologies for sub-Keplerian matter admits two X-type critical points \citep{b26}, and supersonic matter after
%{Liang \& Thompson}{1980}]{b26}
crossing the outer critical point
may undergo shock due to centrifugal pressure \citep{b18,b2,b3,b5}.
%fukue87,Chakrabarti 1989, Chakrabarti 1990, Chakrabarti 1996
The post-shock matter then enters the black hole through the
inner critical point. 
Matter is slowed down in the immediate post-shock region and as entropy
is generated, it makes the post-shock region hot. This hot, slowed down
post-shock region puffs up in
the form of a torus
the CENBOL.
% CTKE96{b6}

Chakrabarti \& Titarchuk (1995) for the first time
%CT95
proposed a disc model which contains both, Keplerian matter and sub-Keplerian
matter. The Keplerian matter is of higher angular momentum and lower
specific energy than the sub-Keplerian matter,
and settles around the {\it equatorial plane} to form the
{\it Keplerian disc} (hereafter KD). Sub-Keplerian matter flanks the cooler
Keplerian disc from the top and bottom,
sandwiching the Keplerian disc, and is known as the
{\it sub-Keplerian halo} (hereafter SKH; also see, \citet{b7}).
%{C97}{}]{b7}
The SKH suffers a shock at a few tens of
Schwarzschild radii. The shock compresses the post-shock
flow making it denser than the pre-shock sub-Keplerian flow.
The hot post-shock flow evaporates the Keplerian
disc and falls to the black hole as a single component, in other
words the shock location ($x_s$) is the outer boundary of CENBOL and the
inner boundary of the KD. A schematic diagram of
such a disc structure is shown in Fig. (1), where KD is shown to be sandwiched
by SKH, and the position of $x_s$, as well as the
central black hole are also shown.
Although the SKH (pre-shock) is optically thin for the
radiations from Keplerian disc, the post-shock
torus is optically slim, in other words the optical depth
of CENBOL is around unity (\ie ${\tau}{\sim}1$; see \citet{b4}, \citet{b6},
\citet{b7} for details).

\citet{b4} showed that, if the Keplerian accretion rate (${\dot M}_{\rm K}$)
is higher than sub-Keplerian accretion rate (${\dot M}_{\rm h}$),
then it supplies more soft photons to cool down the
CENBOL. This results in more power to the soft end of the
accretion disc
spectrum --- a state known as the {\it soft state}.
If on the other hand, ${\dot M}_{\rm h}>{\dot M}_{\rm K}$,
then SKH supplies more hot electrons to the CENBOL than the soft
photons supplied by
the KD. The dearth of soft photons cannot cool down the
CENBOL significantly.
Thus CENBOL remaining puffed up and hot, can intercept a significant fraction
of soft photons produced by KD, and inverse-Comptonize them to produce the hard power-law
tail of the accretion disc spectrum --- a state called {\it hard state}.
This kind of hybrid
disc structure is known as the Two Component Accretion Flow or the TCAF disc
\citep{b4,b6,b17,b7},
%CT95, CTKE96, Ebisawa etal 1996, c97
and has observational support \citep{b36,b37}.
%, \citet{b37}].
% Smith etal 2001, 2002\citet{b36}, \citet{b37}

Chakrabarti and his collaborators have also shown, the unbalanced
thermal gradient force of CENBOL, in the $z$ direction, drives a part of the
in falling matter along the axis of symmetry to form jets
\citep{b8,b9,b15,b16}.
%Chakrabarti, 1998; Chakrabarti, 1999; Das \& Chakrabarti, 1999; Das
%\etal 2001.
There are wide support that the jets are indeed coming out from a region
within $50-100$ Schwarzschild radius of the black hole \citep{b25}.
%(Junor, Biretta \& Livio, 1999){b25}
Similarly, it is believed that jets are produced only in hard states
[see, \citet{b21}, and references therein].
% Gallo, fender, pooley 2003{b21}
Thus it is natural to study interaction of hard radiation from the
CENBOL and the outflowing jets, with the particular interest of studying,
whether momentum deposited to the jet material by these hard photons can
accelerate them to ultra-relativistic speeds.

It is to be remembered that, we are not considering generation mechanism
of jets self-consistently. Since hard radiations are expected to emerge out of
CENBOL, the hard photons `look' directly into the jet vertically above and hence eventually
deposit their momentum into the latter. Furthermore, radiation from a hot CENBOL is
a likely source of pair production and hence the
possibility of radiative momentum deposition
is likely to be higher even for radiations from CENBOL
hitting the outflow at an angle
[see, e.g., \citep{b41}
%{Yamasaki \etal}{1999}]{b41}
for production mechanism of pairs from hot accretion flows].
Thus we consider radiative momentum deposition on
pair dominated jets, which are generated above
the inner surface of CENBOL, \ie within the funnel like region.

\section{Assumptions, Governing Equations and Computation of
Radiative moments from TCAF disc}

We ignore the curvature effects due to the presence of the central
black hole mass. The metric is given by,
$$
d{\tilde s}^2=c^2d{\tilde t}^2-dr^2-r^2d{\phi}^2-dz^2,
\eqno{(1)}
$$
where, $r$, ${\phi}$, and $z$ are the usual coordinates in cylindrical geometry
and $d{\tilde s}$ is the metric in four-space.
The four-velocities are $u^{\mu}$.
The convention we follow is --- the Greek indices signify all four components
and the Latin indices represent only the spatial ones.
The black hole is assumed to be non-rotating and hence the strong
gravity is taken care of by the so-called Paczy\'nski-Wiita potential
\citep{b33}.
%{Paczy\'nski \& Wiita}{1980}]{b33}

In this paper, the generation mechanism of jets is not considered. 
We assume axis-symmetric, steady, rotating jets \ie  
${\partial}/{\partial}{\tilde t}={\partial}/{\partial}{\phi}=0$.
%We also assume 
%the gas pressure is negligible compared to the radiation pressure. This 
%is perhaps the case especially inside the funnel wall close to the axis.
The derivation of the equations of motion of radiation hydrodynamics
for optically thin plasma, was investigated by a number of workers. A detailed
account of such derivation has been presented by \citet{b27} and \citet{b19}
% {Mihalas \& Mihalas}{1984}]{b27}; {Fukue}{1996}]{b19}
[references therein], and are not presented
here.
The equations of motion are;

$$
({\epsilon}+p)\left(u^{\mu}\frac{{\partial}u^r}{{\partial}x^{\mu}}+
\frac{GM_{B}r}{R(R-r_g)^2}-ru^{\phi}u^{\phi} \right)
= -\frac{{\partial}p}{{\partial}r}-u^ru^{\mu}
\frac{{\partial}p}{{\partial}x^{\mu}}+{\rho}\frac{{\sigma}_{T}}{mc}{\Im}^r,
\eqno{(2a)}
$$
$$
({\epsilon}+p)\left(u^{\mu}\frac{{\partial}u^{\phi}}{{\partial}x^{\mu}}
+\frac{2}{r}u^ru^{\phi} \right)
=-u^{\phi}u^{\mu}
\frac{{\partial}p}{{\partial}x^{\mu}}+{\rho}\frac{{\sigma}_{T}}{mcr}{\Im}^{\phi},
\eqno{(2b)}
$$
and
$$
({\epsilon}+p)\left(u^{\mu}\frac{{\partial}u^z}{{\partial}x^{\mu}}+
\frac{GM_{B}z}{R(R-r_g)^2} \right)
= -\frac{{\partial}p}{{\partial}z}-u^zu^{\mu}
\frac{{\partial}p}{{\partial}x^{\mu}}+{\rho}\frac{{\sigma}_{T}}{mc}{\Im}^z.
\eqno{(2c)}
$$
In above equations, ${\epsilon}$, $p$
and ${\rho}$ are the internal energy,
isotropic gas pressure and density measured in the co-moving frame of the
fluid and $R=(r^2+z^2)^{1/2}$. $G$, $M_{B}$, ${\sigma}_{T}$, $m$ and $r_g(2GM_B/c^2)$
are the universal gravitational constant, the mass of the central black hole, Thomson scattering cross-section, mass of the gas particles and
Schwarzschild radius, respectively.
${\Im}^i$ signifies radiative contributions along respective components of the
momentum balance equation.

The radiative contributions are given by;
\begin{eqnarray*}
\hskip 1.0cm
\frac{{\sigma}_T}{m}\frac{{\Im}^r}{c} &  = & \frac{{\sigma}_T}{m}
\left[{\gamma}\frac{F^r}{c}-{\gamma}^2u^rE-u^{j}P^{rj}+u^r
\left(2\frac{{\gamma}}{c}u^{j}F^{j}-u^{j}u^{k}P^{jk}
\right) \right] \\ \nonumber
& = & \left[{\gamma}f^r-{\gamma}^2u^r{\varepsilon}-u^{j}{\wp}^{rj}+
u^r\left(2{\gamma}u^{j}f^{j}-u^{j}u^{k}{\wp}^{jk}
\right) \right]  \hskip 4.11cm (3a)
\end{eqnarray*}

Similarly,
\begin{eqnarray*}
\hskip 1.0cm
\frac{{\sigma}_T}{m}\frac{{\Im}^{\phi}}{c} & = & \frac{{\sigma}_T}{m}
\left[{\gamma}\frac{F^{\phi}}{c}-{\gamma}^2ru^{\phi}E
-u^{j}P^{{\phi}j}+ru^{\phi}
\left(2\frac{{\gamma}}{c}u^{j}F^{j}-u^{j}u^{k}P^{jk}
\right) \right] \\ \nonumber
& = & \left[{\gamma}f^{\phi}-{\gamma}^2ru^{\phi}{\varepsilon}
-u^j{\wp}^{{\phi}j}+ru^{\phi}
\left(2\frac{{\gamma}}{c}u^{j}f^{j}-u^{j}u^{k}{\wp}^{jk}
\right) \right]
\hskip 3.29cm (3b)
\end{eqnarray*}
and
\begin{eqnarray*}
\hskip 1.0cm
\frac{{\sigma}_T}{m}\frac{{\Im}^z}{c} & = & \frac{{\sigma}_T}{m}\left[
{\gamma}\frac{F^z}{c}-{\gamma}^2u^zE
-u^{j}P^{zj}+u^z
\left(2\frac{{\gamma}}{c}u^{j}F^{j}-u^{j}u^{k}P^{jk}
\right) \right] \\ \nonumber
& = & \left [{\gamma}f^z-{\gamma}^2u^z{\varepsilon}-u^{j}{\wp}^{zj}+
u^z\left(2{\gamma}u^{j}f^{j}-u^{j}u^{k}{\wp}^{jk}
\right) \right]  \hskip 4.11cm (3c)
\end{eqnarray*}

In Eqs. (3a-3c), $E(r,{\phi},z)$, $F^i(r,{\phi},z)$, and $P^{ij}(r,{\phi},z)$
are the radiative energy density, three components of radiative flux
and six components of radiative pressure tensors measured in observer frame,
while
${\varepsilon}=\frac{{\sigma}_T}{m}E$, $f^i=\frac{{\sigma}_T}{mc}F^i$,
and ${\wp}^{ij}=\frac{{\sigma}_T}{m}P^{ij}$.
Furthermore, ${\gamma}({\gamma}=u_t)$ is the Lorentz factor. We assume the gas pressure
to be negligible compared to the radiation pressure terms.

\begin{figure}
\vbox{
\vskip 0.0cm
\hskip 0.0cm
\centerline{
\psfig{figure=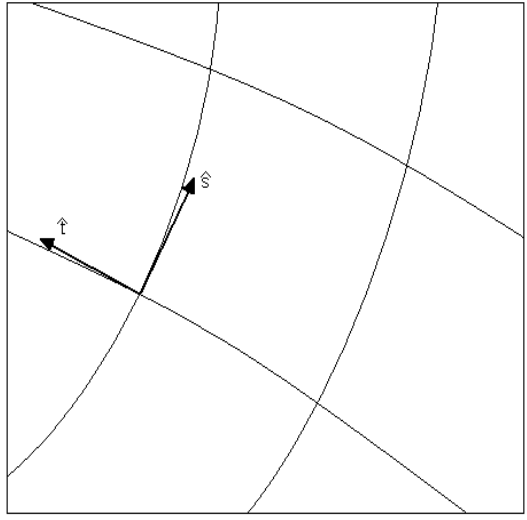,height=7truecm,width=7truecm}}}
\caption[]{Schematic diagram of streamline coordinates
($s$, ${\phi}$, $t$) in the meridional plane.}
\end{figure}

\subsubsection{Streamlines}

In streamline coordinates ($s$, ${\phi}$, $t$),
$s$-axis ($t$-axis) is parallel (perpendicular) to streamlines, where
$ds^2=dr^2+dz^2$, such that $u^{t}$ is zero. In Fig. (2)
we draw a schematic diagram of the streamline coordinates. The basic equations
can then be written as \citep{b19, b20}, \\
%{Fukue}{1996}]{b19}; {Fukue \etal}{2001}]{b20}

%\begin{eqnarray*}
Momentum Balance along the streamline: 
$$
u^{s}\frac{{\partial}u^{s}}
{{\partial}s}=
-\frac{GM_B}{R(R-r_g)^2}\frac{rdr+zdz}{ds}+\frac{(r^2u^{\phi})^2}
{r^3}\frac{dr}{ds}+\frac{{\sigma}_T}{cm} {\tilde {\Im}^s},
\eqno{(4a)}
$$

Angular Momentum conservation:
$$
u^{s}\frac{{\partial}(r^2u^{\phi})}{{\partial}s}=\frac{{\sigma}_T}{cm}
r{\Im}^{\phi},
\eqno{(4b)}
$$
%\end{eqnarray*}

And the streamline equation:

$$
-\frac{GM_B}{R(R-2r_g)^2}\frac{rdz-zdr}{ds}+ru^{\phi}u^{\phi}\frac{dz}{ds}
+(f^r\frac{dz}{ds}-f^z\frac{dr}{ds})=0,
\eqno{(4c)}
$$
 
where,
\begin{eqnarray*}
\hskip 2.0cm   \frac{{\sigma}_T}{m}\frac{{\tilde {\Im}^s}}{c} & = & {\gamma}f^{s}-{\gamma}^2{\varepsilon}u^s-({\wp}^{ss}u^s+
r{\wp}^{s{\phi}}u^{\phi})
+u^s{\{}2{\gamma}(f^su^s+rf^{\phi}u^{\phi}) \\ \nonumber 
& - & ({\wp}^{ss}u^su^s+2r{\wp}^{s{\phi}}
u^su^{\phi}+r^2{\wp}^{{\phi}{\phi}}u^{\phi}u^{\phi}){\}}. \hskip 5.3cm (4d)
\end{eqnarray*}

In Eq. (4d), 
$$
f^s=f^r\frac{dr}{ds}+f^z\frac{dz}{ds}
$$

$$
{\wp}^{ss}={\wp}^{rr}\left(\frac{dr}{ds} \right)^2+2{\wp}^{rz}\frac{dz}{ds}\frac{dr}{ds}
+{\wp}^{zz}\left(\frac{dz}{ds} \right)^2
$$
and,
$$
{\wp}^{s{\phi}}={\wp}^{r{\phi}}\frac{dr}{ds}+{\wp}^{z{\phi}}\frac{dz}{ds}
$$

Equations (4a-4c), can be simplified in terms of $v$ and ${\lambda}$
and can be expressed in geometric units ($r_g=c=M_B=1$),
for simplicity we will keep the same symbols representing various quantities
defined so far. \\
We now define a three velocity measured by the static observer in geometrical
units defined above,
$$
v^2_s=-\frac{u_su^s}{u_{\tilde t}u^{\tilde t}},
\eqno{(5a)}
$$
while the angular-momentum and angular velocity are being defined as,
$$
{\lambda}=-\frac{u_{\phi}}{u_{\tilde t}} \hskip 2.0cm {\&} \hskip 2.0cm
{\omega}=\frac{u^{\phi}}{u^{\tilde t}}.
\eqno{(5b)}
$$
We further define the 3-velocity measured by co-rotating observer,
$$
v^2=\frac{v^2_s}{1-{\omega}{\lambda}}.
\eqno{(5c)}
$$
So that ${\gamma}^2=1/(1-v^2_s-{\omega}{\lambda})=1/{\{}(1-v^2)
(1-{\omega}{\lambda}){\}}={\gamma}^2_v{\gamma}^2_{\lambda}$, $u^s={\gamma}_vv$,
and $u^{\phi}=({\gamma}{\lambda})/r^2$.

Equations (4a-4c) are re-written as,
$$
\frac{dv}{dz}=\frac{N_1}{D_1},
\eqno{(6a)}
$$
where,

\begin{eqnarray*}
\hskip 1.0cm N_1 & = & [{\gamma}{\cal F}^s-{\gamma}^2{\gamma}_vv{\cal E}
-{\gamma}_vv{\cal P}^{ss}-{\gamma}\frac{\lambda}{r}
{\cal P}^{s{\phi}}
+2{\gamma}({\gamma}^2_vv^2{\cal F}^s+{\gamma}{\gamma}_vv\frac{\lambda}{r}
{\cal F}^{\phi})-{\gamma}^3_vv^3{\cal P}^{ss} \\ \nonumber
& - & 2{\gamma}{\gamma}^2_vv^2 \frac{\lambda}{r}{\cal P}^{s{\phi}}
-{\gamma}^2{\gamma}_vv
\frac{\lambda^2}{r^2}{\cal P}^{\phi\phi}]-\frac{1}{2R(R-1)^2}
\left(r\frac{dr}{dz}+z \right)+\frac{({\gamma}{\lambda})^2}{r^3}\frac{dr}{dz},
\hskip 1.3cm (6b)
\end{eqnarray*}

and

$$
D_1={\gamma}^4v.
\eqno{(6c)}
$$

$$
\frac{d{\lambda}}{dz}=\frac{N_2}{D_2}
\eqno{(6d)}
$$

\begin{eqnarray*}
\hskip 1.0cm N_2 & = & [{\gamma}{\cal F}^{\phi}-{\gamma}^3
{\frac{\lambda}{r}}
{\cal E}-{\gamma}_vv{\cal P}^{s{\phi}}-{\gamma}\frac{\lambda}{r}
{\cal P}^{{\phi}{\phi}}+{\gamma}
\frac{\lambda}{r}{\{}2{\gamma}({\gamma}_vv{\cal F}^s+{\gamma}\frac{\lambda}
{r}{\cal F}^{\phi})-({\gamma}^2_vv^2{\cal P}^{ss} \\ \nonumber
& + & 2{\gamma}{\gamma}_vv\frac{\lambda}{r}{\cal P}^{s{\phi}}+
{\gamma}^2\frac{{\lambda}^2}{r^2}{\cal P}^{{\phi}{\phi}}){\}}]r
-{\gamma}{\gamma}^3_vv^2{\lambda}\frac{dv}{dz}+{\gamma}{\gamma}_v
{\gamma}^2_{\lambda}v\frac{{\lambda}^3}{r^3}\frac{dr}{dz}
\hskip 4.2cm (6e) 
\end{eqnarray*}

$$
D_2={\gamma}{\gamma}_v\left(1+\frac{({\gamma}_{\lambda}{\lambda})^2}{r^2}
\right)v
\eqno{(6f)}
$$
and,
$$
\frac{dr}{dz}=\frac{-r/[2R(R-1)^2]+{\gamma}f^r
+({\gamma}{\lambda})^2/r^3}
{-{z}/[{2R(R-1)^2}]+{\gamma}f^z}
\eqno{(6g)}
$$

Quantities defined in Eqs. (6b) and (6e) are:

$$
{\cal F}^s=f^r\frac{dr}{dz}+f^z
\eqno{(7a)}
$$
$$
{\cal F}^{\phi}=f^{\phi}\frac{ds}{dz}
\eqno{(7b)}
$$
$$
{\cal E}={\varepsilon}\frac{ds}{dz}
\eqno{(7c)}
$$
$$
{\cal P}^{ss}={\wp}^{rr}\frac{dr}{dz}\frac{dr}{ds}+2{\wp}^{rz}\frac{dr}{dz}
+{\wp}^{zz}\frac{dz}{ds}
\eqno{(7d)}
$$
$$
{\cal P}^{s{\phi}}={\wp}^{r{\phi}}\frac{dr}{dz}+{\wp}^{z{\phi}}
\eqno{(7e)}
$$
$$
{\cal P}^{{\phi}{\phi}}={\wp}^{{\phi}{\phi}}\frac{ds}{dz}.
\eqno{(7f)}
$$

We now have to solve Eqs. (6a), (6d) and (6g), for given radiation
field (${\varepsilon}$, $f^i$, ${\wp}^{ij}$), specified by disc
parameters.
It is to be noted that for motion along the axis ${\lambda}=0$
and $\frac{dr}{dz}=\frac{d{\lambda}}{dz}=0$, then Eq. (6a)
reduces to Eq. (6) of \citet{b14}.
%paper-I {b14}

\subsection{Computation of radiative moments from TCAF disc}

The radiation reaching each point within the funnel like region
and the region above it, is coming from two
parts of the disc, namely, the CENBOL and Keplerian disc, hence all
the radiative
moments should have both the contributions. 

\begin{figure}
\hskip 3.0cm
\includegraphics[scale=0.6]{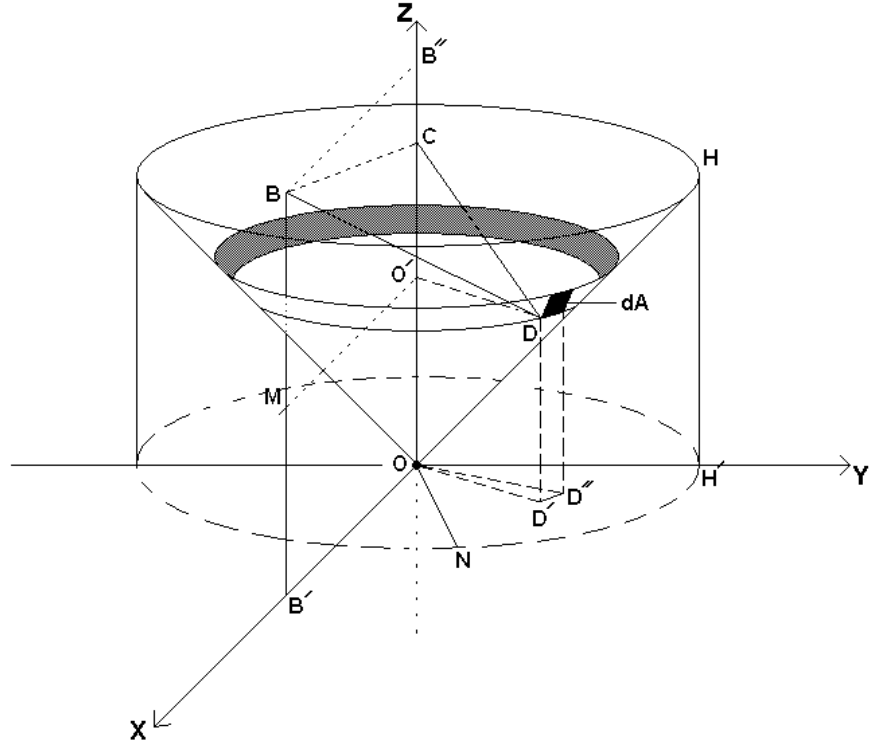} 

\caption[]{Cartoon diagram of CENBOL. The outer KD and 
SKH are not shown. $B$($r$, $0$, $z$) is the field point
and $D$($x$,${\phi}$, $y$) is the source point on the inner surface
of the CENBOL. The position of the black hole is at $O$.
The inner surface of the CENBOL is assumed conical.
The position of the black hole is $O$. $ON=x_s$, and
$HH^{\prime}=h_s$.
${\angle B^{\prime}OD^{\prime}}={\phi}$ \&  
${\angle D^{\prime}OD^{{\prime}{\prime}}}=d{\phi}$. And $d{\cal A}$
is the differential area at $D$. Only the top half is shown.}
\end{figure}

In Fig. (3) a schematic diagram of the CENBOL is presented.
The inner surface of the CENBOL is assumed conical (described by,
rotating $OH$ about the axis of symmetry), and the outer surface
of CENBOL is cylindrical ($HH^{\prime}$). $XOY$ describes the
equatorial plane. $O$ is the position of the black hole.
$B$($r$, $0$, $z$) is the field point where the various radiative moments
are to be calculated. $D$($x$, $\phi$, $y$) is the source point on the
CENBOL inner surface. $ON=OH^{\prime}=x_s$ is the location of the shock.
And $HH^{\prime}=h_s$ is the shock height. The local normal at $D$
is $\frac{\bf DC}{|{\bf DC}|}$. The differential area about $D$ is
marked as $d{\cal A}$. The shock height $h_s$ depends on the shock location,
and is expressed as $h_s{\sim}0.6(x_s-1)$ [see, \citet{b14}].

By definition, the radiative moments (in natural units) at $B$ are;
$$
E=\frac{1}{c}\int Id{\Omega}=\frac{1}{c}\left({\int}_{C}
I_{C}d{\Omega}_{C}+{\int}_{K}I_{K}d{\Omega}_{K}
\right),
\eqno{(8a)}
$$
$$
\frac{F^i}{c}=\frac{1}{c}\int Il^id{\Omega}=\frac{1}{c}\left({\int}_{C}
I_{C}l^i_{C}d{\Omega}_{C}+{\int}_{K}I_{K}l^i_{K}d{\Omega}_{K} \right),
\eqno{(8b)}
$$
and
$$
P^{ij}=\frac{1}{c}\int Il^il^jd{\Omega}=\frac{1}{c}\left({\int}_{C}
I_{C}l^i_{C}l^j_{C}d{\Omega}_{C}+{\int}_{K}I_{K}l^i_{K}l^j_{K}d{\Omega}_{K}
\right).
\eqno{(8c)}
$$

In Eqs. (8a-8c), $I$, $d{\Omega}$, $l^i$ are the frequency integrated
intensity
from the disc, differential solid angle at $B$, and the direction
cosines at $B$ w.r.t $D$, respectively. 
Suffix $C$ and $K$ represent quantities linked to CENBOL and the
KD respectively.
The expressions of solid angles subtended at $B$ from $D$ is given by
$$
d{\Omega}_{C}=\frac{(d{\cal A}){\rm cos}{\angle CDB}}{BD^2}=
\frac{x{\hskip 0.1cm}{\rm cosec}{\theta}
{\hskip 0.1cm}dx{\hskip 0.1cm}d{\phi}{\hskip 0.1cm}
{\rm cos}{\angle CDB}}{BD^2},
\eqno{(8d)}
$$
where, ${\rm cos}{\angle CDB}=\frac{(BD^2+CD^2-BC^2)}{2(BD)(CD)}$,
and $BD^2=r^2+x^2+(z-y)^2-2{\hskip 0.1cm}r{\hskip 0.1cm}x{\hskip 0.1cm}
{\rm cos}{\phi}$, $CD=x{\hskip 0.1cm}{\rm sec}{\theta}$,
$BC^2=r^2+(z-x{\hskip 0.1cm}{\rm tan}{\theta}-y)^2$
and ${\theta}[=tan^{-1}(x_s/h_s)]$ is the semi-vertical angle of the CENBOL inner surface.
Similarly it is easy to find from,
Fig. (2), the direction cosines are given by, \\
$l^r_C=(r-x{\hskip 0.1cm}{\rm cos}{\phi})/BD$;
$ l^{\phi}_C=-(x{\hskip 0.1cm}{\sin}{\phi})/BD$;
and $l^z_C=(z-y)/BD$.

The frequency averaged CENBOL intensity is given by,
$$
I_C=\frac{I_{C0}}{(1+z_{\rm red})^4}=\frac{L_{\rm C}}{{\pi}{\cal A}(1+z_{\rm red})^4},
\eqno{(8e)}
$$
where, $L_{\rm C}$ and ${\cal A}$ is the total CENBOL luminosity
and total CENBOL area, $z_{\rm red}$ is the red-shift factor taken up to the
first order of the disc velocity. The inner edge of CENBOL \ie
also the TCAF disc is taken up to $x_{in}=2r_g$, within which
the general relativistic effect has to be considered. $I_{C0}$ is assumed
uniform for simplicity, as was explained in \citet{b14}.
%Paper-I

We are not solving the disc-equations simultaneously,
so we, in principle, cannot consider the Doppler shift of the 
photons coming out of CENBOL. Nonetheless, not considering the disc
motion, robs us of a vital element of physics. The rotational
velocity of the disc generates the $\phi$ component of the radiative flux.
So we make an estimate of the post-shock disc motion. At the
shock, the in-falling matter is virtually stopped. We solve the
geodesic equation, starting from $x_s$ with a very small
velocity (${\approx}0.01c$). We assume the solution (${\tilde u}_{st}$)
[Appendix A]
to be the 3-velocity of matter along the surface of the CENBOL.
The angular momentum of the post shock matter is sub-Keplerian and
almost constant
[\eg \citet{b4,b5}]; as the infall time-scale is much smaller than the viscous
time-scale. So it is assumed to be constant, and
is also
the initial specific angular momentum
(${\lambda}_{in}=1.7$ in the geometrical units defined above)
of the jet. Under such assumptions,
the radial, azimuthal and axial 3-velocity of the matter on the CENBOL
surface is given by
(expressed in dimensionless units described above),
${\tilde u}_x={\tilde u}_{st}{\rm sin}{\theta}$,
${\tilde u}_{\phi}=(1-1/x)^{1/2}({\lambda}_{in}/x)$
and ${\tilde u}_y=u_{st}{\rm cos}{\theta}$. Therefore $(1+z_{red})=
(1-{\tilde u}_il^i)$, where 
$$
{\tilde u}_il^i=\frac{({\tilde u}_x{\hskip 0.1cm}{\rm cos}{\phi}-
{\tilde u}_{\phi}{\hskip 0.1cm}{\rm sin}{\phi})(r-x{\rm cos}{\phi})}{BD}-
\frac{({\tilde u}_x{\hskip 0.1cm}{\rm sin}{\phi}+{\tilde u}_{\phi}
{\hskip 0.1cm}{\rm cos}{\phi})(x{\hskip 0.1cm}{\rm sin}{\phi})}
{BD}+\frac{{\tilde u}_y(z-y)}{BD}.
\eqno{(8f)}
$$

\begin{figure}

\hskip -0.2cm
\includegraphics[scale=0.36]{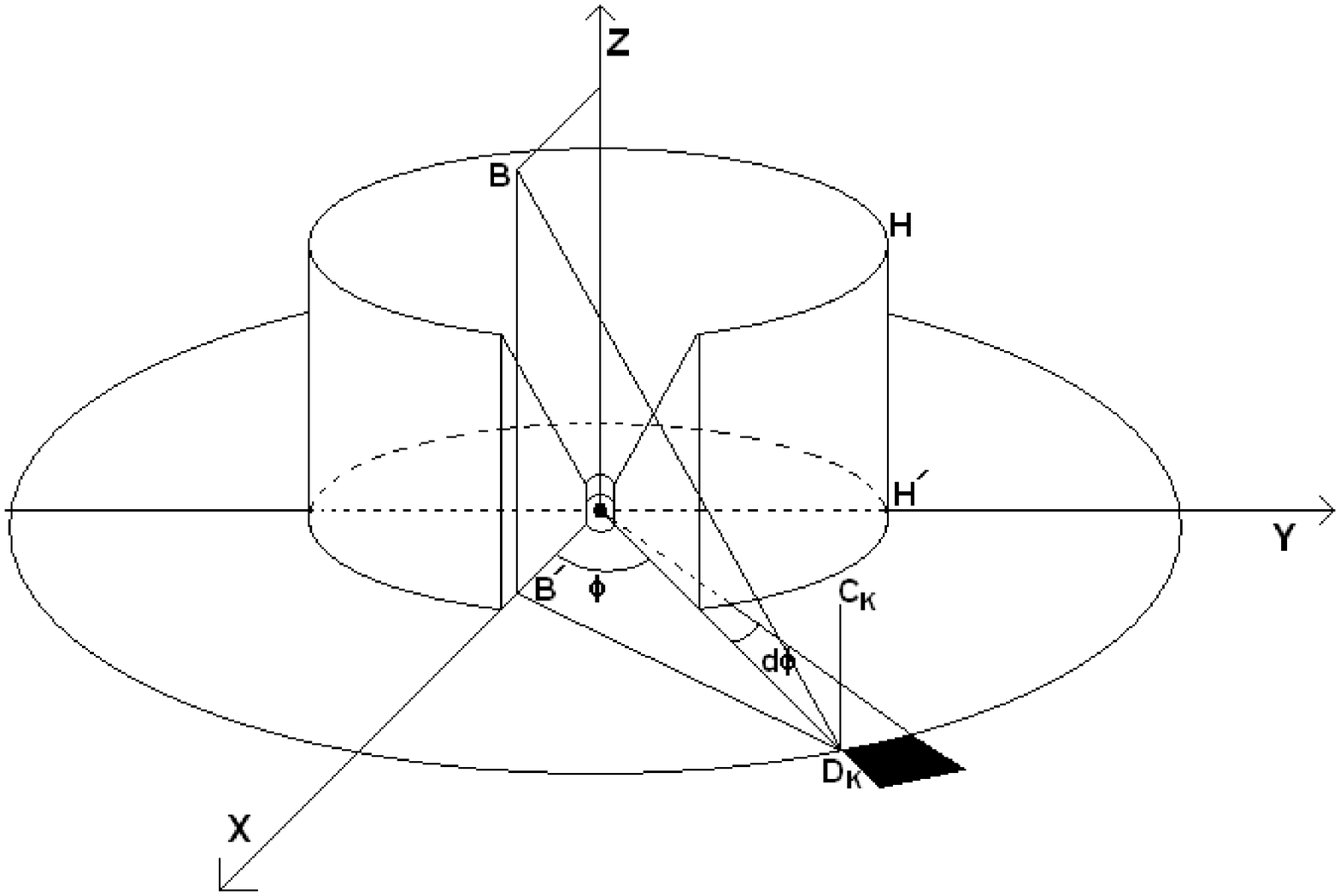}
\hskip -4.0cm (a)
\vskip -5.8cm
\hskip 7.5cm 
\includegraphics[scale=0.39]{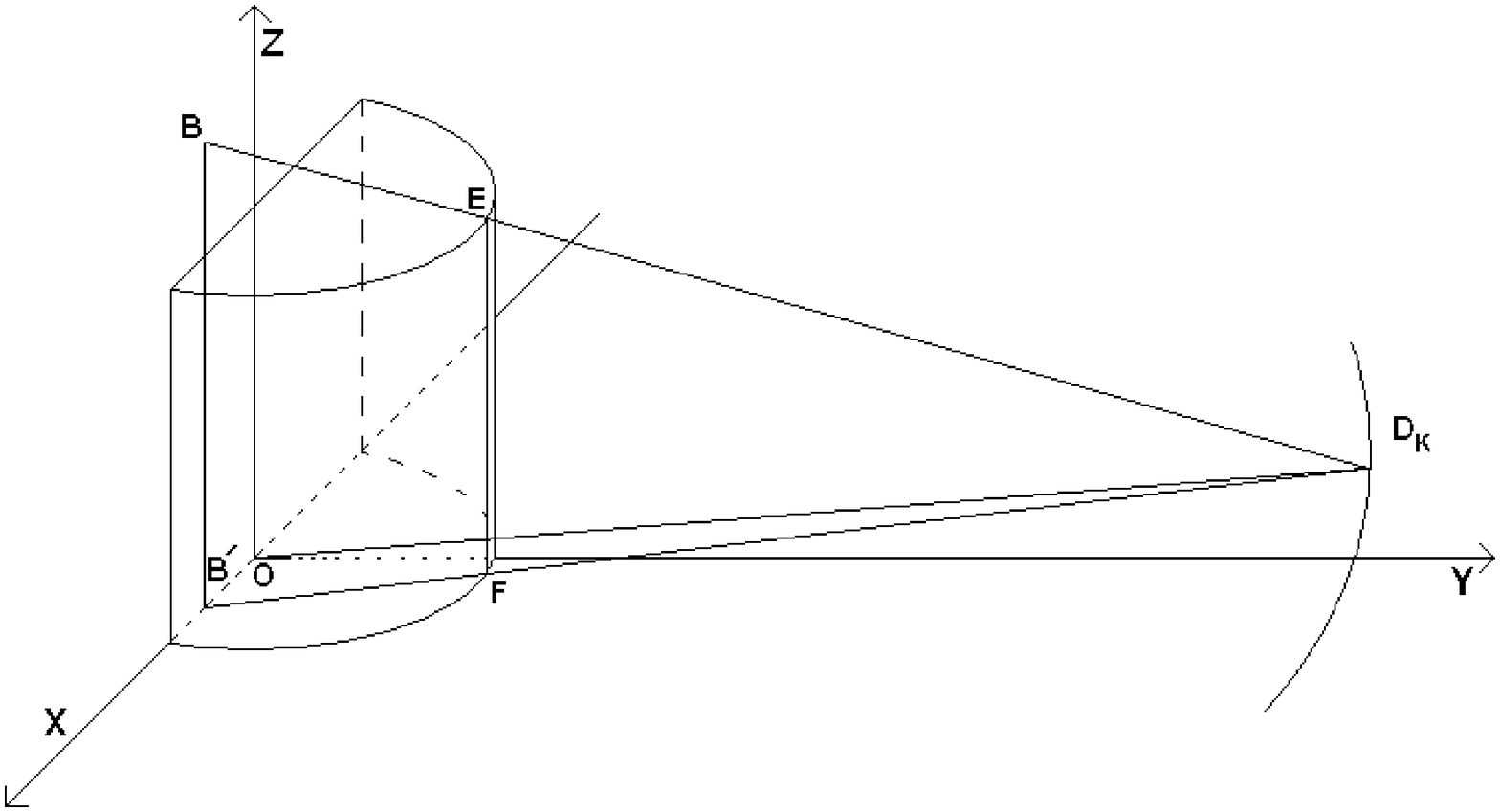}

\vskip -0.1cm \hskip 10.0cm (b)
\vskip 1.0cm
\caption[]{(a) Cartoon diagram of CENBOL. The outer Keplerian disc and 
sub-Keplerian halo are not shown. $B$($r$, $0$, $z$) is the field point
and $D$($x$,${\phi}$, $y$) is the source point on the inner surface
of the CENBOL. The inner surface of the CENBOL is assumed conical.
The position of the black hole is also shown. $ON=x_s$, and $HH^{\prime}=h_s$.
${\angle B^{\prime}OD^{\prime}}={\phi}$ \&  
${\angle D^{\prime}OD^{{\prime}{\prime}}}=d{\phi}$. And $d{\cal A}$
is the differential area.
(b) Cartoon diagram of CENBOL \& Keplerian disc (KD) geometry. The
SKH is not shown. $B$($r$, $0$, $z$) is the field point
and $D_K$($x_K$,${\phi}$, $0$) is the source point on the KD.
The CENBOL is represented by the cylinder.
The black spot represents $O$ the position of the black hole.
$OH^{\prime}=x_s$, and $HH^{\prime}=h_s$.
And $d{\cal A}_K$
is the differential area at $D_K$ (shaded).}
\end{figure}

In Fig. (4a-b), we represent a cartoon diagram of CENBOL and Keplerian
disc (KD). In Fig. (4a),
$D_K(x_K, {\phi}, 0)$ is the source point on the KD, and as in previous
figure, $B(r, 0, z)$ is field point where the various moment are computed.
In Fig. (4b), only one half of the CENBOL/KD geometry above the equatorial
plane is shown. $B$ is still the field point, but $D_K(x_K,{\phi}_f,0)$ is
the limit, up to which $B$ can see the annulus on KD defined by radius ($x_K$),
in the positive ${\phi}$ direction. So ${\phi}_f$ is the limit of
integration for ${\phi}$. 
From Fig. (4a), one can easily find out,
$$
d{\Omega}_K=\frac{d{\cal A}_K}{BD^2_K}{\rm cos}{\angle CDB}
=\frac{z{\hskip 0.1cm}x_K{\hskip 0.1cm}d{\phi}{\hskip 0.1cm}dx_K}{BD^3_K},
\eqno{(8g)}
$$

where,
$BD^2_K=r^2+z^2+x^2_K-2{\hskip 0.1cm}r{\hskip 0.1cm}x_K{\hskip 0.1cm}{\rm cos}{\phi}$, and the direction cosines are given by, \\
$l^r_K=(r-x_K{\hskip 0.1cm}{\rm cos}{\phi})/BD_K$; $l^{\phi}_K=-(x_K
{\hskip 0.1cm}{\rm sin}{\phi})/BD_K$;
and $l^z_K=z/BD_K$.

The frequency averaged KD intensity (see, NT73) is given by,
$$
I_K=\frac{I_{K0}}{(1+{\xi}_{red})^4}=\frac{(3GM_B{\dot M}_{\rm K})/(8{\pi}^2r^3_g)\left(x^{-3}_K-{\surd {3}}x^{-7/2}_K
\right)}{(1+{\xi}_{red})^4},
\eqno{(8h)}
$$

Now, the Doppler shift term is given by $1+{\xi}_{red}=1-{\varpi}_il^i_K$,
where,
$$
{\varpi}_il^i_K=-{\tilde u}_K{\hskip 0.1cm}{\rm sin}{\phi}\frac{r-x_K
{\hskip 0.1cm}{\rm cos}{\phi}}{BD_K}
-{\tilde u}_K{\hskip 0.1cm}{\rm cos}{\phi}\frac{x_K{\hskip 0.1cm}
{\rm sin}{\phi}}{BD_K}.
\eqno{(8i)}
$$

In the above equation ${\tilde u}_K={\sqrt {1/{\{}2(x_K-1){\}}}}$
--- the
Keplerian velocity around a non-rotating black hole (in geometrical units).

\noindent{\it Shadow effect of CENBOL on the jets}: 

As the jets are produced within the funnel like region of the TCAF disc,
up to certain $z$ the radiation from KD to the jet material, is blocked by the
presence of CENBOL. It is easy to find from Figs. (3-4), that
for material at a particular $r$ ($<x_s$, as jets are produced from the
CENBOL region), the radiations from KD is completely blocked
for $z{\leq}h_s(x_o-r)/(x_o-x_s)$. But even for
$z>h_s(x_o-r)/(x_o-x_s)$, the jet cannot `see' the entire Keplerian disc. 
From Fig. (4b), it is clear that the jet material at $B$ cannot see,
the whole of the annular area defined by radius $x_K$, as part of it
is blocked by the CENBOL. If for a particular $r$,
$h_s(x_o-r)/(x_o-x_s){\leq}z{\leq}h_s(x_K+r)/(x_K-x_s)$,
then, from Fig. (4b), one can find a expression
for ${\phi}_f$,
$$
{\phi}_f={\rm cos}^{-1}\left(\frac{x^2_K+x^2_s-FD^2_K}{2x_Kx_s} \right)
+{\rm cos}^{-1}\left(\frac{r^2+x^2_s-(B^{\prime}D_K-FD_K)^2}{2rx_s} \right),
\eqno{(8j)}
$$

\noindent where, $B^{\prime}D^2_K=r^2+x^2_K-2rx_K{\rm cos}{{\phi}_f}$,
and $FD_K=(h_sB^{\prime}D)/r$. ${\phi}_f$ is to be computed numerically
from Eq. (8j). For, $z>h_s(x_K+r)/(x_K-x_s)$,
${\phi}_f=2{\pi}$. Eqs. (8a-8c) are integrated, with the help of
Eqs. (8d-8j), to get the expressions of various radiative moments
of a TCAF disc.

Let us now multiply, ${\sigma}_T/m$, with Eqs. (8a-8c),
to get,

\begin{eqnarray*}
\hskip 2.5cm {\varepsilon} & = & {\varepsilon}_{K0}{\int}^{x_o}_{x_s}
[{\int}^{{\phi}_f}_0\frac{z(x^{-2}_K-{\sqrt {3}}x^{-5/2}_K)d{\phi}^{\prime}
}{(r^2+z^2+x^2_K-2rx_K{\rm cos}{\phi}^{\prime})^{3/2}
(1+{\xi}_{red})^4} \\ \nonumber
& + & {\int}^{2{\pi}}_{{\phi}_f}\frac{z(x^{-2}_K-{\sqrt {3}}x^{-5/2}_K)d{\phi}^{\prime}
}{(r^2+z^2+x^2_K-2rx_K{\rm cos}{\phi}^{\prime})^{3/2}
(1+{\xi}_{red})^4}]dx_K
\\ \nonumber
& + & {\varepsilon}_{C0}{\int}^{x_s}_{x_{in}}{\int}^{2{\pi}}_0
\frac{(x^2+y^2)^{1/2}{\rm cos}{\angle {CDB}}d{\phi}^{\prime}
dx}{(r^2+z^2+x^2-2rx{\rm cos}{\phi}^{\prime})^{3/2}(1+z_{red})^4} \\ \nonumber
& = & {\varepsilon}_{K0}{\tilde E}_K(r,z,x_s,x_o)
+{\varepsilon}_{C0}{\tilde E}_C(r,z,x_s,x_o) \\ \nonumber
& = & {\varepsilon}_K + {\varepsilon}_C
\hskip 10.2cm (9a)
\end{eqnarray*}

\begin{eqnarray*}
\hskip 2.5cm f^i & = & f_{K0}{\int}^{x_o}_{x_s}
[{\int}^{{\phi}_f}_0
\frac{z(x^{-2}_K-{\sqrt {3}}x^{-5/2}_K){\hskip 0.1cm}
l^i_Kd{\phi}^{\prime}}
{(r^2+z^2+x^2_K-2rx_K{\rm cos}{\phi}^{\prime})^{3/2}
(1+{\xi}_{red})^4} \\ \nonumber
& + & {\int}^{2{\pi}}_{{\phi}_f}
\frac{z(x^{-2}_K-{\sqrt {3}}x^{-5/2}_K){\hskip 0.1cm}
l^i_Kd{\phi}^{\prime}}
{(r^2+z^2+x^2_K-2rx_K{\rm cos}{\phi}^{\prime})^{3/2}
(1+{\xi}_{red})^4}]dx_K   \\ \nonumber
& + & f_{C0}{\int}^{x_s}_{x_{in}}{\int}^{2{\pi}}_0
\frac{(x^2+y^2)^{1/2}{\rm cos}{\angle {CDB}}l^id{\phi}^{\prime}
dx}{(r^2+z^2+x^2-2rx{\rm cos}{\phi}^{\prime})^{3/2}(1+z_{red})^4}
\\ \nonumber
& = & f_{K0}{\tilde F}^i_K(r,z,x_s,x_o)+f_{C0}{\tilde F}^i_C
(r,z,x_s,x_o) \\ \nonumber
& = & f^i_K+f^i_C
\hskip 10.2cm (9b)
\end{eqnarray*}

\begin{eqnarray*}
\hskip 2.5cm {\wp}^{ij} & = & {\wp}_{K0}{\int}^{x_o}_{x_s}
[{\int}^{{\phi}_f}_0
\frac{z(x^{-2}_K-{\sqrt {3}}x^{-5/2}_K){\hskip 0.1cm}
l^i_K{\hskip 0.1cm}l^j_Kd{\phi}^{\prime}}
{(r^2+z^2+x^2_K-2rx_K{\rm cos}{\phi}^{\prime})^{3/2}
(1+{\xi}_{red})^4} \\ \nonumber
& + & {\int}^{2{\pi}}_{{\phi}_f}
\frac{z(x^{-2}_K-{\sqrt {3}}x^{-5/2}_K){\hskip 0.1cm}
l^i_K{\hskip 0.1cm}l^j_Kd{\phi}^{\prime}}
{(r^2+z^2+x^2_K-2rx_K{\rm cos}{\phi}^{\prime})^{3/2}
(1+{\xi}_{red})^4}] dx_K\\ \nonumber
& + & {\wp}_{C0}{\int}^{x_s}_{x_{in}}{\int}^{2{\pi}}_0
\frac{(x^2+y^2)^{1/2}{\rm cos}{\angle {CDB}}{\hskip 0.1cm}
l^i{\hskip 0.1cm}l^jd{\phi}^{\prime}
dx}{(r^2+z^2+x^2-2rx{\rm cos}{\phi}^{\prime})^{3/2}(1+z_{red})^4}
\\ \nonumber
& = & {\wp}_{K0}{\tilde P}^{ij}_K(r,z,x_s,x_o)+{\wp}_{CO}
{\tilde P}^{ij}_C(r,z,x_s,x_o) \\ \nonumber
& = & {\wp}^{ij}_K+{\wp}^{ij}_C
\hskip 10.0cm (9c)
\end{eqnarray*}

In Eqs. (9a-9c), the space-dependent part of $\varepsilon$,
$ f^i $, and ${\wp}^{ij}$ are expressed as ${\tilde E}$,
${\tilde F}^i$ and ${\tilde P}^{ij}$. Suffix `$K$' and `$C$'
signify Keplerian and CENBOL contributions.
If the moments are expressed in dimensionless units then
the constants in Eqs. (9a-9c), are given by,

$$
{\varepsilon}_{C0}={f}_{C0}
={\wp}_{C0}
=\frac{1.3{\times}10^{38}{\ell}_c{\sigma}_{T}}{2{\pi}cm{\cal A}GM_{\odot}}
\eqno{(9d)}
$$
and
$$
{\varepsilon}_{KO}={f}_{K0}
={\wp}_{K0}
=\frac{4.32{\times}10^{17}{\dot m_k}{\sigma}_Tc}{32{\pi}^2mGM_{\odot}},
\eqno{(9e)}
$$

where, ${\ell}_c=L_{\rm C}/L_{\rm Edd}$ ($L_{\rm C}$ --- CENBOL luminosity \&
$L_{\rm Edd}$ --- Eddington luminosity), ${\dot m}_k={\dot M}_{\rm K}/
{\dot M}_{\rm Edd}$ (${\dot M}_{\rm K}$ --- Keplerian accretion rate,
${\dot M}_{\rm Edd}$ --- Eddington Accretion rate), and
${\cal A}$ is the CENBOL surface.

For simplicity, we will not compute the shock location  $x_s$ or the
the CENBOL luminosity ($L_{\rm C}$) -- instead, we will supply them as free
parameters. They can be easily computed from accretion parameters 
\citep{b2,b3,b4,b16,b13}.\\
%(\eg C89, CT95, Das \etal 2001, Chattopadhyay \etal 2003).
To obtained all the components of radiation field, we supply the following disc-parameters, \\
(a) the inner radius of the CENBOL $x_{in}(=2r_g,
\mbox{ as explained in \S 3.1 })$,
(b) the shock location $x_s$,
(c) the CENBOL luminosity ${\ell}_c$ (in units of $L_{\rm Edd}$), 
(d) the Keplerian accretion rate 
${\dot m}_k$.

The expression of Keplerian luminosity was given in \citet{b14}, and
is;
$$
L_{K}=r^2_g{\int}^{x_o}_{x_s}2{\pi}I_{K0}2{\pi}x_{K}dx_{K}
=\frac{3}{4}{\dot m}_k\left [-\frac{1}{x_{K}}+\frac{2}{3x_{K}}
{\sqrt{\frac{3}{x_{K}}}} \right]^{x_o}_{x_s}L_{\rm Edd}={\ell}_kL_{\rm Edd}
\eqno{(10)}
$$

It is to be remembered, that as jets are only observed in intermediate
to hard spectral states of the accretion disc, so we will
constrain our analysis in the domain ${\ell}_k/{\ell}_c{\sim}1$
to ${\ell}_k/{\ell}_c<1$.

\subsection{The components of radiation field}

\begin{figure}
\includegraphics[scale=0.32]{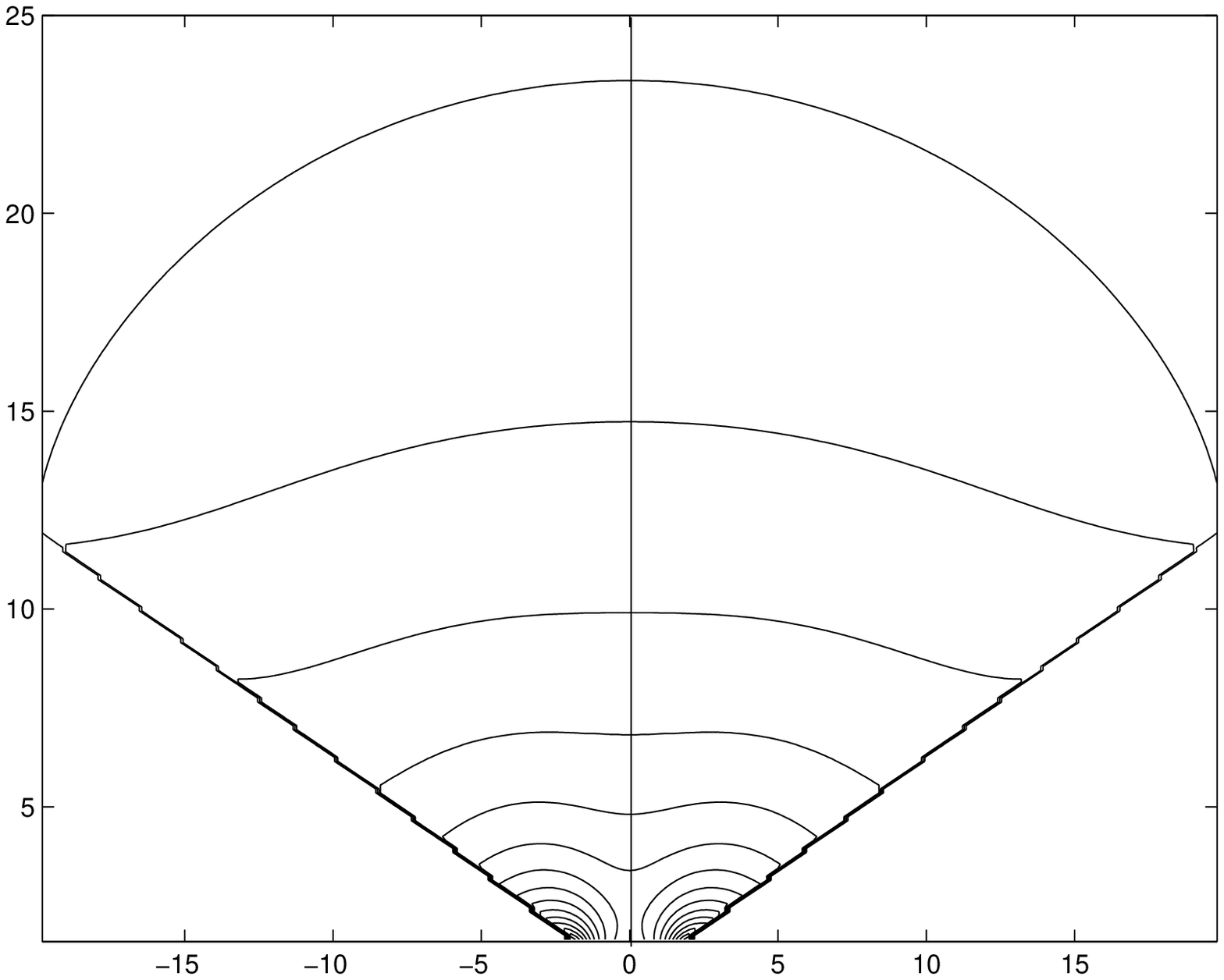}
\includegraphics[scale=0.32]{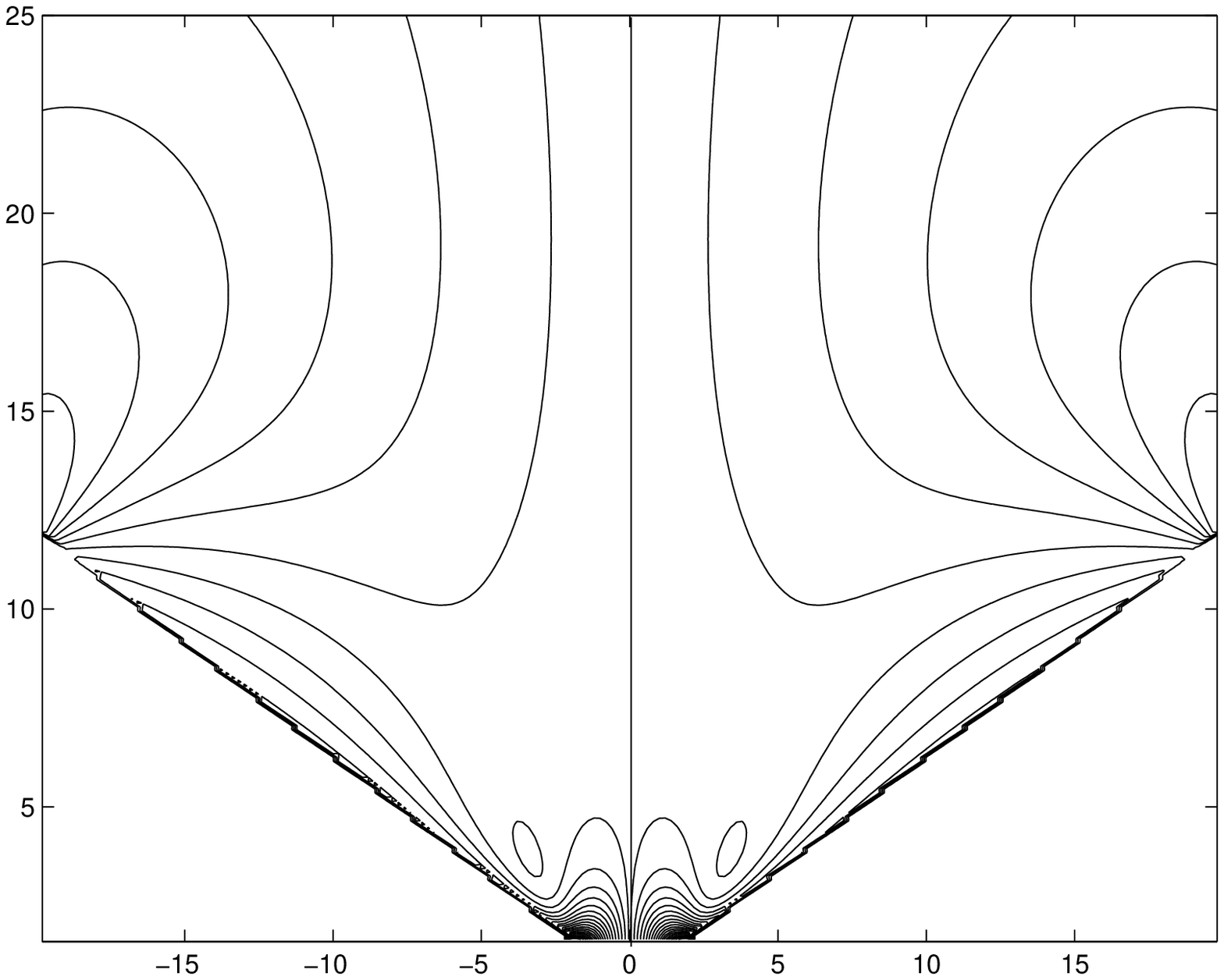}
\includegraphics[scale=0.32]{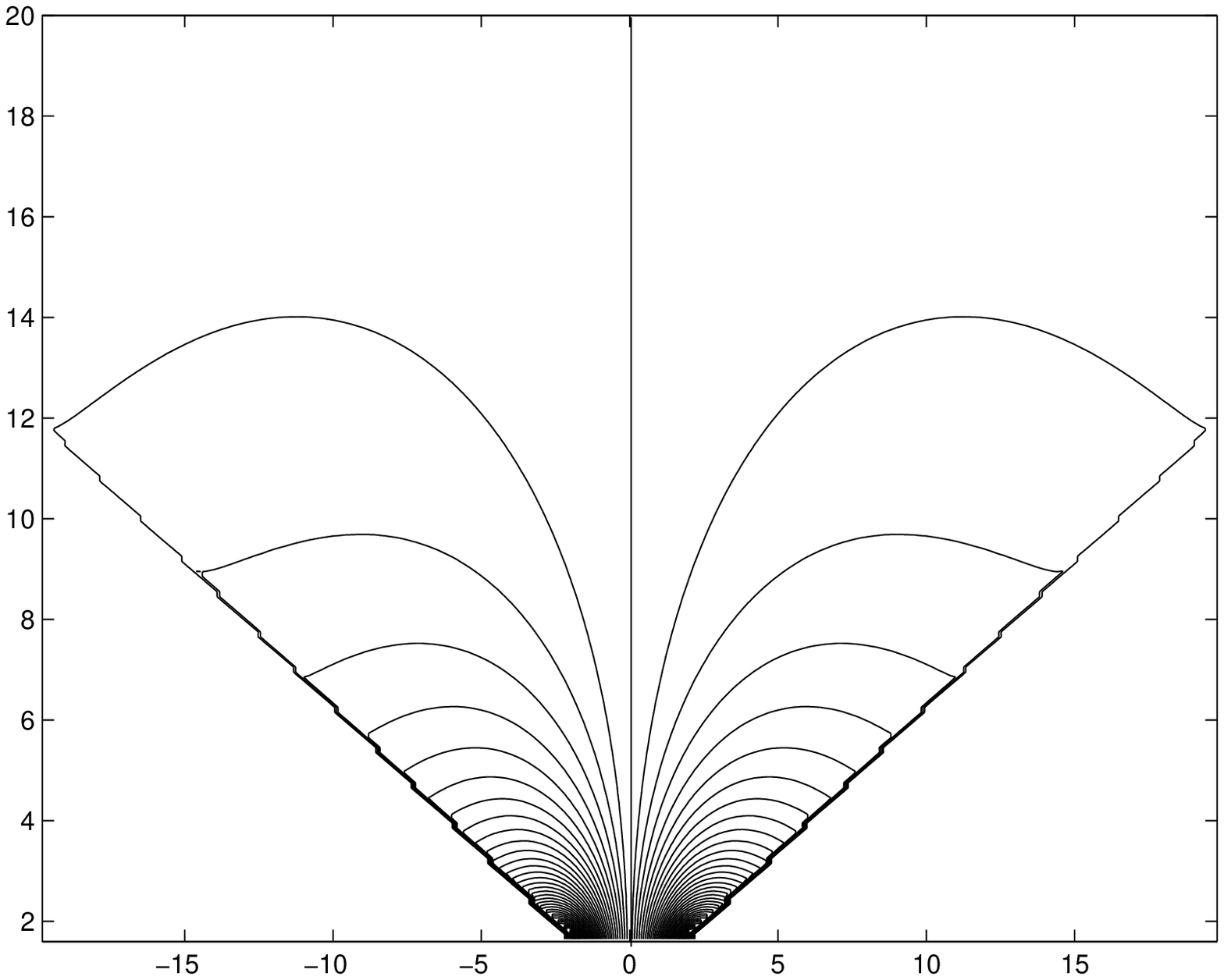}
\hskip -1.0cm (a)  \hskip 6.0cm (b) \hskip 6.0cm (c)  
\vskip 0.5cm
\includegraphics[scale=0.32]{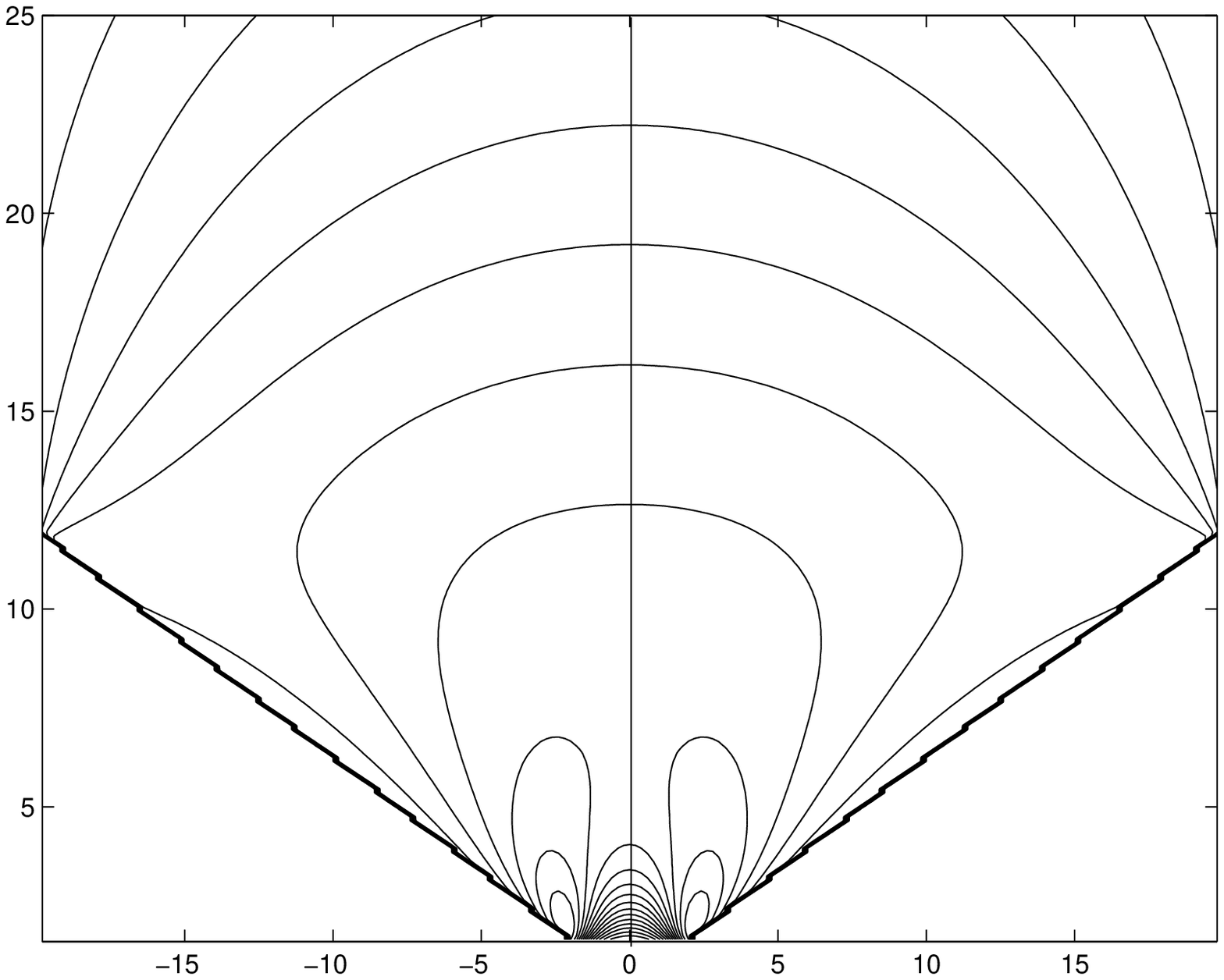}
\includegraphics[scale=0.32]{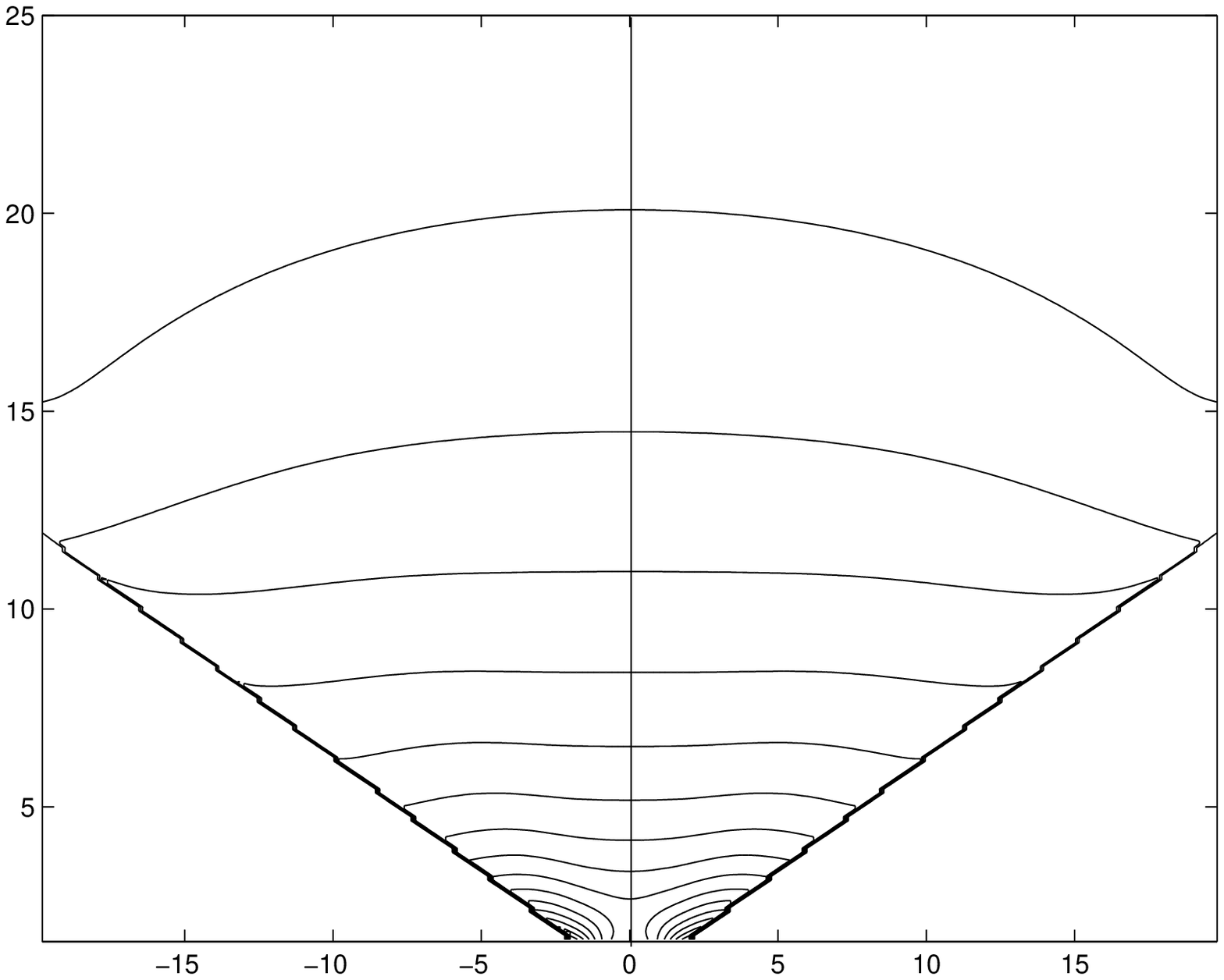}
\includegraphics[scale=0.32]{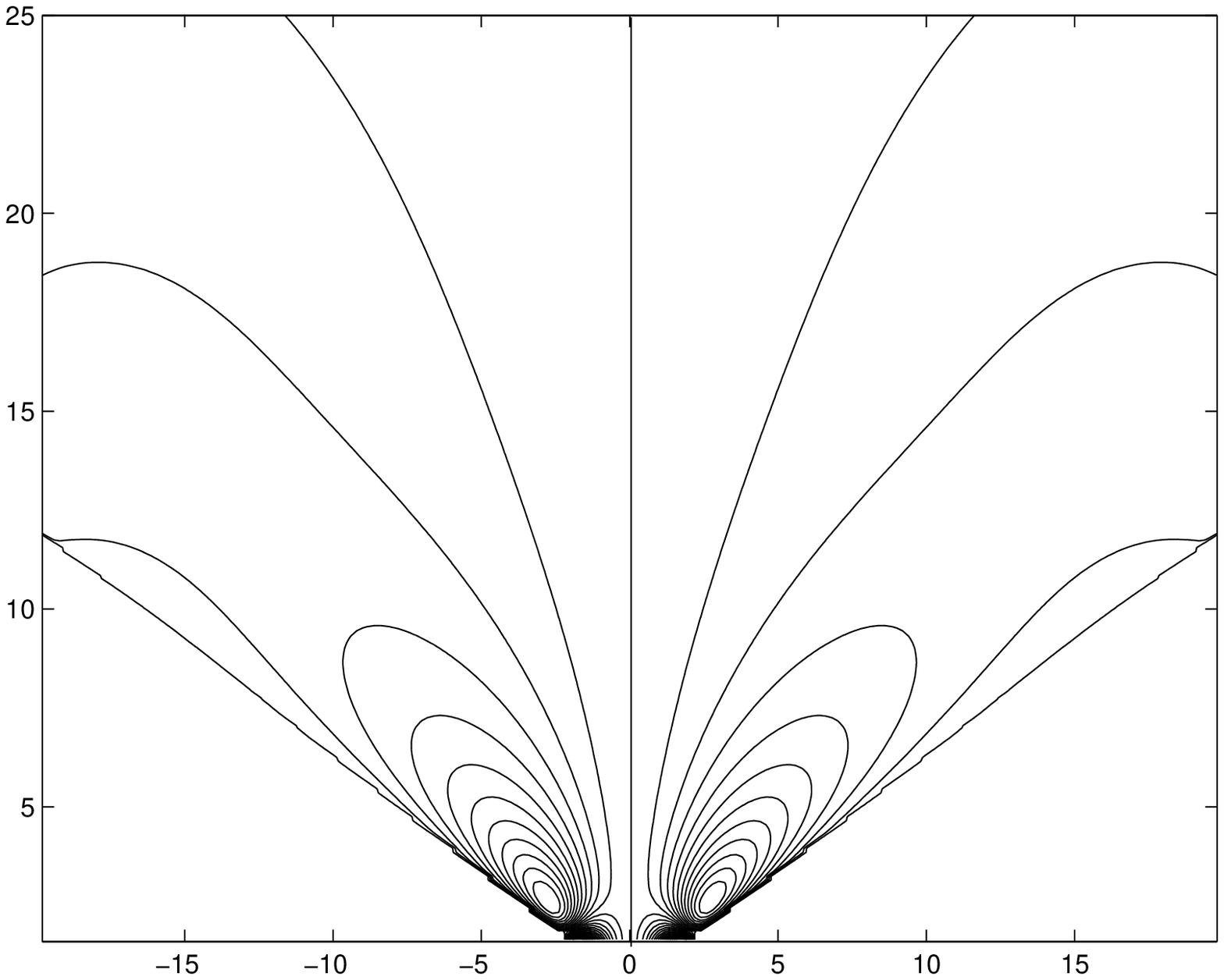}
\hskip -1.0cm (d) \hskip 6.0cm (e)  \hskip 5.0cm (f)
\vskip 0.5cm
\includegraphics[scale=0.32]{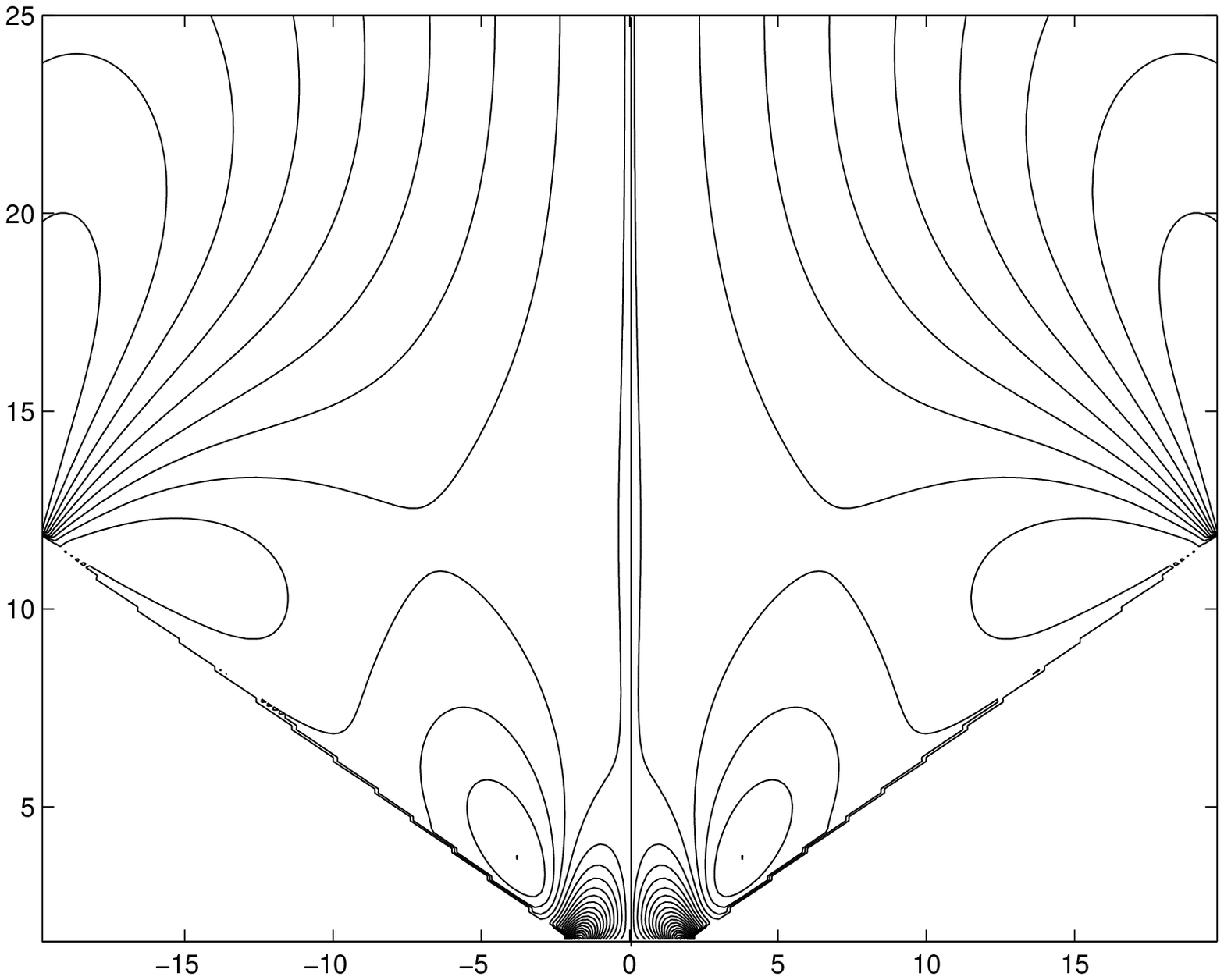}
\includegraphics[scale=0.32]{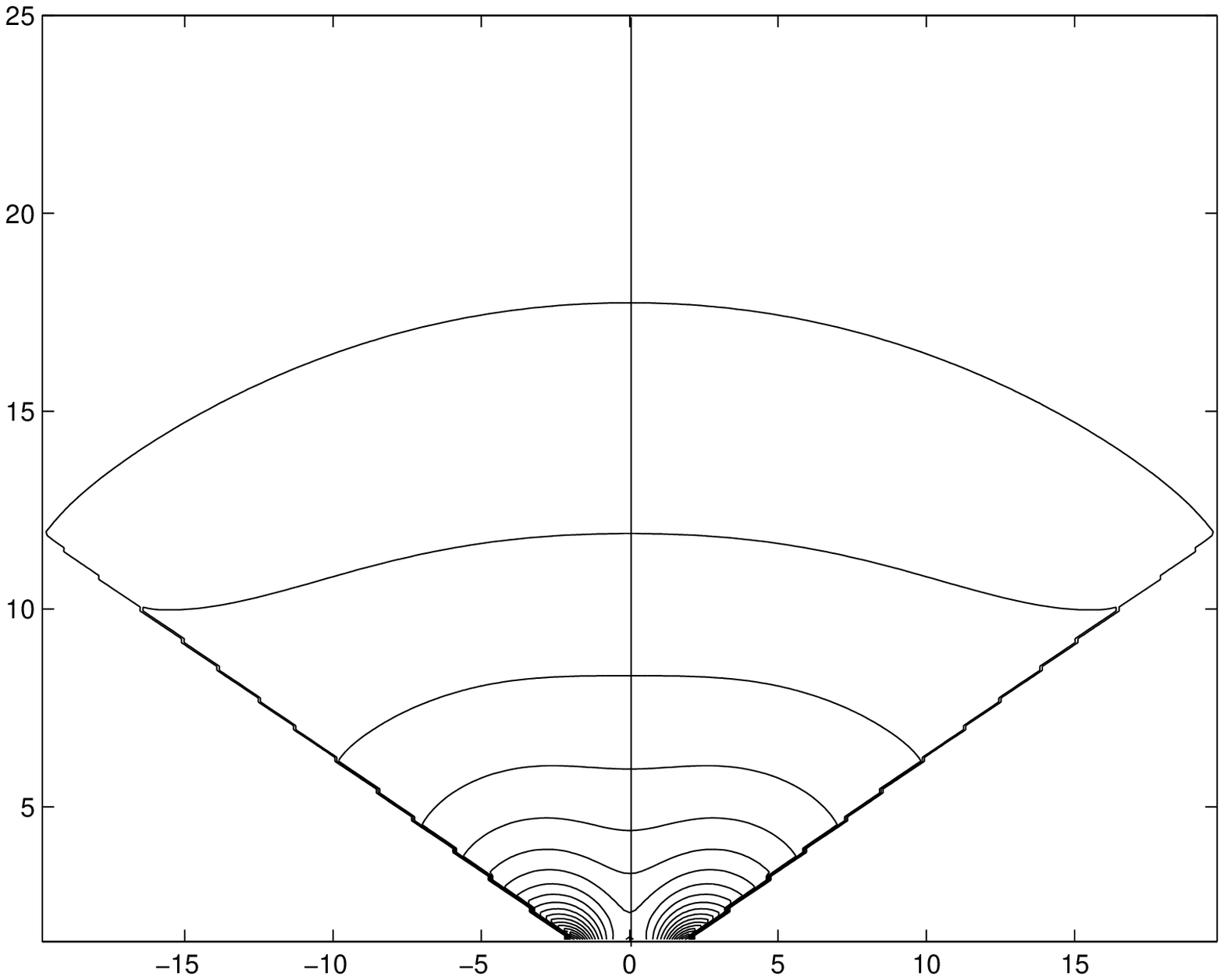}
\includegraphics[scale=0.32]{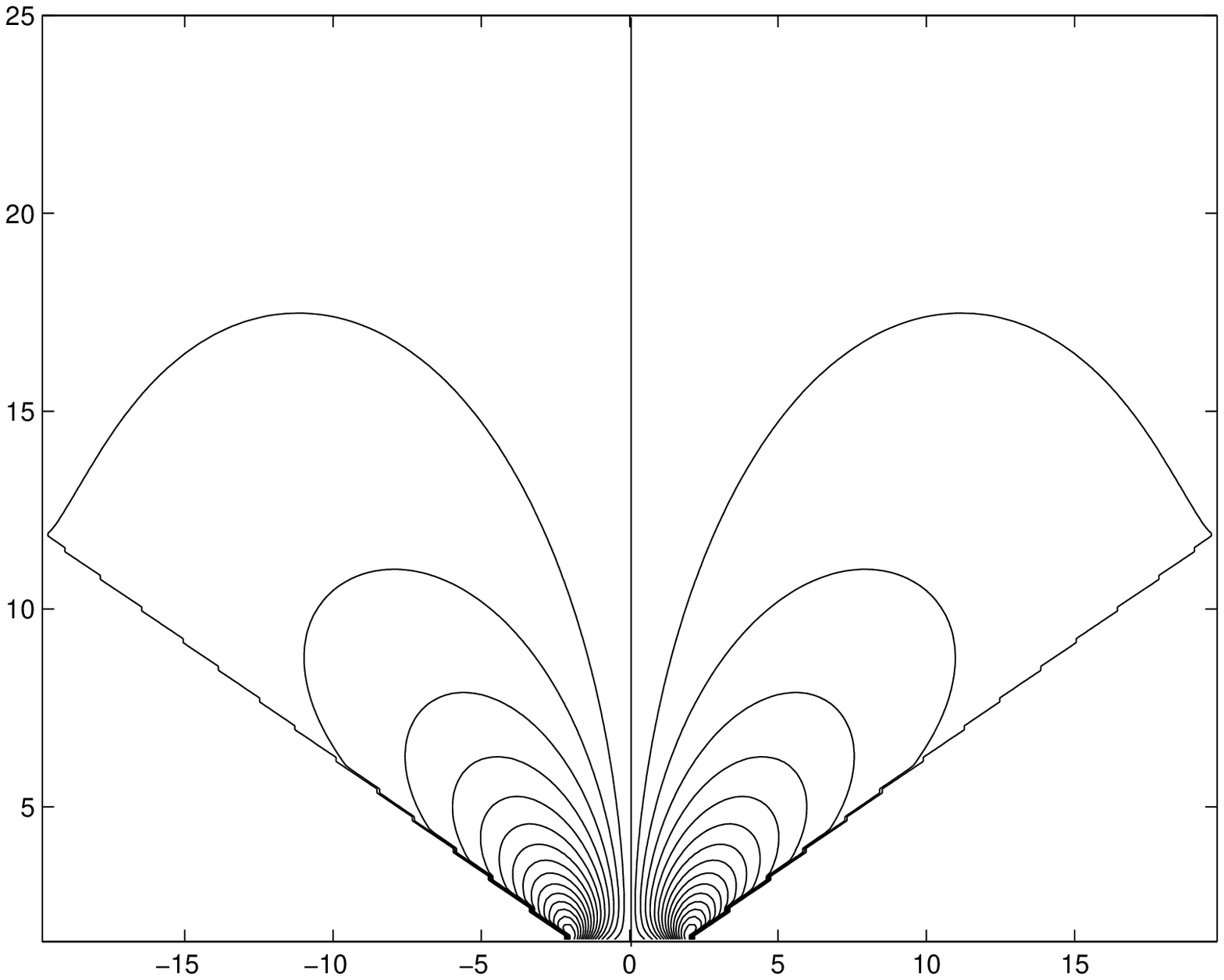}
\hskip -1.0cm (g)  \hskip 6.0cm (h)  \hskip 5.0cm (i)
\vskip 0.5cm
\includegraphics[scale=0.32]{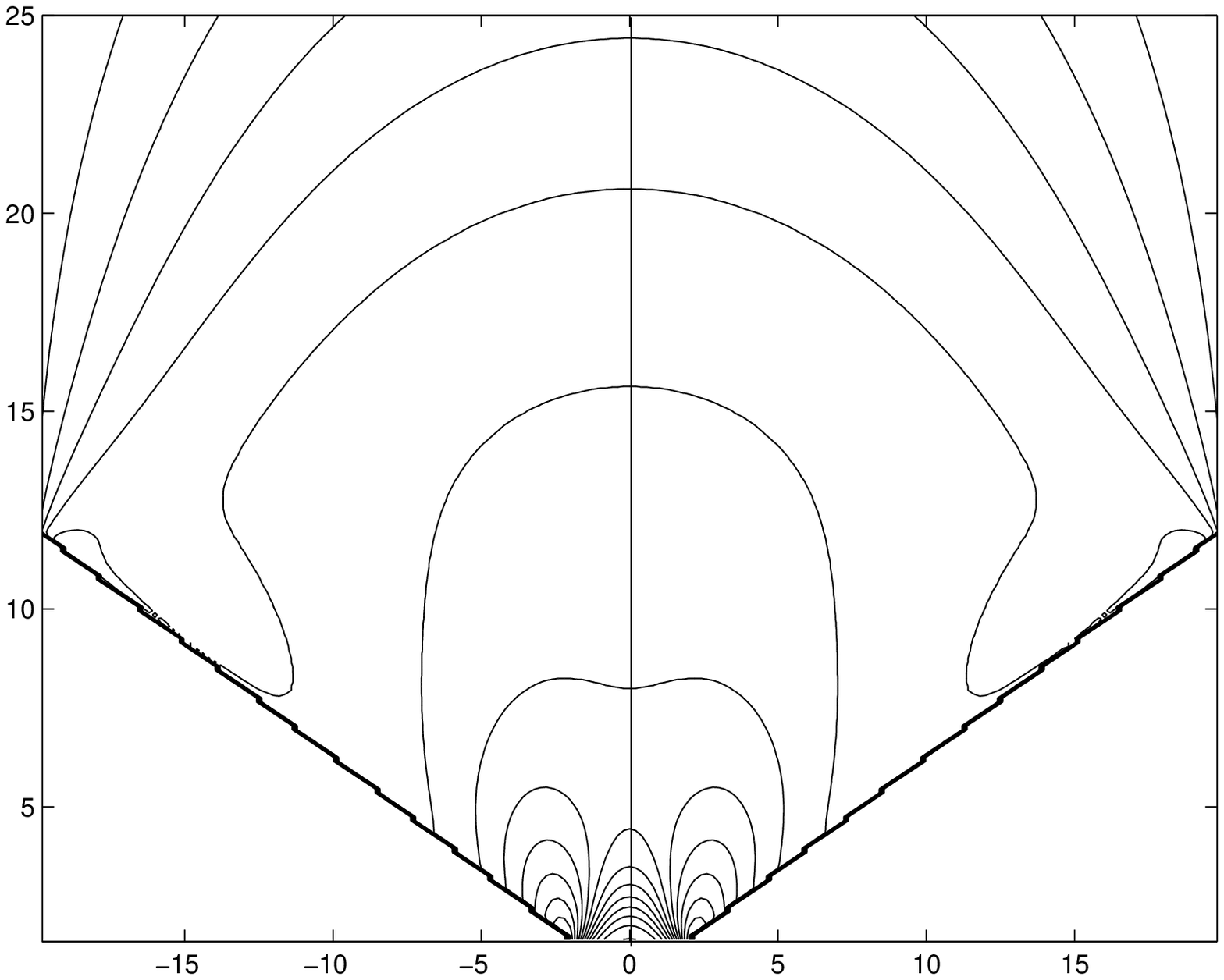}
\vskip 0.1cm \hskip -0.1cm (j)
\vskip 0.5cm 
\caption[]{The space-dependent part of the ten independent components of
radiation field due to the CENBOL. (a) ${\tilde E}_C(r,z)$,
(b) ${\tilde F}^r_C(r,z)$,
(c) ${\tilde F}^{\phi}_C(r,z)$, (d) ${\tilde F}^z_C(r,z)$,
(e) ${\tilde P}^{rr}_C(r,z)$, (f) ${\tilde P}^{r{\phi}}_C(r,z)$, (g) ${\tilde P}^{rz}_C(r,z)$,
(h) ${\tilde P}^{{\phi}{\phi}}_C(r,z)$,
(i) ${\tilde P}^{{\phi}z}_C(r,z)$, (j) ${\tilde P}^{zz}_C(r,z)$.
The disc parameters are $x_s=20r_g$ \& $x_{in}=2r_g$. }
\end{figure}

In Fig. (5), we show the contour plots of  the space-dependent
part of 
various moments of radiation 
due to the CENBOL. The shock location is $x_s=20r_g$.
The space variation of various moments due to the CENBOL
is strongest within the funnel like region of the disc,
so we only plot them within and above the funnel like region.
Further as the jets are likely to be produced in this region, so
this is the region that matters for our purpose.
Contour plots of various moments are ${\tilde E}_C(r,z)$
[Fig (5a); max. value: 50.55], ${\tilde F}^r_C(r,z)$ [Fig. (5b);
max/min value: 1.7/-5.42], ${\tilde F}^{\phi}_C(r,z)$
[Fig. (5c); max. value: 28.94], ${\tilde F}^z_C(r,z)$ [Fig. (5d);
max/min values: 5.2/-1.19], 
${\tilde P}^{rr}_C(r,z)$ [Fig. (5e); max. value: 14.62],
${\tilde P}^{r{\phi}}_C(r,z)$ [Fig. (5f); max/min values: 0.98/-1.66],
${\tilde P}^{rz}_C(r,z)$ [Fig. (5g); max/min values: 0.67/-1.39],
${\tilde P}^{{\phi}{\phi}}_C(r,z)$ [Fig. (5h); max value: 29.8],
${\tilde P}^{{\phi}z}_C(r,z)$ [Fig. (5i); max value: 4.4] and
${\tilde P}^{zz}_C(r,z)$ [Fig. (5j); max value: 3.88].

\begin{figure}
\includegraphics[scale=0.32]{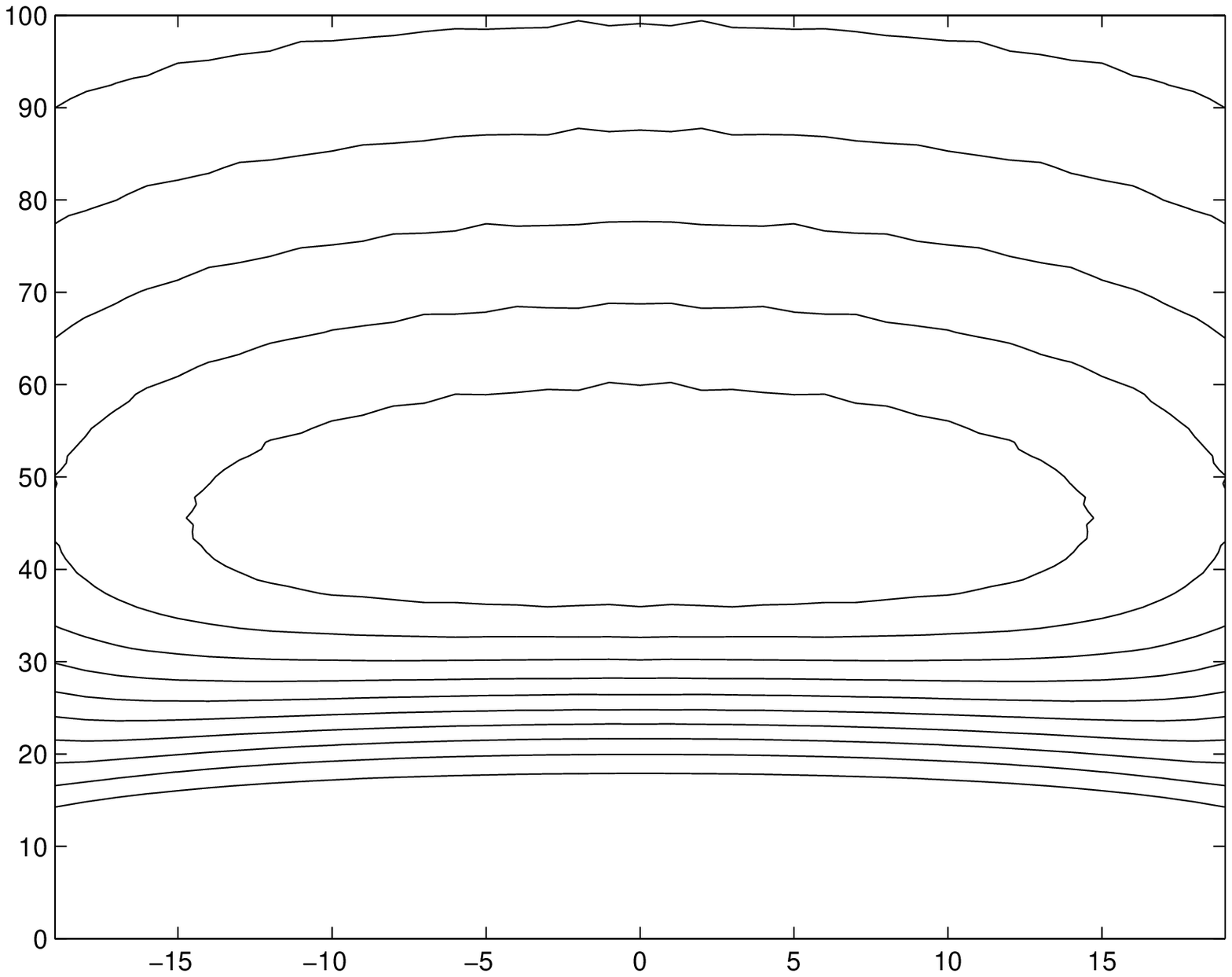}
\includegraphics[scale=0.32]{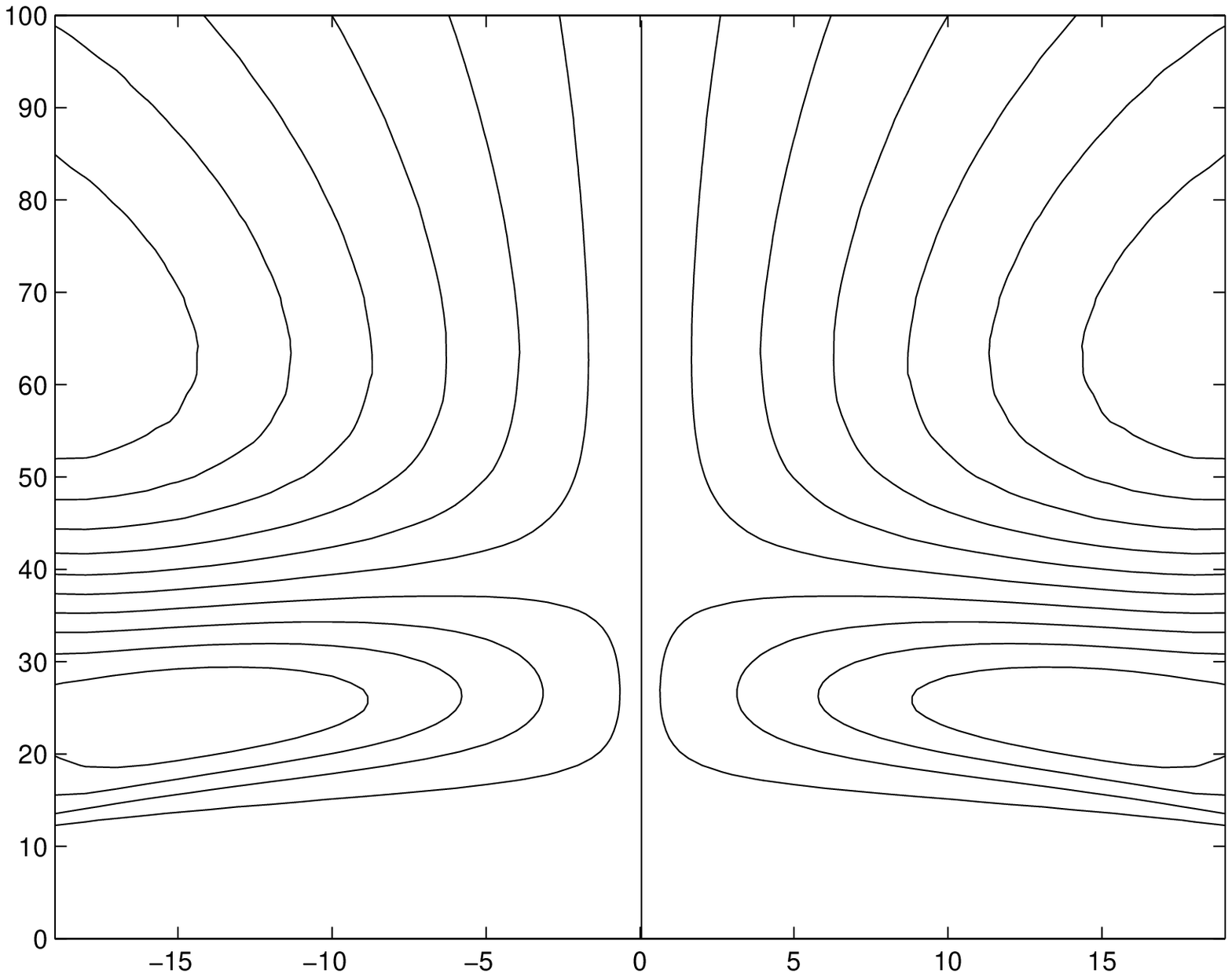}
\includegraphics[scale=0.32]{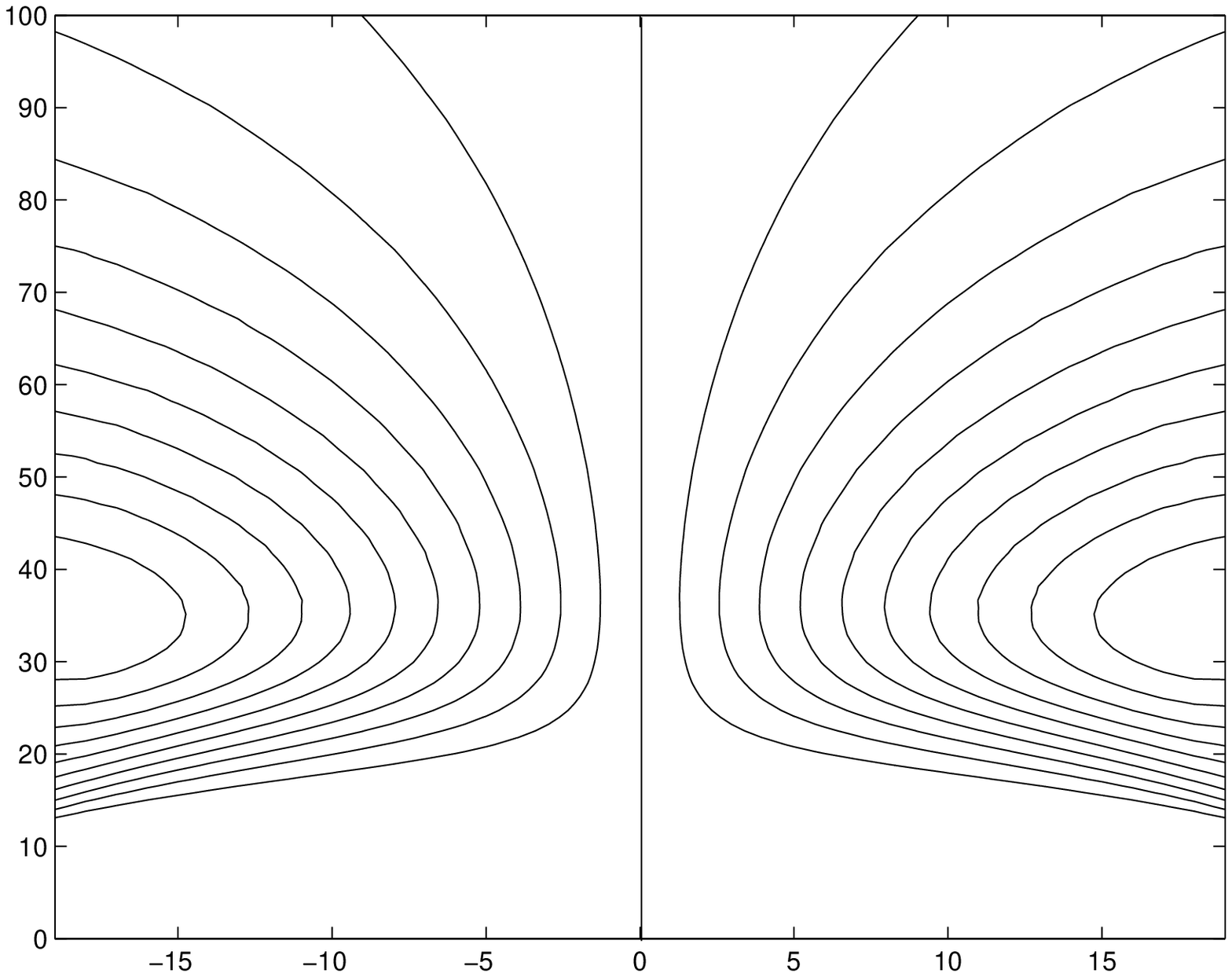}
\hskip -1.0cm (a)  \hskip 6.0cm (b) \hskip 6.0cm (c)  
\vskip 0.5cm
\includegraphics[scale=0.32]{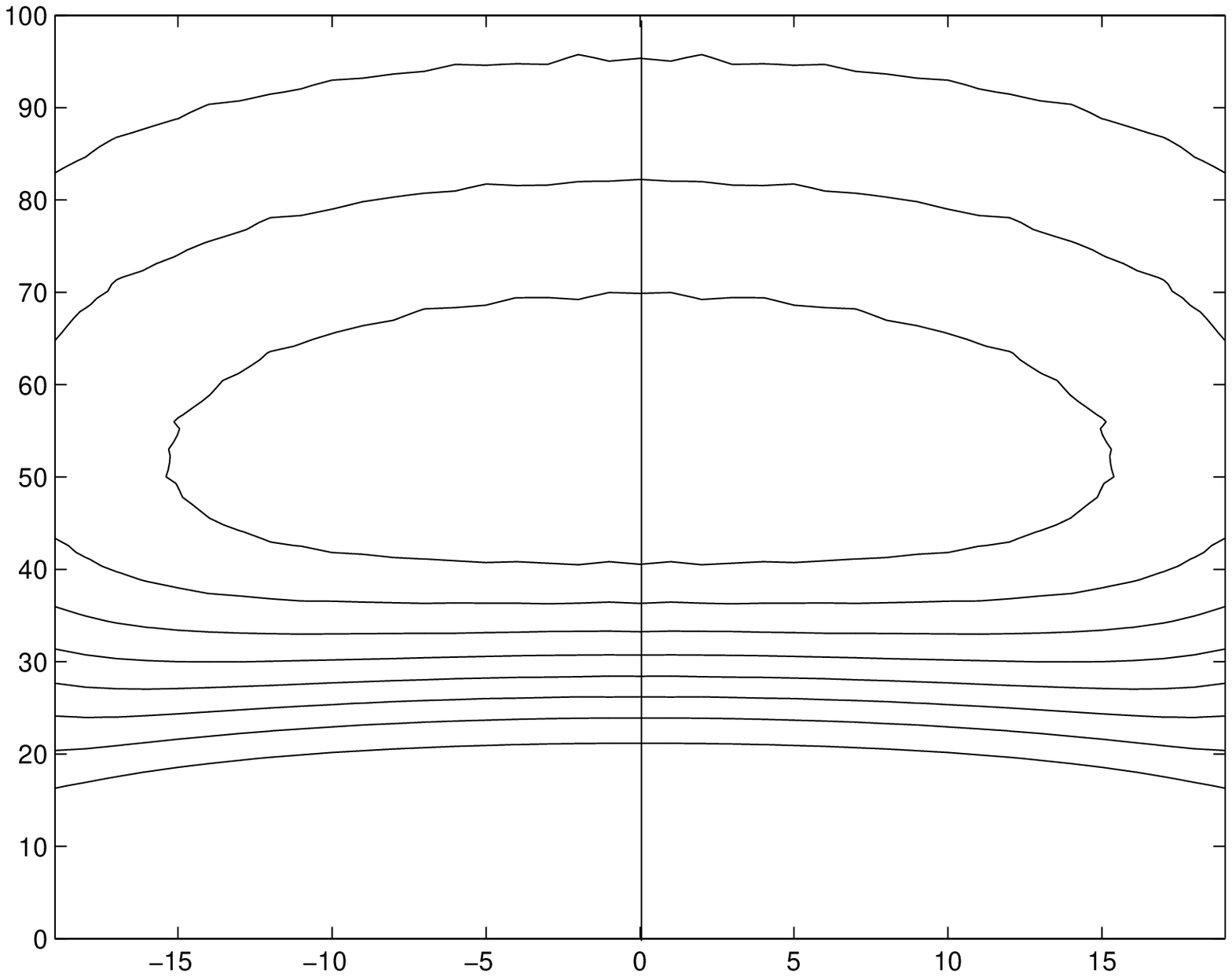}
\includegraphics[scale=0.32]{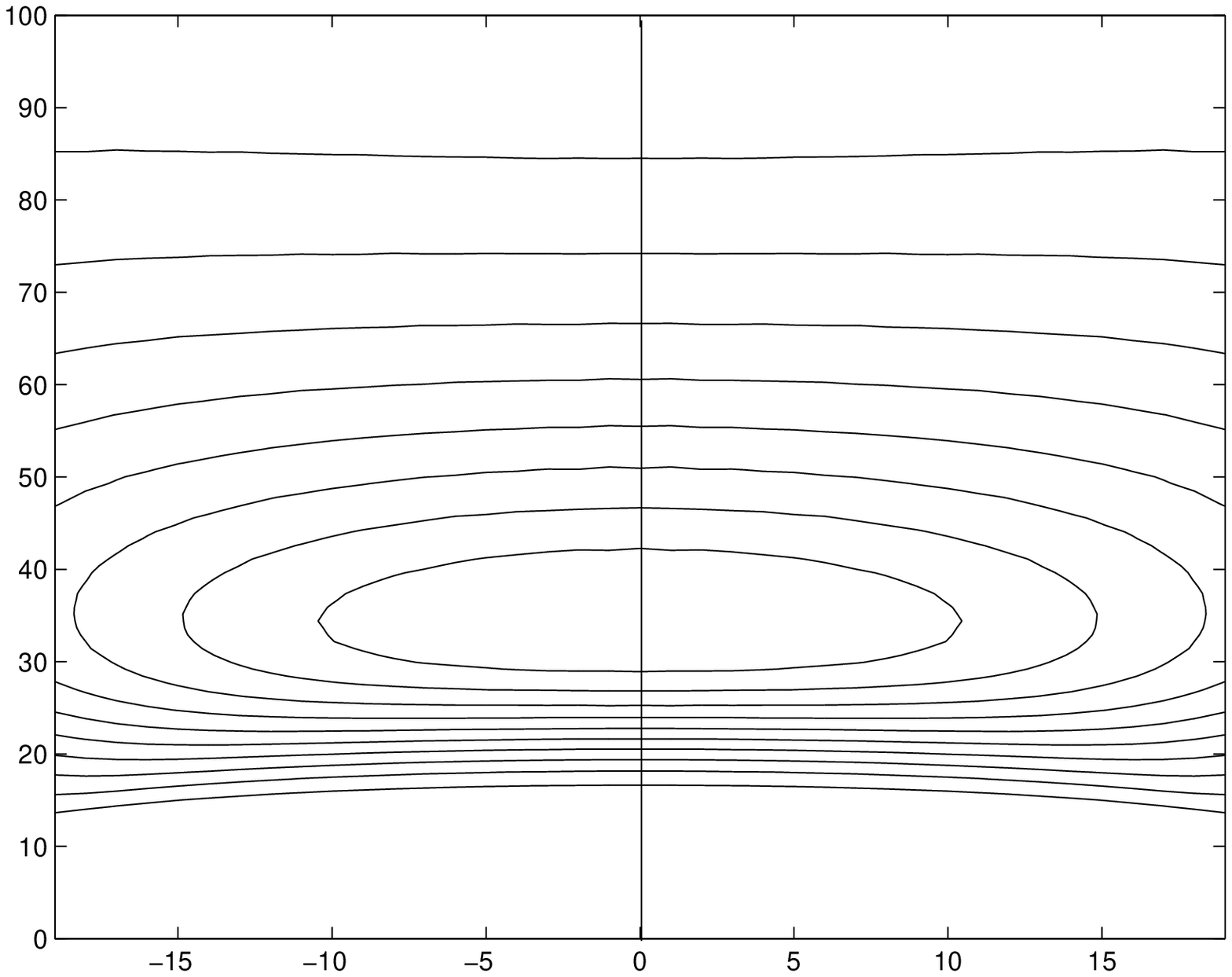}
\includegraphics[scale=0.32]{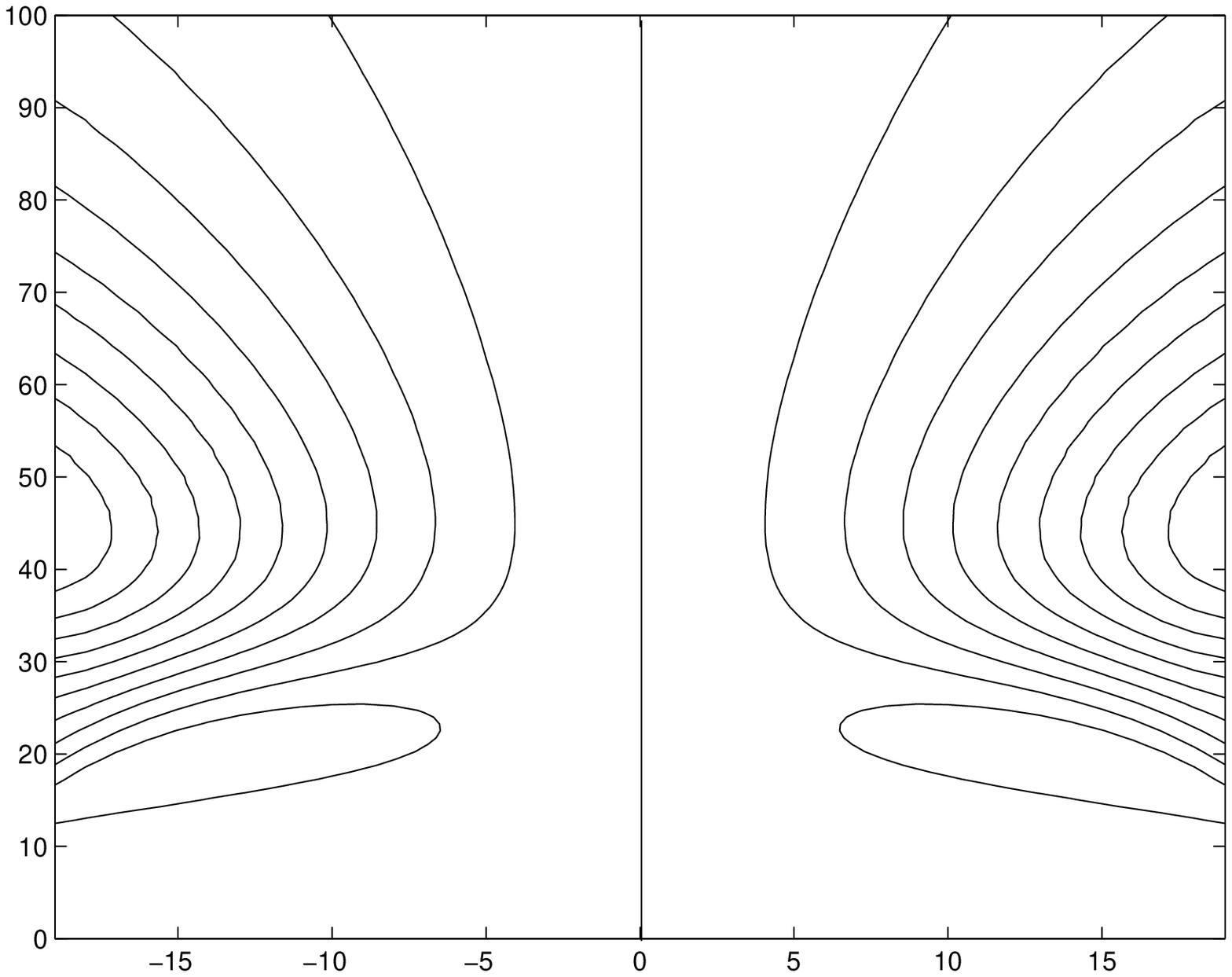}
\hskip -1.0cm (d) \hskip 6.0cm (e)  \hskip 5.0cm (f)
\vskip 0.5cm
\includegraphics[scale=0.32]{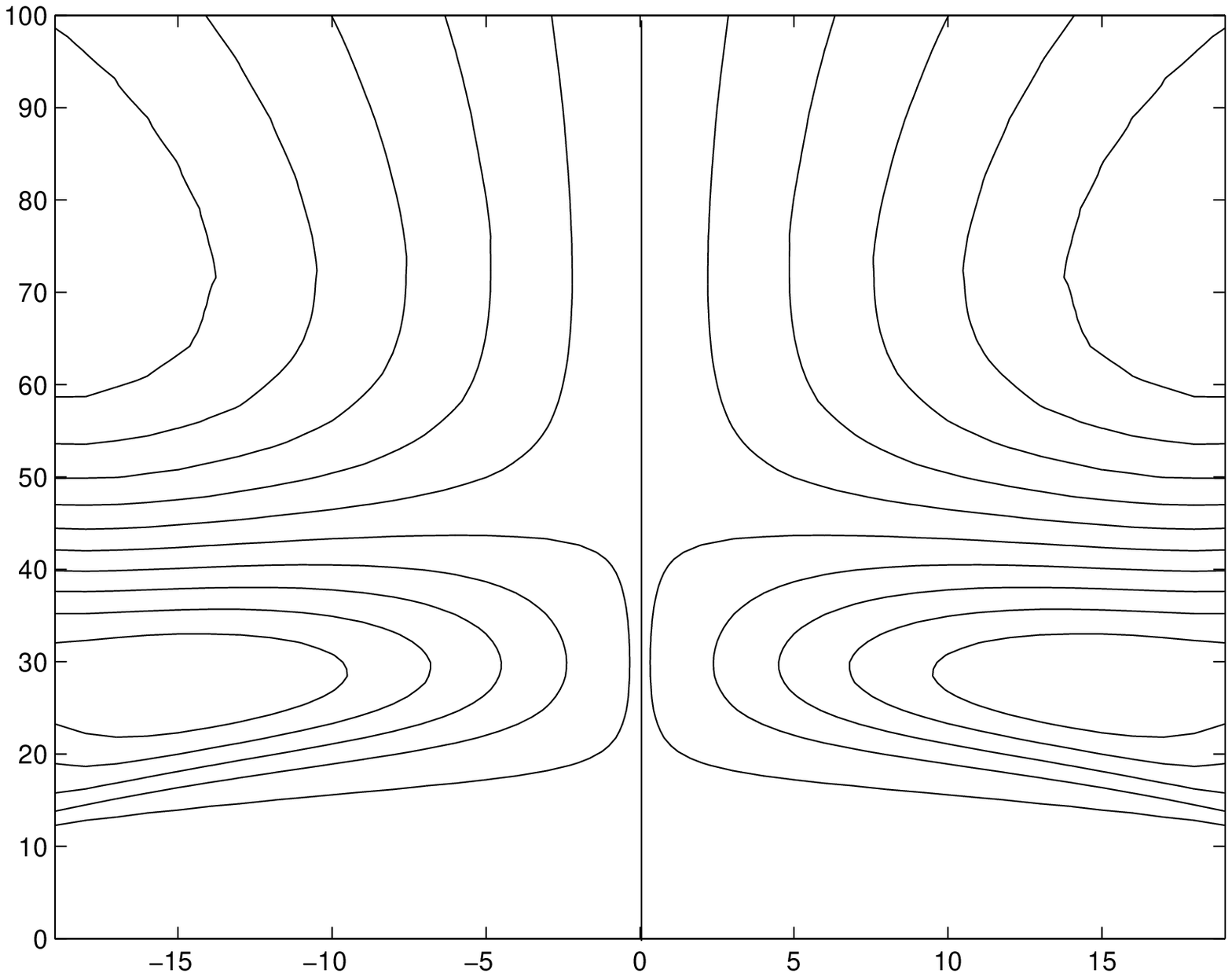}
\includegraphics[scale=0.32]{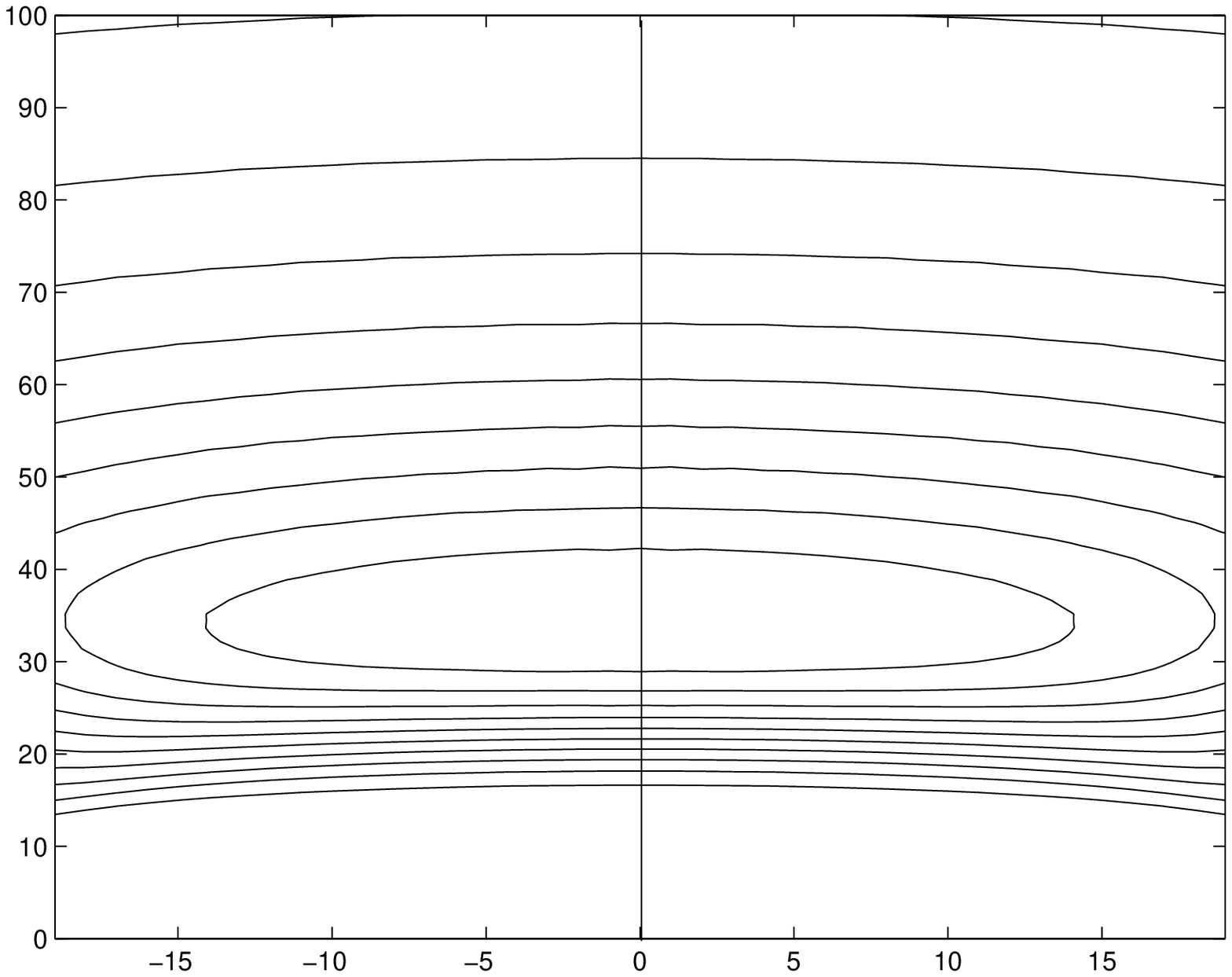}
\includegraphics[scale=0.32]{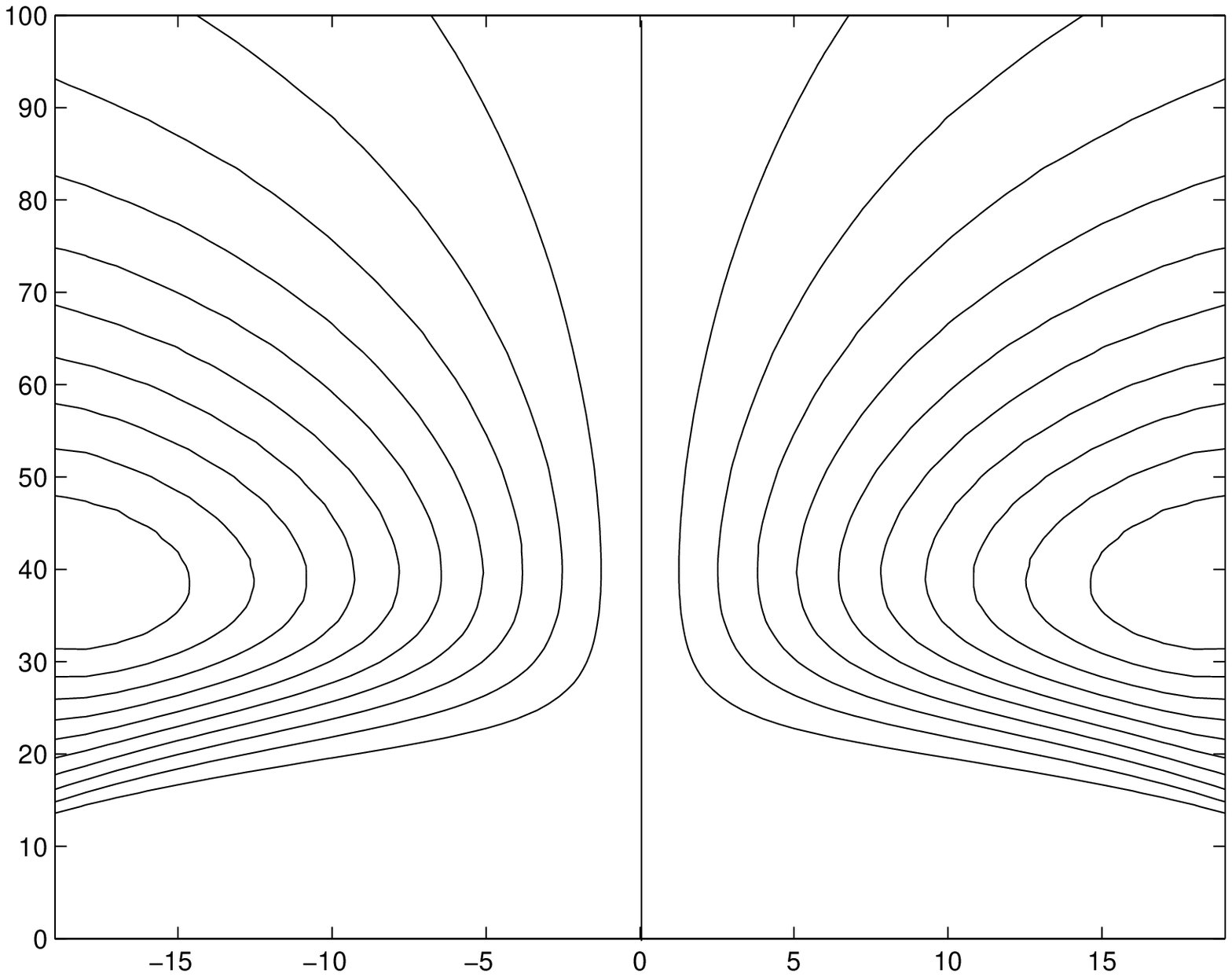}
\hskip -1.0cm (g)  \hskip 6.0cm (h)  \hskip 5.0cm (i)
\vskip 0.5cm
\includegraphics[scale=0.32]{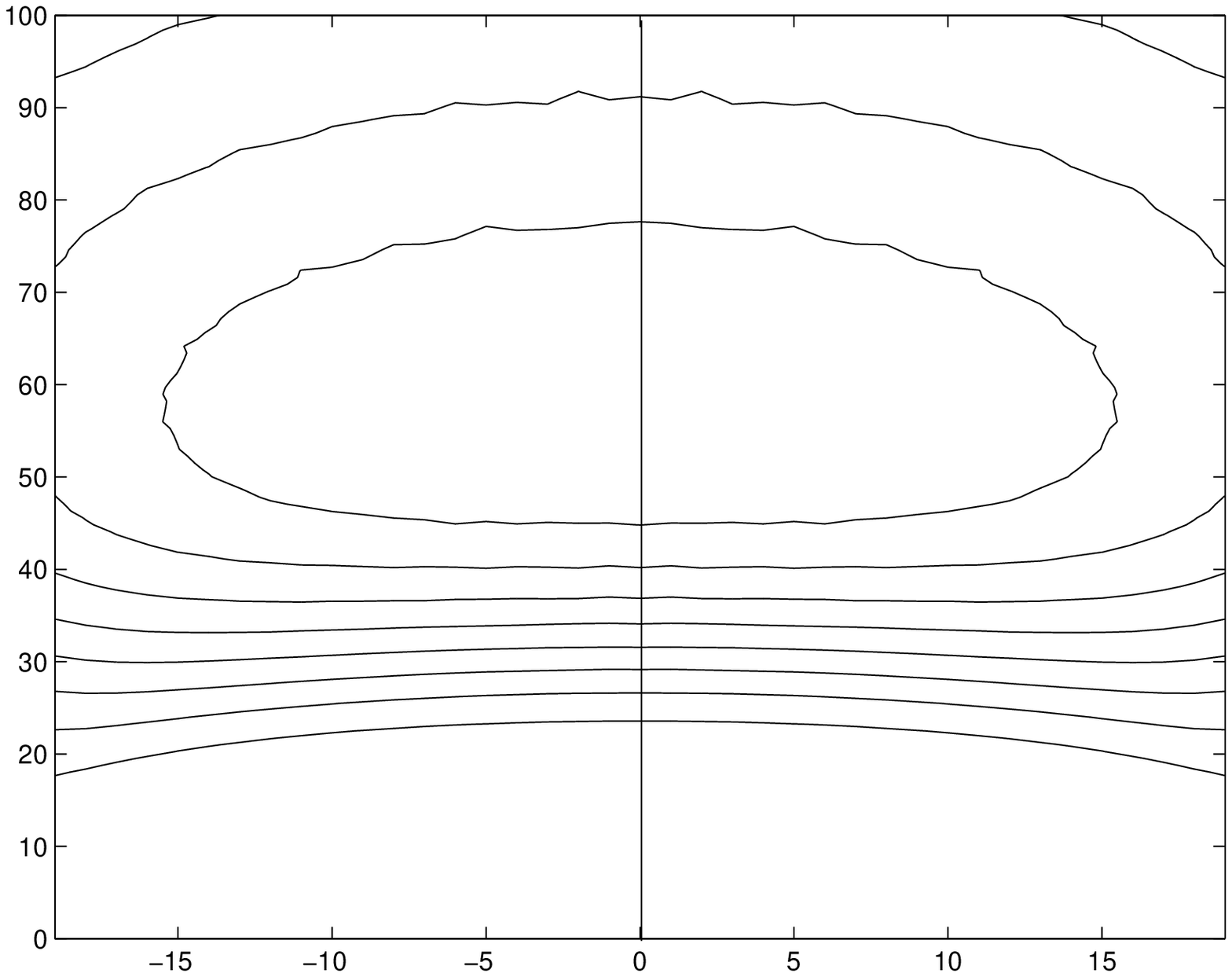}
\vskip 0.1cm \hskip -0.1cm (j)
\vskip 0.5cm 
\caption[]{The space-dependent part of the ten independent components of
radiation field due to the CENBOL. (a) ${\tilde E}_K(r,z)$,
(b) ${\tilde F}^r_K(r,z)$,
(c) ${\tilde F}^{\phi}_K(r,z)$, (d) ${\tilde F}^z_K(r,z)$,
(e) ${\tilde P}^{rr}_K(r,z)$, (f) ${\tilde P}^{r{\phi}}_K(r,z)$, (g) ${\tilde P}^{rz}_K(r,z)$,
(h) ${\tilde P}^{{\phi}{\phi}}_K(r,z)$,
(i) ${\tilde P}^{{\phi}z}_K(r,z)$, (j) ${\tilde P}^{zz}_K(r,z)$.
The disc parameters are $x_s=20r_g$ \& $x_o=500r_g$. }
\end{figure}

There are few features that are to be noted from the various moments
of radiation, computed from the CENBOL:
\begin{enumerate}
  \item The radiation field is highly anisotropic within the funnel like region
  of the CENBOL. 
  \item All the moments maximizes and have very sharp gradients around the inner edge $x_{in}$.
  \item Within the funnel (\ie for $r<x_s$ \& $z{\lsim}h_s$)  ${\tilde F}^r_C<0$,
        but at moderate values of $r$ \& $z$ away from the black hole,
        ${\tilde F}^r_C>0$.
  \item ${\tilde F}^z_C<0$ very close to black hole.
  \item ${\tilde F}^{\phi}_C>0$ for $r>0$ \& $z>0$.
  \item At $r=0$, ${\tilde F}^{\phi}_C={\tilde F}^r_C=0$.
  \item At $r{\sim}x_{in}$ \& $z{\rightarrow}$small, ${\tilde F}^{\phi}_C>
        {\tilde F}^z_C>{\tilde F}^r_C$.
  \item At $r{\sim}x_{in}$ \& $z{\rightarrow}$large, ${\tilde F}^{\phi}_C
        {\approx}{\tilde F}^r_C$ but less than ${\tilde F}^z_C$.
  \item ${\tilde E}_C$ is the most dominant of all the moments.
  \item $P^{rr}_C{\lsim}P^{\phi \phi}_C$ \& $P^{rr}_C(P^{\phi \phi}_C)>P^{zz}_C$.
  \item $P^{r \phi}_C{\approx}P^{rz}_C$.
  \item For $x_s{\sim}10$---$20r_g$ \& $R{\rightarrow}100r_g$, the radiation field      
        due to CENBOL approaches that due to a point source \citep{b14}.
\end{enumerate} 

It is evident that, as $z$ becomes small,
${\tilde F}^{\phi}_C>{\tilde F}^z_C>{\tilde F}^r_C$, still 
higher values of ${\tilde E}_C$ \& ${\tilde P}^{\phi j}_C$ will ensure that the
${\lambda}$ gained by ${\tilde F}^{\phi}_C$ will be less than that
reduced by the drag terms of Eq. (6d). It is also evident that as ${\tilde F}^r_C$
changes from $<0$ to $>0$, as one goes away from the black hole
and the axis of symmetry, which
means closer to the axis and the black hole, radiation from CENBOL
would push the jet material towards the axis and further away it
will tend to spread the jet.
This doesnot mean that, more luminous the CENBOL, more
is the  spreading, since radiation drag along $r$-direction will limit the radial
expansion of jet. 
  
In Fig. (6), the contour maps of various moments of the radiation field
due to the outer Keplerian disc is plotted. The shock location is same
as Fig. (5). In contrast to the contribution due to the CENBOL,
the Keplerian contribution is zero within the domain $r{\leq}x_s$,
$z{\leq}h_s(x_o-r)/(x_o-x_s)$. The anisotropic nature of the
radiation field extends to a region
at much larger distances above the funnel like region of the CENBOL.
The contour map of various moments from the Keplerian disc are,
${\tilde E}_K(r, z)$ [Fig. (6a); max. value: $2.23{\times}10^{-5}$],
${\tilde F}^r_K(r, z)$ [Fig. (6b); max/min values:
$1.77{\times}10^{-6}$/$-1.12{\times}10^{-6}$],
${\tilde F}^{\phi}_K(r, z)$
[Fig. (6c); max. value: $8.62{\times}10^{-7}$],
${\tilde F}^z_K(r, z)$ [Fig. (6d); max. value: $1.7{\times}10^{-5}$],
${\tilde P}^{rr}_K(r,z)$ [Fig. (6e); max. value: $5.75{\times}10^{-6}$],
${\tilde P}^{r \phi}_K(r,z)$ [Fig. (6f); max/min values: $9.49{\times}
10^{-8}/-1.37{\times}10^{-8}$], ${\tilde P}^{rz}_K(r,z)$
[Fig. (6g); max/min values: $1.25{\times}10^{-6}/-8.6{\times}10^{-7}$],
${\tilde P}^{\phi \phi}_K(r,z)$ [Fig. (6h); max. value: $5.75{\times}
10^{-6}$], ${\tilde P}^{\phi z}_K(r,z)$ [Fig. (6i); max value: $5.9
{\times}10^{-7}$] and ${\tilde P}^{zz}_K(r,z)$ [Fig. (6j); max. value:
$1.4{\times}10^{-5}$].

As in the previous figure, there are few features to be noted
in Fig. (6) too. They are:
\begin{enumerate}
\item Each of the moments of the radiation field
due to the KD is few orders of magnitude less than the corresponding
ones due to CENBOL.
\item Max$({\tilde F}^z_K)>$ max$({\tilde F}^r_K)>$ max$({\tilde F}^{\phi}_K)$.
That means spreading of jets will be less.
\item The gradients of moments from the KD are lesser compared to that from the
CENBOL.
\item ${\tilde F}^z_K>0$.
\item ${\tilde F}^r_K<0$ in a larger domain $\Rightarrow$ pushing the jet materials
towards axis in a larger domain above the funnel of the CENBOL, and the angular momentum gained will also be less as $({\tilde F}^{\phi}_K)$ is smallest of the
three components of flux.
\item ${\tilde P}^{rr}_K{\approx}{\tilde P}^{\phi \phi}_K{\approx}
{\tilde P}^{zz}_K$.
\end{enumerate}

It is quite evident that in the hard state, Keplerian contribution to the
radiative momentum will be marginal compared to the CENBOL contribution.
But as ${\tilde F}^r_C$ is weakest amongst all the components of CENBOL flux,
and that ${\tilde F}^r_K$ is directed towards the axis in a larger
domain, thus for ${\ell}_k/{\ell}_c{\lsim}1$ it is possible to observe
greater collimation.  

It is to be noted in Figs. (5a-6j), that the various components of
radiative moments computed in this paper are quantitatively different
from earlier works on thin disc \citep{b39,b40}, as well as, on slim disc
\citep{b43}. When comparing radiation fields from thin disc
and that from TCAF disc model, the first point to be noticed, is the
geometry
of the two disc model is different, secondly the disc motions are different,
and thirdly the spatial variation of disc intensity ($I_{CO}$ and $I_{KO}$)
in TCAF model is different from that of purely thin disc. Even the radiative contribution from KD of the TCAF
model to various radiative moments [Figs. (6a-6j)], differs from that
computed by \citet{b40}. The reasons are that (a) in \citet{b40},
the inner radius of the thin disc is $3r_g$, while in this paper
the inner radius of KD is $x_s$ (${\sim}$few${\times}10r_g$), so
the jet does not `see' the most luminous part of the KD, instead `sees'
the luminous CENBOL, and (b) the shadow effect of CENBOL.
The shadow effect of CENBOL on jet material is extensively
discussed in \S 3.1,
we will just point out here that for jet material at $z{\leq}h_s(x_o-r)/(x_o-x_s)$,
radiation from KD is completely blocked and for $z>h_s(x_o-r)/(x_o-x_s)$,
the jet material only `sees' a fraction of the outer rim of the KD.
With increasing $z$, the jet `sees' more and more inner part of the KD.
In \citet{b40}, such shadow effect was not considered and hence the difference
in spatial variation of various radiative components between these two papers,
namely between Figs. (6a-6j) of this paper and Figs. (2,3,4) of \citet{b40}.

In \citet{b43}, radiative moments are calculated for slim disc model,
for three height to disc-radius ratio (H/R), (a) H/R${\sim}0.05$,
(b) H/R${\sim}0.4$ and (c) H/R${\sim}0.56$. In the present paper,
height to radius ratio of the CENBOL is $h_s/x_s{\sim}0.57$ (for
$x_s=20r_g$, but is always $<0.6$ for any higher $x_s$),
so one might be tempted to think that,
cases (b) and (c) of \citet{b43} should be similar
to Figs. (5a-5j) of this paper. Components of radiative moments
are not solely determined by disc geometry, but also on the disc
dynamics and its radiative properties, and the CENBOL and the slim disc
models of \citet{b43} differ on both these counts.
The radial velocity component of the slim disc considered
was $v_r{\sim}c_1/r^{1/2}$ (${\equiv}{\tilde u}_{st}$ in our case)
and the azimuthal velocity to be
$v_{\phi}{\sim}c_2/r^{1/2}$ (${\equiv}{\tilde u}_{\phi}$, and $c_1$, $c_2$ depends on advection parameter, viscosity parameter, the ratio of specific heats,
etc). The radial velocity of the CENBOL has no explicit
analytical expression [computed in Appendix (A)], and, is not proportional to
$r^{-1/2}$.
In the immediate post-shock region, the radial velocity (${\tilde u}_{st}$) of matter in CENBOL is much less than the radial velocity of matter in models
(b) and (c) of \citet{b43}, but close to the black hole it is higher.
Even the nature of ${\tilde u}_{\phi}$ (in CENBOL)
and $v_{\phi}$ (in slim disc) are different.
The higher velocity components close to $x_{in}$,
enhances the CENBOL intensity close to the black hole,
and ultimately enhancing each of the radiative moments
near the inner edge of the CENBOL compared to that achieved in slim disc.
 
This implies that the stream line velocity (${\sim}10^{-4}$) at the base of the jet above
the CENBOL, will experience greater driving force but as the jet velocity is small there,
the jet
will experience less drag force. The opposite will be true for angular momentum:
the drag forces will reduce angular momentum more than the jet may gain from
the radiation, as ${\lambda}_{in}$ is high at the jet base.
  
The other major difference
in the two models is the difference in radiative property.
The CENBOL radiation in the rest frame is uniform, while that
of the slim disc model falls like $r^{-2}$.
Thus the intensity at the outer edge of CENBOL
is comparatively more significant than the slim disc case.
This makes the radial flux directed inwards in a larger part
of the domain above the CENBOL, compared to the region above the slim disc.
Inspite of the quantitative differences between the moments computed, by CENBOL and 
the slim disc, the overall qualitative similarity between
Figs. (5a-5j) of this paper, and Figs. (4) of \citet{b43}
is quite evident.

\section{Results}
The results are obtained by integrating Eqs. (6a, 6d, 6g), in the radiation
field of the TCAF disc given in Eqs (9a-9e). Apart from the disc-parameters
supplied to calculate the radiation field from the disc, we also
supply the injection radius of the jet $r_i$. As the jets
are launched from the inner surface of the funnel like region of the CENBOL,
the injection height ($z_i$) should be just above the CENBOL inner surface.
With no loss of generality we take $z_i=r_i(h_s/x_s)+0.1$ in units of $r_g$.
We assume that the outflow is made up of purely electron-positron pair
plasma. We are interested to study the dependence of terminal speed as well
as the relative spread of jets on disc parameters such as the ${\ell}_c$,
${\dot m}_k$, $x_s$, $r_{in}$. We define $r_{\infty}/z_{\infty}$ as the
relative spread of the jets, where $r_{\infty},z_{\infty}$ are the cylindrical radial and 
axial coordinates at which $v{\rightarrow}v_{\infty}$. If $r_{\infty}/z_{\infty}<0.1$, we define
the jet as
well collimated, $0.1<r_{\infty}/z_{\infty}<0.2$ as fairly collimated,
$0.2<r_{\infty}/z_{\infty}<0.3$ as poorly collimated and so on.

\subsection{Dependence on ${\ell}_c$ \& ${\dot m}_k$:}

In Fig. (7a), $v_s$ is plotted with $log(z)$, where the CENBOL luminosity is
increased from ${\ell}_c=0.2$ (long-dashed), ${\ell}_c=0.3$ (dashed),
${\ell}_c=0.4$ (solid) in units of $L_{\rm Edd}$, while
the Keplerian luminosity corresponds to ${\dot m}_k=6$ (in units of
${\dot M}_{\rm Edd}$) and $x_s=20r_g$. The {\it parameters
which are kept constant through out the paper are,
$v_{in}=10^{-4}$, 
${\lambda}_{in}=1.7$,
$x_{in}=2r_g$ and $x_o=500r_g$
}.
The injection radius for the jet
is $r_{in}=3r_g$
in Figs. (7a-b).
We see that the streamline velocity
$v_s$ increases with increasing ${\ell}_c$, but the amount of
increase in $v_s$ for equivalent increase in ${\ell}_c$, decreases.

\begin{figure}
\vbox{
\vskip 0.0cm
\hskip 0.0cm
\centerline{
\psfig{figure=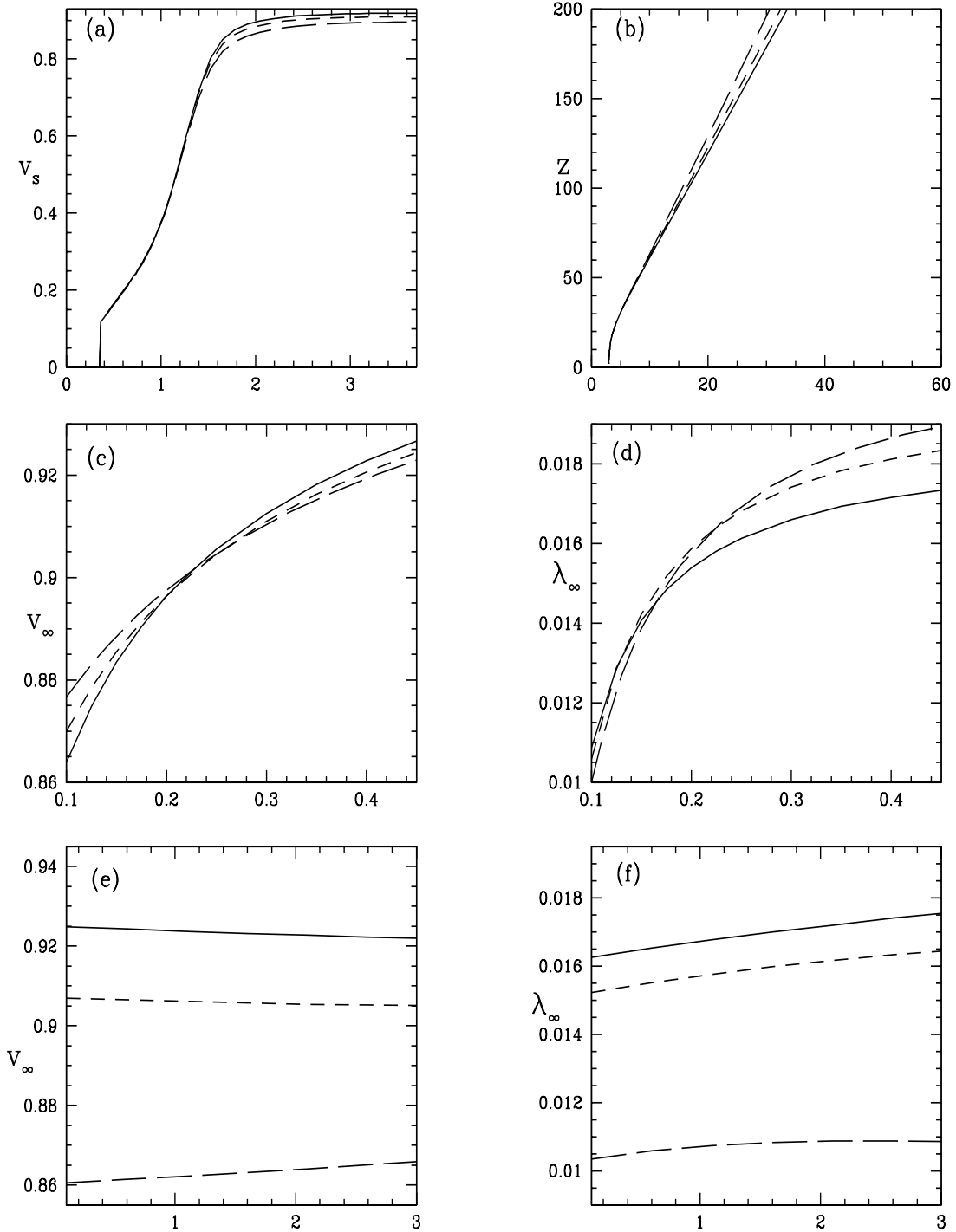,height=19truecm,width=16truecm}}}
\vskip 0.5cm
\hskip 0.0cm
\caption[]{ (a) Variation of $v_s$ with $log(z)$. Each of the curves
represents ${\ell}_c=0.2$ (long-dashed), ${\ell}_c=0.3$ (dashed)
and ${\ell}_c=0.4$ (solid) in units of $L_{\rm Edd}$. (b) The streamlines
of the jet solution in the $z-r$ plane, corresponding to ${\ell}_c=0.2$ (long-dashed), ${\ell}_c=0.3$ (dashed) and ${\ell}_c=0.4$ (solid). For both the figures ${\dot m}_k=6$,
$r_{in}=3r_g$.
(c) Variation of terminal speed $v_{\infty}$ with ${\ell}_c$.
Each curve represents ${\dot m}_k=2$ (solid), ${\dot m}_k=5$ (dashed),
and ${\dot m}_k=8$ (long-dashed),
where $r_{in}=2r_g$.
(d) Variation of terminal specific angular momentum ${\lambda}_{\infty}$ with
${\ell}_c$. Each curve represents ${\dot m}_k=2$ (solid), ${\dot m}_k=5$ (dashed), and ${\dot m}_k=8$ (long-dashed), where $r_{in}=2r_g$.
(e) Variation of terminal speed $v_{\infty}$ with ${\dot m}_k$.
Each curve represents ${\ell}_c=0.1$ (long-dashed), ${\ell}_c=0.25$
(dashed) and ${\ell}_c=0.4$ (solid) in units of $L_{\rm Edd}$.
The injection radius of the jet is $r_{in}=2r_g$.
(f) Variation of terminal sp. angular momentum ${\lambda}_{\infty}$ with
${\dot m}_k$. Each curve represents ${\ell}_c=0.1$ (long-dashed), ${\ell}_c=0.25$
(dashed) and ${\ell}_c=0.4$ (solid) in units of $L_{\rm Edd}$.
The injection radius of the jet is $r_{in}=2r_g$.
For all the figures $v_{in}=10^{-4}$, ${\lambda}_{in}=1.7$, $x_{in}=2r_g$,
$x_s=20r_g$ and $x_o=500r_g$. Values of $v_{in}$, ${\lambda}_{in}$, $x_{in}$
and $x_o$ is kept constant through out the paper.}
\end{figure}

This is to be expected as it was well documented in Paper-I,
that CENBOL radiation is a good accelerator. Let us look at Fig. (7b),
to see how the corresponding streamlines behave.

In Fig. (7b), the streamlines ($z$ vs $r$) are plotted for
${\ell}_c=0.2$ (long-dashed), ${\ell}_c=0.3$ (dashed),
${\ell}_c=0.4$ (solid), while ${\dot m}_k=6$ is kept fixed.
Other disc parameters are $x_s=20r_g$
and $r_{in}=3r_g$. As the CENBOL luminosity is increased,
the streamlines are spreading.
Though the spreading is decreasing for equivalent increase in
${\ell}_c$, as in the case of streamline velocities.
Before going into the reason for this let us probe into
further details.

In Fig. (7c), terminal speed $v_{\infty}$
is plotted with ${\ell}_c$ for ${\dot m}_k=2$ (solid), ${\dot m}_k=5$ (dashed)
${\dot m}_k=8$ (long-dashed), other disc parameters being $r_{in}=2r_g$
and $x_s=20r_g$. Similar to Fig. (7a), we see that $v_{\infty}$
increases appreciably with ${\ell}_c$. For lower values of ${\ell}_c$,
$v_{\infty}$ increases with ${\dot m}_k$, but decreases with
increasing ${\dot m}_k$ for higher values of ${\ell}_c$.
Although the change in $v_{\infty}$ due to the change in ${\dot m}_k$
is small compared to the change due to ${\ell}_c$. This phenomenon was also
reported in \citet{b14}. As this is rotating jet, we would naturally
try to see how the jet ${\lambda}$ behaves at large distances.

In Fig. (7d), ${\lambda}_{\infty}$ is plotted with ${\ell}_c$ for ${\dot m}_k=2$ (solid), ${\dot m}_k=5$ (dashed)
${\dot m}_k=8$ (long-dashed), other disc parameters being $r_{in}=2r_g$
and $x_s=20r_g$. 
We see that generally ${\lambda}_{\infty}$ increases
with ${\ell}_c$. Within the funnel shape region of the CENBOL,
the radiation field produces strong drag terms
along $\phi$ direction which removes ${\lambda}$
more than ${\lambda}$ increased by $f^{\phi}$, but
further away from black hole $\lambda$ reduces so much that the drag
terms [${\propto}{\lambda}(r,z)$] become ineffective, and
the jet gains angular momentum. So as the CENBOL becomes more and more luminous
the jet gains angular momentum. Furthermore, for higher values of
${\ell}_c$, ${\lambda}_{\infty}$ increases with
${\dot m}_k$. This shows that KD radiation do not remove
angular momentum any better than the CENBOL radiation.

In Fig. (7e), $v_{\infty}$ is plotted with 
${\dot m}_k$, parametrized by ${\ell}_c=0.4$ (solid), ${\ell}_c=
0.25$ (dashed) and ${\ell}_c=0.1$ (long-dashed). As was reported in
\citet{b14}, $v_{\infty}$ has a very weak dependence
on ${\dot m}_k$. In Fig. (7f), ${\lambda}_{\infty}$ is plotted with
${\dot m}_k$, parametrized by ${\ell}_c=0.4$ (solid), ${\ell}_c=
0.25$ (dashed) and ${\ell}_c=0.1$ (long-dashed). It is evident that 
${\lambda}_{\infty}$ has a very weak dependence on ${\dot m}_k$.

Let us now see the effect of radiation on spreading the jet.
In Figs. (8a-b) we plot $r_{\infty}/z_{\infty}$.
In Fig. (8a), $r_{\infty}/z_{\infty}$ is plotted with ${\ell}_c$,
parametrized for ${\dot m}_k=2$ (solid), ${\dot m}_k=5$ (dashed)
${\dot m}_k=8$ (long-dashed) [same parameters as Fig. (7c-d)], while
in Fig. (8b), $r_{\infty}/z_{\infty}$ is plotted with 
${\dot m}_k$, parametrized by ${\ell}_c=0.4$ (solid), ${\ell}_c=
0.25$ (dashed) and $0.1$ (long-dashed) [same parameters as Fig. (7e-f)].
The most remarkable contrast is that the jets are more collimated
with increasing KD luminosity, while it tends to spread with increasing
CENBOL luminosity.
Though one must notice that spreading do not increase monotonically
with ${\ell}_c$, but tends to decrease with equivalent increase
in ${\ell}_c$. There are two features which are quite intriguing,
(i) contrasting nature of CENBOL and KD radiation field in terms of
collimation of the jet, and (ii) KD radiation seems to have nominal
effect on determining $v_{\infty}$ and ${\lambda}_{\infty}$, but
plays relatively a greater role in collimation.

\begin{figure}
\vbox{ 
\vskip 0.0cm
\hskip 0.0cm
\centerline{
\psfig{figure=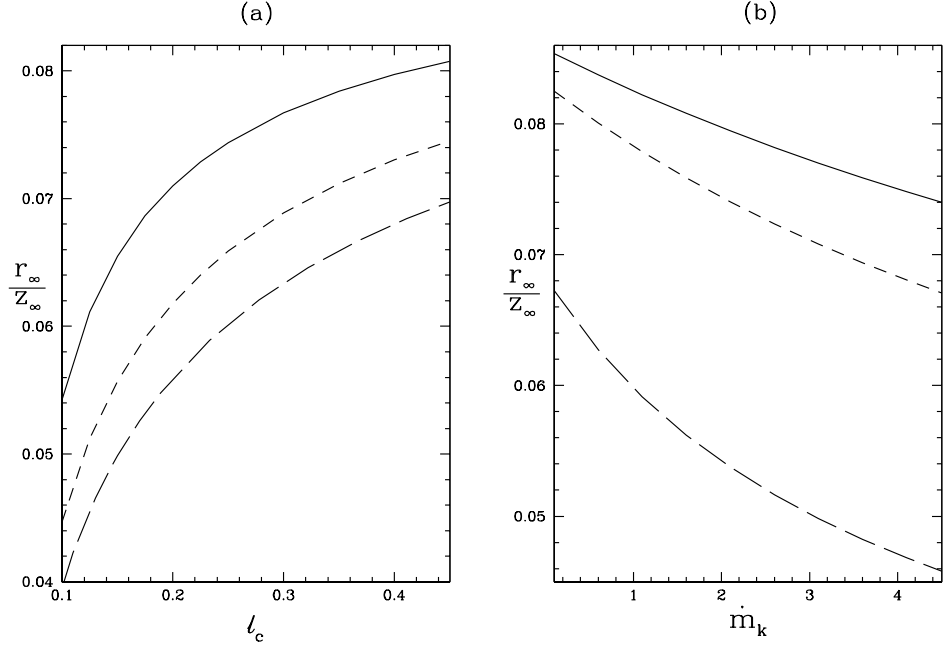,height=13truecm,width=13truecm}}}
\vskip -3.0cm
\hskip 0.0cm
\caption[]{(a) Variation of $r_{\infty}/z_{\infty}$ with ${\ell}_c$.
Each curve represents ${\dot m}_k=2$ (solid), ${\dot m}_k=5$
(dashed). (b) Variation of $r_{\infty}/z_{\infty}$ with ${\dot
m}_k$. Each curve represents ${\ell}_c=0.4$ (solid), ${\ell}_c=
0.25$ (dashed) and ${\ell}_c=0.1$ (long-dashed).
The curves are drawn for $x_s=20r_g$. }
\end{figure}

The reason that radiation from the KD can collimate the jets better
than radiations from CENBOL,
can be understood from Figs. (5-6) and Eq. (6g).
We know that rotating matter tends to move away from the
axis due to centrifugal force. If the angular momentum is reduced
by the drag forces then the spreading may be arrested.
Close to the CENBOL surface the radiation field is dominated
by the CENBOL radiation and it produces large drag forces along $\phi$.
Further out as the jet starts to `see' KD radiation, $\lambda$ is reduced
to the extent the drag force becomes marginal and the jet starts to
gain some angular momentum. From Figs. (5-6), it is evident that
the contributions from KD to the total radiation field is much less than
that of the CENBOL so, angular momentum removed or added to the jet
by the radiation from KD
is marginal [\eg Fig. (7f)]. From Eq. (6g), it is clear that
the spreading of jets depends as much on centrifugal force,
as on $f^r$. From the discussion bellow Figs. (5-6), we know that,
${\tilde F}^r_C$ is weakest amongst all the ${\tilde F}^i_C$'s,
while ${\tilde F}^r_K{\lsim}{\tilde F}^z_K$ but
${\tilde F}^r_K>{\tilde F}^{\phi}_K$,
and ${\tilde F}^r_K$ is towards the axis in a larger domain.
Thus when ${\ell}_c$
increases it doesnot help in collimation, but as ${\dot m}_k$
increases the higher negative values of ${\tilde F}^r_K$, makes
$f^r$ more and more negative, and so it pushes the jet towards
the axis of symmetry and helps in collimation, inspite of not removing angular
momentum to the extent as CENBOL radiation does.

\begin{figure}

\includegraphics[scale=0.35]{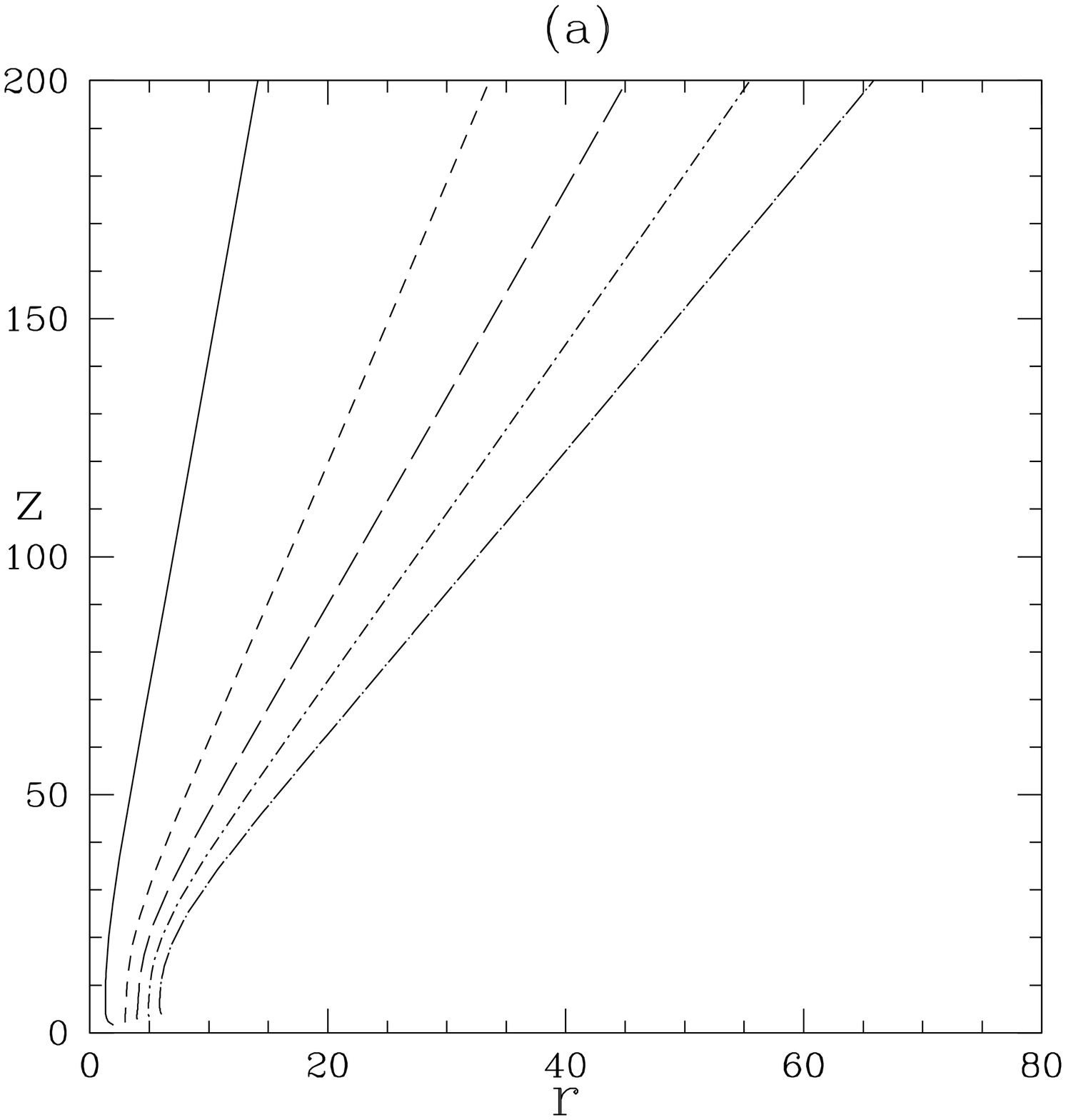}
\hskip 1.0cm
\includegraphics[scale=0.35]{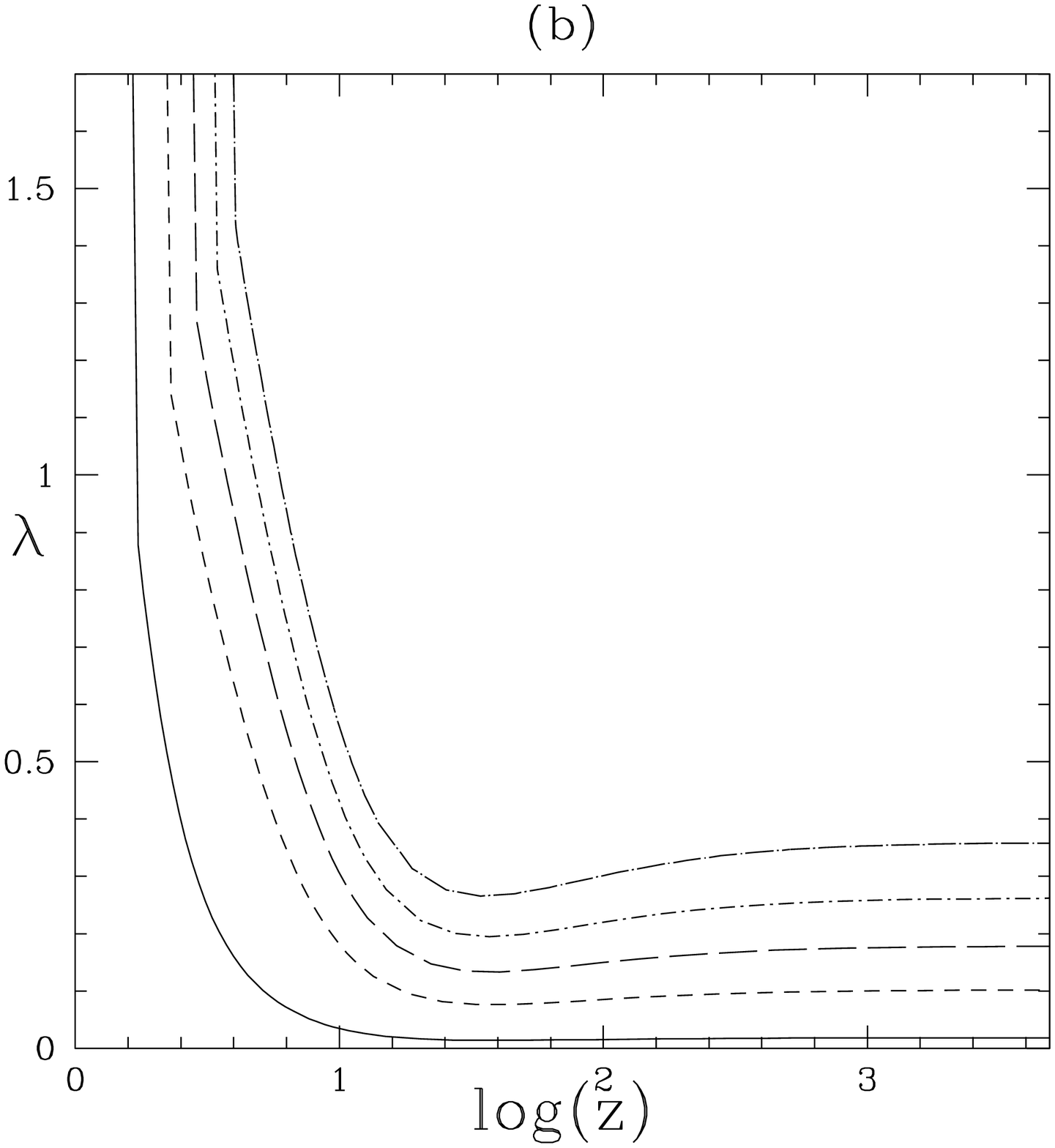}
\vskip 0.2cm
\includegraphics[scale=0.35]{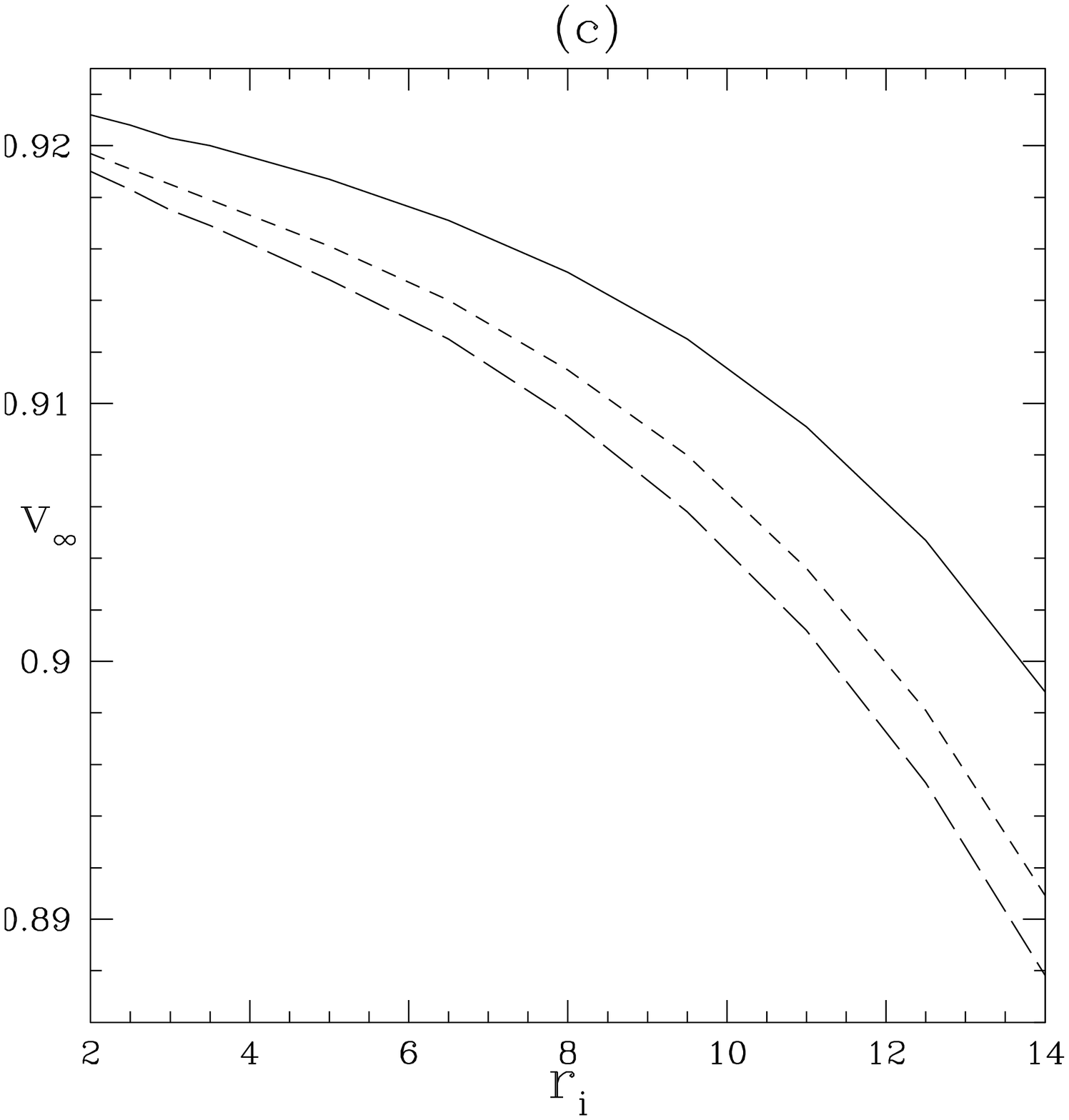}
\hskip 1.0cm
\includegraphics[scale=0.35]{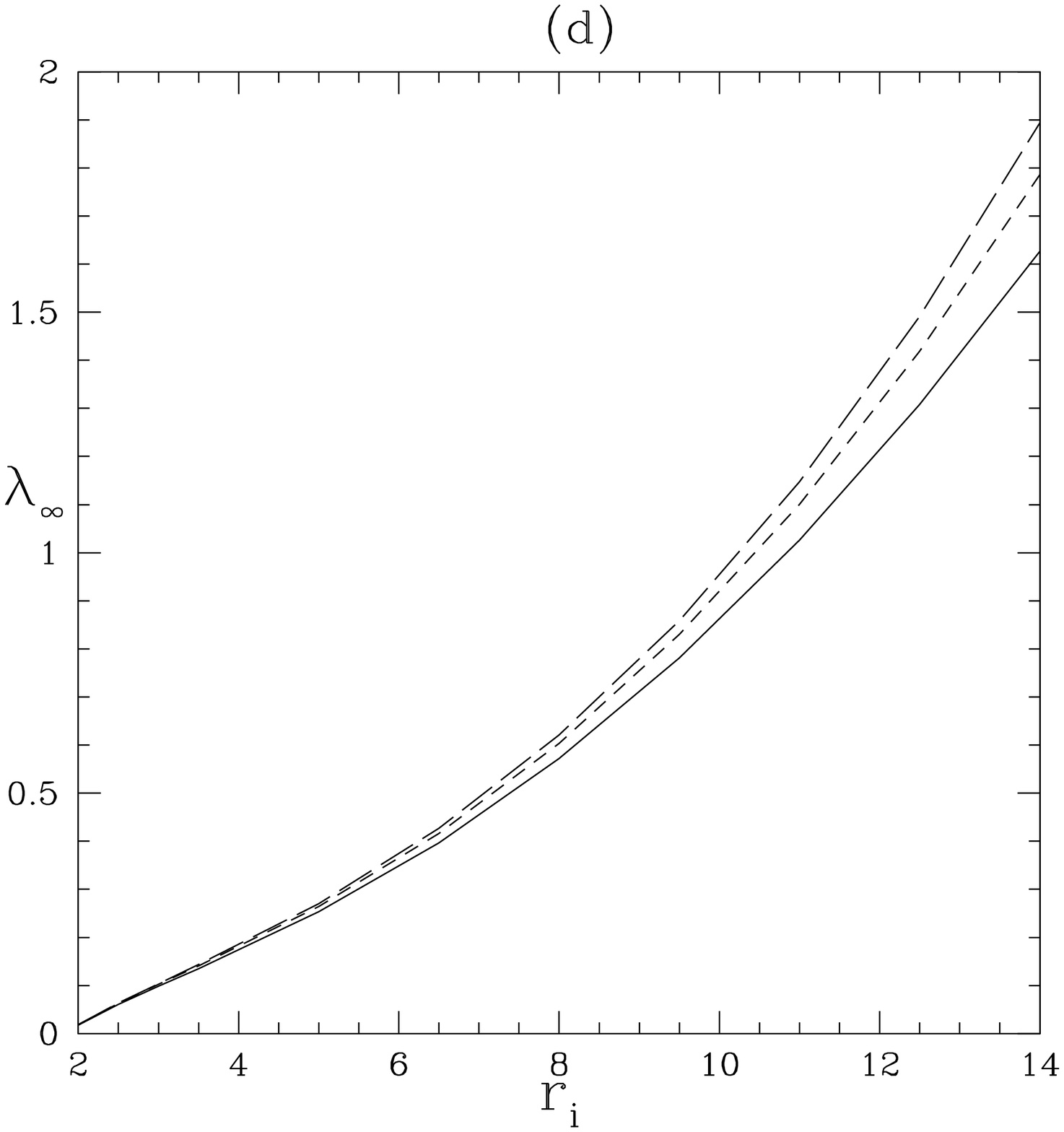}
\vskip 0.2cm
\includegraphics[scale=0.35]{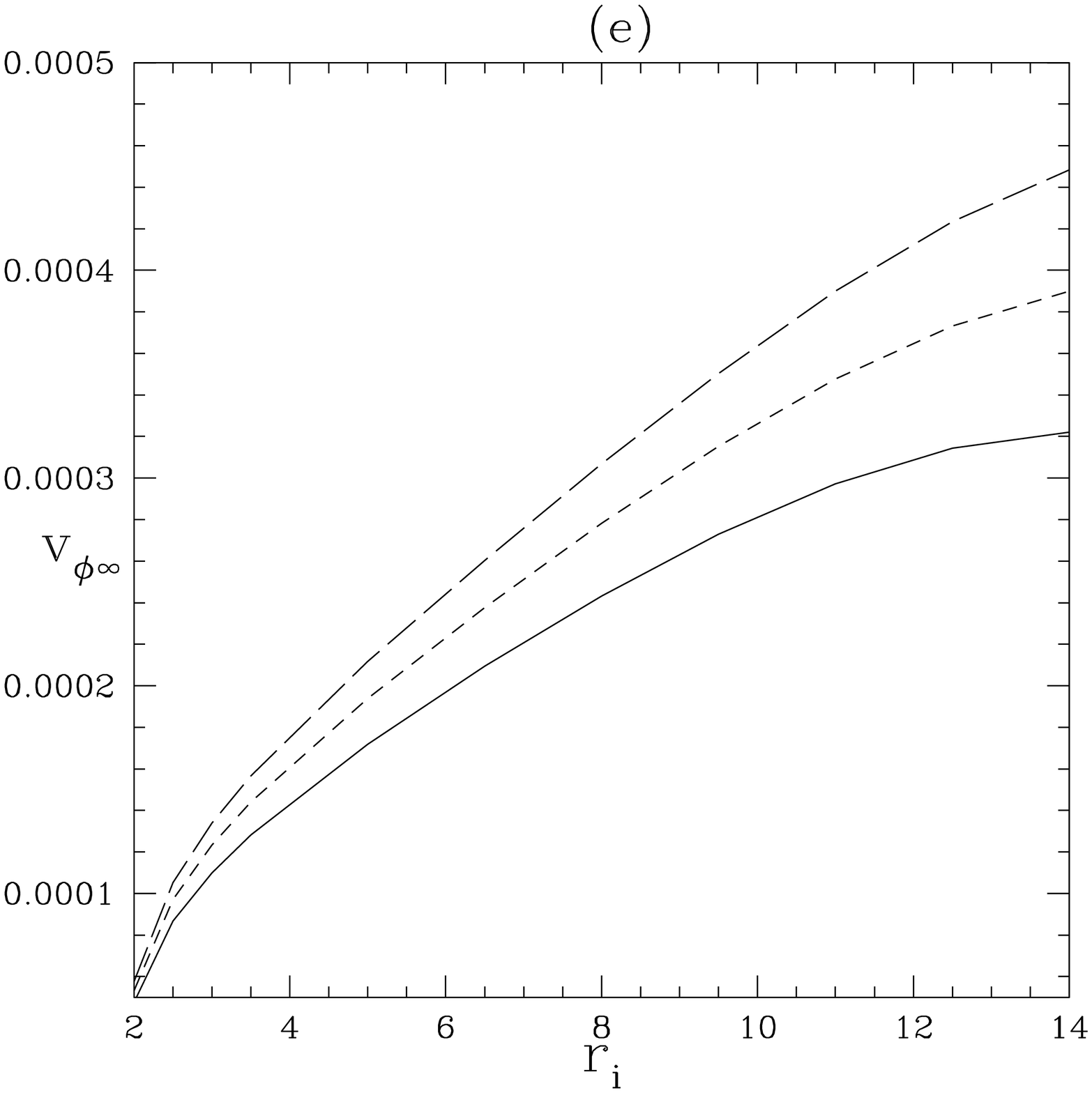}
\hskip 1.0cm
\includegraphics[scale=0.35]{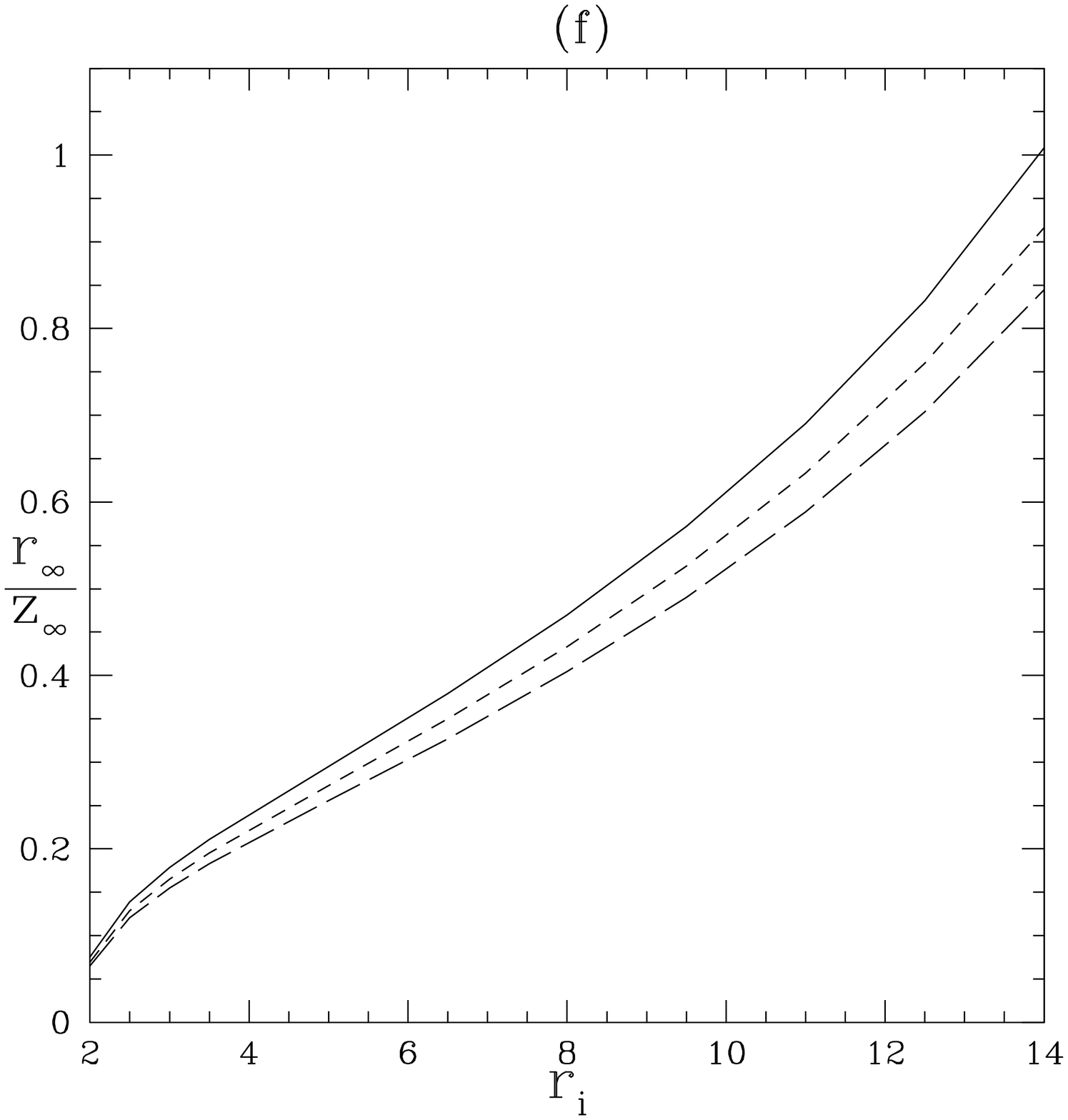}

\vskip 0.0cm
\hskip 0.0cm
\caption[]{ (a) Streamlines for $r_i=2r_g$ (solid), $r_i=3r_g$ (dashed), $r_i=4r_g$
(long dashed), $r_i=5r_g$ (dashed-dotted) \& $r_i=6r_g$ (long dashed-dotted);
${\ell}_c=0.4$, ${\dot m}_k=6$. (b) ${\lambda}$ vs $log(z)$, corresponding to the previous figure. (c) $v_{\infty}$ vs $r_i$, (d) ${\lambda}_{\infty}$ vs
$r_i$, (e) $v_{{\phi}{\infty}}$ vs $r_i$, and (f) $r_{\infty}/z_{\infty}$
vs $r_i$; for ${\dot m}_k=4$ (solid), ${\dot m}_k=7$ (dashed) \& ${\dot m}_k=10$ (long dashed). For all the figures $x_s=20r_g$.}
\end{figure}

\subsection{Dependence on $r_i$:}

In Fig. 9, we show the effect of injection radius of the jets.
In Fig. (9a), streamlines are plotted for $r_i=2r_g$ (solid),
$r_i=3r_g$ (dashed), $r_i=4r_g$ (long-dashed), $r_i=5r_g$ (dashed-dotted),
and $r_i=6r_g$ (long dashed-dotted), where ${\ell}_c=0.4$, ${\dot m}_k=
6$ and $x_s=20r_g$. In Fig. (9b), corresponding $\lambda$ distribution
is shown along the streamlines for $r_i=2r_g$ (solid),
$r_i=3r_g$ (dashed), $r_i=4r_g$ (long-dashed), $r_i=5r_g$ (dashed-dotted),
and $r_i=6r_g$ (long dashed-dotted), where ${\ell}_c=0.4$, ${\dot m}_k=
6$ and $x_s=20r_g$. From both the figures we see that the injection height
$z_i$ changes with $r_i$ as has been discussed at the start of this section.
It is evident that the as the injection radius is varied, the angular momentum of the jets are higher and it spreads further. The angular
momentum of the jets at different $r_i$ is same, but the radiative moments
just above the CENBOL surface decreases with $r$. In other words the
jets will gain less $\lambda$ from the radiation field, but at the
same time they will lose less $\lambda$ due to the drag terms.
The net effect is that with increasing $r_i$ the jets are of higher angular
momentum and thus these jets are less collimated.

On the other hand, jets are generated with very low velocity at the base
(${\sim}r_i$), so the drag term is negligible, but as all the radiative 
moments decreases with $r$ above the CENBOL surface, the driving that
the jets get due to $f^r$ \& $f^z$ is much less as $r_i$ is increased.
In Figs. (9c-f), various terminal values are plotted with $r_i$, for
${\dot m}_k=4$ (solid), ${\dot m}_k=7$ (dashed) and ${\dot m}_k=10$
(long-dashed), other parameters being ${\ell}_c=0.4$, $x_s=20r_g$.
In Fig. (9c), we see that $v_{\infty}$ decreases with $r_i$,
but for fixed values of $r_i$, it increases with lower value
of ${\dot m}_k$ or in other words higher value of ${\ell}_c/{\ell}_k$.
It has been observed in \citet{b14} and also in Fig. (7c) of this paper,
that for ${\ell}_c{\gsim}0.22$, $v_{\infty}$ decreases with increasing
${\dot m}_k$.
In Fig. (9d), on the other hand shows that ${\lambda}_{\infty}$
increasing with increasing $r_i$.
In Fig. (9e), the rotational velocity [$v^2_{\phi}=-u_{\phi}u^{\phi}/
u_tu^t$] at infinity ($v_{\phi \infty}$) is plotted with $r_i$.
Similar to ${\lambda}_{\infty}$, $v_{\phi \infty}$ also increases with
$r_i$.
A interesting feature is seen if one compares Fig. (9f),
with the previous two.
In Fig. (9f), the relative spread $r_{\infty}/z_{\infty}$ is
shown to increase with $r_i$, but the interesting feature is,
as one increases ${\dot m}_k$, $r_{\infty}/z_{\infty}$ decreases,
but both ${\lambda}_{\infty}$ as well as $v_{\phi \infty}$ increases.
This vindicates Eq. (6g), which says, $f^r$ and $f^z$ determines
the streamlines along with the centrifugal force.

\begin{figure}
\vbox{
\vskip 0.0cm
\hskip 0.0cm
\centerline{
\psfig{figure=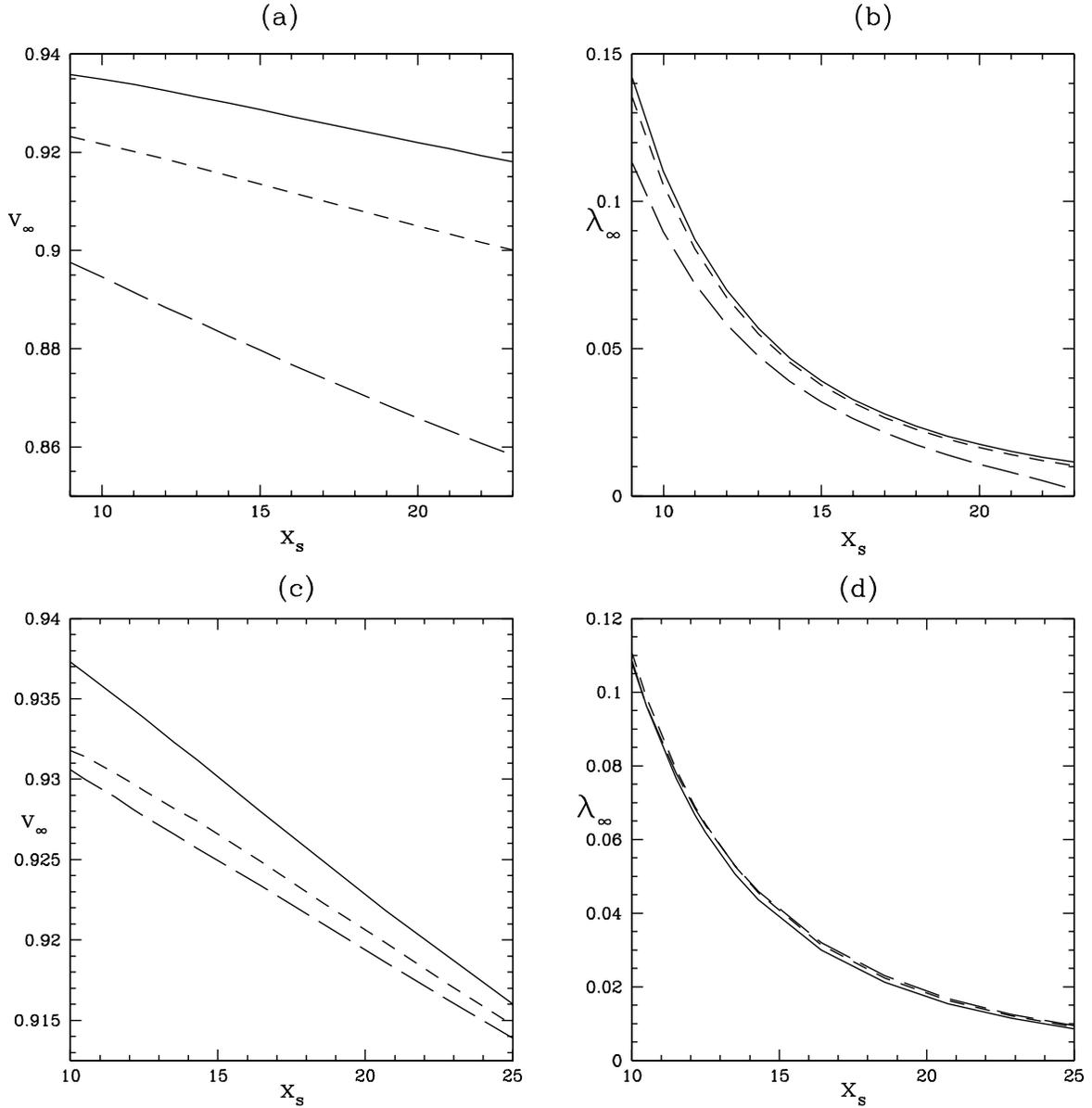,height=16truecm,width=16truecm}}}
\vskip 0.0cm
\hskip 0.0cm
\caption[]{(a) Variation of $v_{\infty}$ with $x_s$, ${\ell}_c=0.4$ (solid),
${\ell}_c=0.25$ (dashed), ${\ell}_c=0.1$ (long-dashed), for constant
${\dot m}_k=3$. (b) Variation of ${\lambda}_{\infty}$
with $x_s$, ${\ell}_c=0.4$ (solid),
${\ell}_c=0.25$ (dashed), ${\ell}_c=0.1$ (long-dashed), for constant
${\dot m}_k=3$.
(c) Variation of $v_{\infty}$ with $x_s$, ${\dot m}_k=2$ (solid), ${\dot m}_k=5$ (dashed),
${\dot m}_k=8$ (long-dashed) for ${\ell}_c=0.4$.
(d) Variation of ${\lambda}_{\infty}$
with $x_s$. ${\dot m}_k=2$ (solid), ${\dot m}_k=5$ (dashed),
${\dot m}_k=8$ (long-dashed) for ${\ell}_c=0.4$.  For all the figures $r_i=2r_g$.}
\end{figure}

\subsection{Dependence on $x_s$:}

Until now we have investigated the solutions for $x_s=20r_g$.
We now concentrate on the dependence of jet solutions on $x_s$.

In Fig. (10a), $v_{\infty}$ is plotted with $x_s$, for
${\ell}_c=0.4$ (solid), ${\ell}_c=0.25$ (dashed) and ${\ell}_c=0.1$
(long-dashed), where ${\dot m}_k=3$ and $r_i=2r_g$ are kept fixed.
With increasing values of $x_s$, the CENBOL intensity decreases
and the driving force goes down, producing lesser terminal speed.
We plot ${\lambda}_{\infty}$, corresponding to the cases in Fig. (10a),
in Fig. (10b). Similar to $v_{\infty}$, ${\lambda}_{\infty}$ also
decreases with $x_s$. The reason for this is same as in the previous
case, \ie the CENBOL intensity decreases with increasing $x_s$.
Interestingly, as ${\ell}_c$ is increased, $v_{\infty}$ increases
appreciably but ${\lambda}_{\infty}$ is increased marginally.
Close to the injection radius, the ${\lambda}$ of the jet decreases rapidly [see, Fig. (9b)],
and then when at large distances from the disc, ${\lambda}$ 
becomes too low so that the drag in the azimuthal
direction becomes negligible, the jet starts to gain some angular
momentum from the radiation. So increasing ${\ell}_c$ will
produce higher ${\lambda}_{\infty}$, but as the gain in ${\lambda}$
occurs at distances farther away from the disc, the radiative moments
falls off anyway, so the increase in ${\lambda}_{\infty}$ is small.

In Fig. (10c), $v_{\infty}$ is plotted with $x_s$, for
${\dot m}_k=2$ (solid), ${\dot m}_k=5$ (dashed) and ${\dot m}_k=8$ (long-dashed), with consant ${\ell}_c=0.4$. The general conclusion that
$v_{\infty}$ decreases with increasing $x_s$ is still valid, though
increasing ${\dot m}_k$ decreases $v_{\infty}$.
Another difference we do notice is, 
the curve of $v_{\infty}$ widens with increasing $x_s$, for
higher values of ${\ell}_c$ in Fig. (10a), 
while in Fig. (10c), $v_{\infty}$ curves converges with increasing
$x_s$. The reason is one and the same, \ie with increasing $x_s$,
the KD contribution to all the components of radiative moments
decreases and the CENBOL contribution dominates.
As CENBOL radiation is a good accelerator, so for higher $x_s$,
increasing ${\ell}_c$ produces higher $v_{\infty}$ relative to
the case where ${\dot m}_k$ is increased.

In Fig. (10d), ${\lambda}_{\infty}$ is plotted with $x_s$,
for ${\dot m}_k=2$ (solid), ${\dot m}_k=5$ (dashed) and ${\dot m}_k=8$ (long-dashed), with consant ${\ell}_c=0.4$. ${\lambda}_{\infty}$
decreases with $x_s$ but has almost
no dependence on KD radiation. This is because the azimuthal
component of the radiative flux produced by KD is smaller compared
to the other components (see \S 3.2).

In Fig. (11a), relative spread $r_{\infty}/z_{\infty}$ is plotted
with $x_s$, for the same parameters as Fig. (10a-b), \ie for ${\ell}_c=0.4$ (solid), ${\ell}_c=0.25$ (dashed) and ${\ell}_c=0.1$
(long-dashed), where ${\dot m}_k=3$ and $r_i=2r_g$ are kept fixed.
We find $r_{\infty}/z_{\infty}$ decreases with increasing $x_s$.
As ${\lambda}_{\infty}$ decreases with increasing $x_s$
[see, Fig. (10b)], so it is not surprising that there would be greater
collimation. Still for fixed values of $x_s$, the difference in
${\lambda}_{\infty}$ is marginal, while on the other hand
the difference in $r_{\infty}/z_{\infty}$ is larger!
If we turn our attention to Fig. (11b), which is plotted for
${\dot m}_k=2$ (solid), ${\dot m}_k=5$ (dashed) and ${\dot m}_k=8$ (long-dashed), with consant ${\ell}_c=0.4$ [same case as Fig. (10c-d)],
it is again seen that, $r_{\infty}/z_{\infty}$ generally decreases
with $x_s$, at the same time, decrement of $r_{\infty}/z_{\infty}$
for fixed values of $x_s$ with increasing ${\dot m}_k$
is larger , although ${\lambda}_{\infty}$ is almost indistinguishable.
From Figs. (10a-d), we know why $v_{\infty}$ and ${\lambda}_{\infty}$
decreases with increasing $x_s$. But increasing collimation is partly
due to decreasing ${\lambda}_{\infty}$, otherwise variation of
$r_{\infty}/z_{\infty}$ would have just mirrored the variation of
${\lambda}_{\infty}$. Increasing $x_s$ makes $f^r_C$ negative in a larger part of the funnel like region. Simultaneously,
though radiative contributions by KD becomes weaker, none the less
it makes $f^r_K<0$ in a still larger part of the domain.
Hence the combination of decreasing angular momentum
as well as $f^r<0$ in a larger region around the axis,
collimates the jets to a greater degree. 

Thus we conclude from Figs. (10a-d, 11a-b),
if ${\ell}_c>0.2$, $x_s>20r_g$ then jets with terminal properties
$v_{\infty}{\gsim}0.9c$ and $r_{\infty}/z_{\infty}<0.1$ is possible.

\begin{figure}
\vbox{
\vskip -1.0cm
\hskip 0.0cm
\centerline{
\psfig{figure=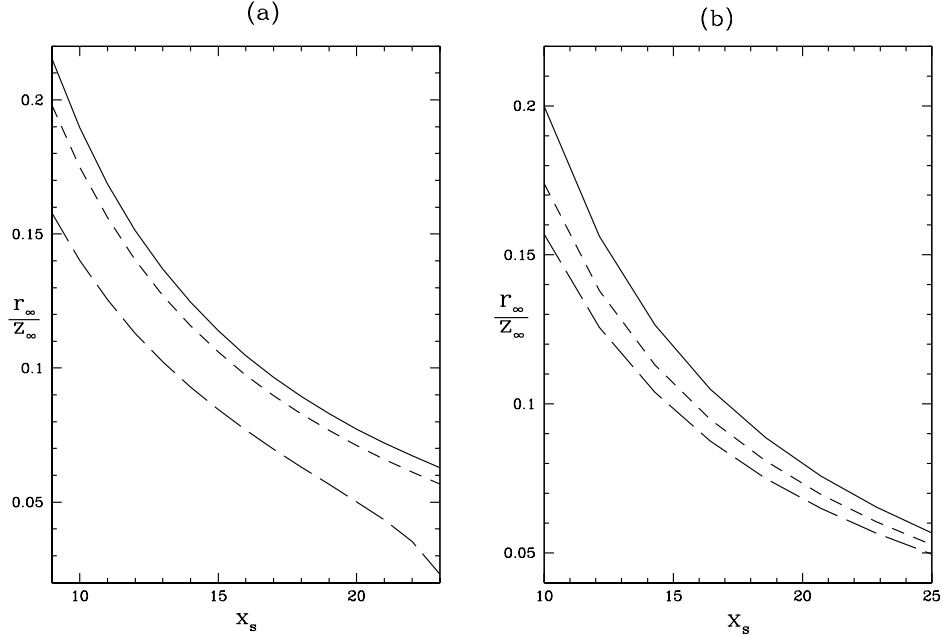,height=13truecm,width=13truecm}}}
\vskip -3.5cm
\hskip 0.0cm
\caption[]{ Variation of $r_{\infty}/z_{\infty}$ with $x_s$.
(a) ${\ell}_c=0.4$ (solid),
${\ell}_c=0.25$ (dashed), ${\ell}_c=0.1$ (long-dashed), for constant
${\dot m}_k=3$. (b) ${\dot m}_k=2$ (solid), ${\dot m}_k=5$ (dashed),
${\dot m}_k=8$ (long-dashed) for ${\ell}_c=0.4$. Variation of ${\lambda}_{\infty}$
with $x_s$. For all the figures $r_i=2r_g$.}
\end{figure}

It will be interesting if these results are contrasted with earlier works.
It is natural that these results would differ from earlier works, since the
radiation field of TCAF disc is different from either thin \citep{b40}
or slim disc \citep{b43}. For both the disc models thin and slim, radiation field generally spreads the jet, and due to radiation drag the radiation field suppresses
the motion along $z$ direction. In another model \citep{b20},
%fukue etal 2001 {b20}
where the authors considered inner non-luminous disc and outer luminous disc,
the jets got collimated with increasing disc luminosity and the angular momentum got decreased too. The reason for collimation is that the injection radius of the jet
$r_{i}$ ($r_0$ in their nomenclature) is less the inner boundary of the disc $x_{in}$ ($r_B$ in their nomenclature).
This makes the radial flux $f^r<0$ in a larger region above (and bellow) the disc
surface, and it pushes the jet material towards the axis.
Along with this, radiation drag reduces angular momentum, so the jets are collimated.
In both the earlier papers \citet{b40} and \citet{b43}, the injection radius
is greater than the inner radius of the disc, which produces $f^r>0$ which
pushes the jets outwards, at the same time the maximum of $f^{\phi}$ is in and around the injection radius, which manages to spin up the jet further, these
two effects combines to increase the angular momentum of the jets even more.
{\it So the crucial point for collimation is whether or not} $f^r<0$.

TCAF disc model has two radiation sources the CENBOL and the outer KD,
and $r_i>x_{in}$.
Jets from TCAF model starts with very low streamline velocity ($v_{in}=10^{-4}$),
and moderadtely high initial angular momentum (${\lambda}_{in}=1.7{\equiv}$
angular momentum of the CENBOL).
As the radiative energy density (${\varepsilon}$) and the pressure components (${\wp}^{ij}$) are quite intense close to the inner edge
of the disc, the higher value of ${\lambda}_{in}$ ensures very high
drag force along $\phi$. As a result, the jet $\lambda$ is reduced appreciably,
close to the injection radius [\eg Fig. (9b)]. Apart from
reduction of the angular momentum of the jets,
the radial flux $f_r$ is negative, just above the CENBOL surface, and this collimates the
jets.
On the other hand, as $v_{in}$ is very small (i.e. drag along
streamline is small), resulting in the jets being accelerated
very fast, shooting upto $30-40r_g$
almost vertically, or sometimes initially pushed towards the axis
[\eg Figs. (7b, 9a)].  At above $30-40r_g$ above the CENBOL
surface $f_r>0$, and the drag along
$\phi$ is weakened [as, discussed in \S 4.1], these facts tend to spread the jet,
although the spreading is arrested by KD luminosity, as well as, by the weakening
of the $f^{\phi}_C$. 

One must also remember that, in jets around TCAF model, $r_i>x_{in}$.
For $r_i{\approx}x_{in}$, radiations from the CENBOL region $x_{in}-r_i$
is negligible, and hence $f^r$ is more negative near $r_i$, but for $r_i>x_{in}$,
radiations from the CENBOL region $x_{in}-r_i$ cannot be neglected, making
$f^r$ is less negative near $r_i$. Thus we see for smaller values of $r_i$
collimation and acceleration of jets are better [\eg Figs. (9a-9f)].

Although ${\lambda}_{\infty}$ increases with ${\ell}_c$, but one must notice that
${\lambda}_{\infty}{\ll}{\lambda}_{in}$ [\eg Figs. (7d, 10b, 10c)], which 
points to the fact, that indeed ${\lambda}$ is reduced by drag force.
One must also
notice, that jets are more collimated by increasing outer
KD luminosity [\eg Fig. (8b)], or increasing $x_s$ [\eg Fig. (11b)],
basically means that jets are basically collimated by $f^r<0$,
as was indicated by \citet{b20}. 

\subsection{Dependence on ${\ell}_k/{\ell}_c$:}

Until now in this section, we have seen that increasing
${\ell}_c$ produces faster jets, while increasing ${\dot m}_k$
and $x_s$ makes the jets more collimated. We have also seen that
increasing $r_i$ produces less $v_{\infty}$ and the collimation is 
worse. The information of CENBOL radiation is provided by its
luminosity, the information of KD radiation is supplied by
${\dot m}_k$ and $x_s$, so to get a better understanding, we now study how the relative
proportions of CENBOL and KD luminosity affect the jet solutions.

\begin{figure}

\includegraphics[scale=0.4]{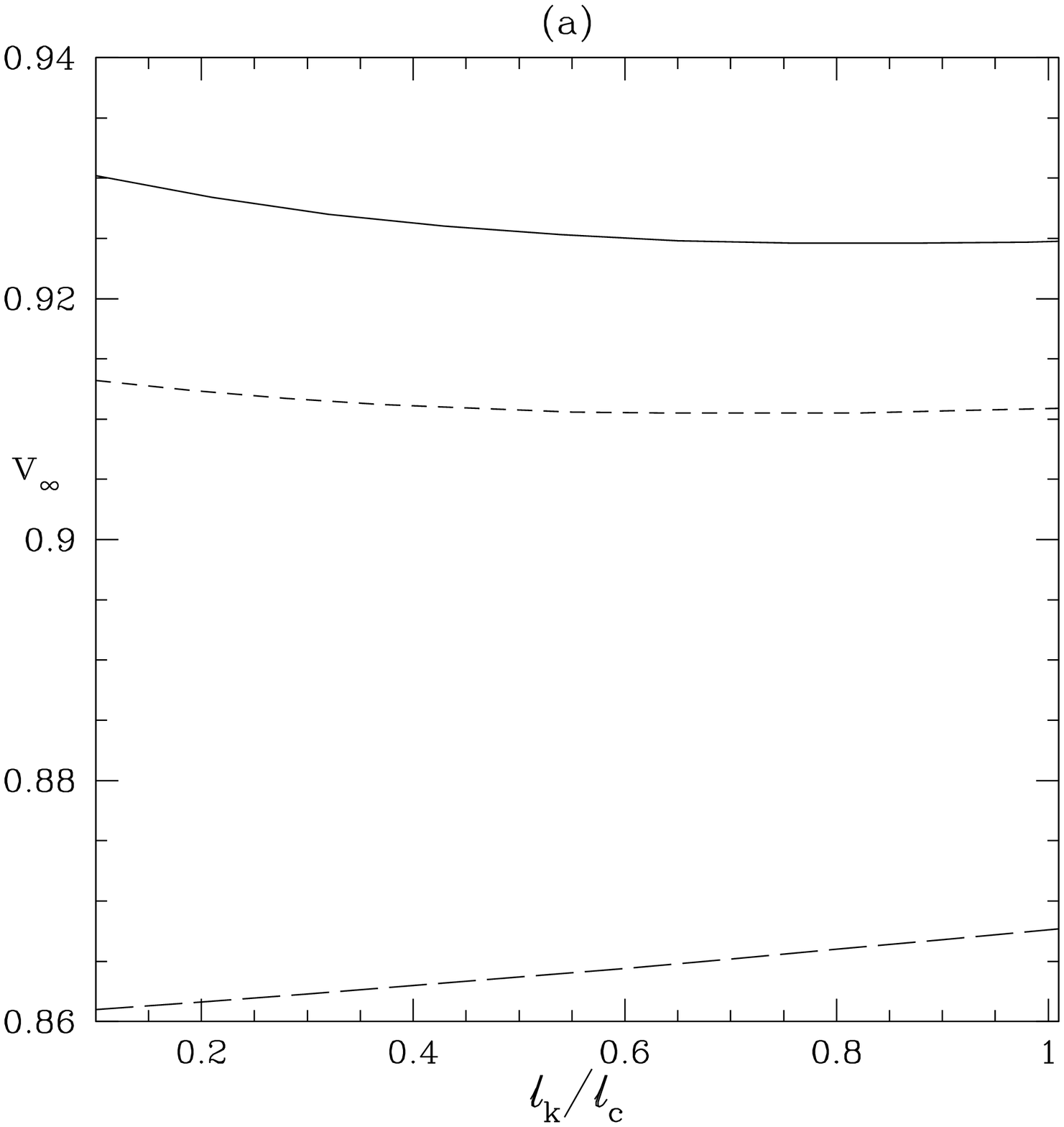}
\hskip 1.0cm
\includegraphics[scale=0.4]{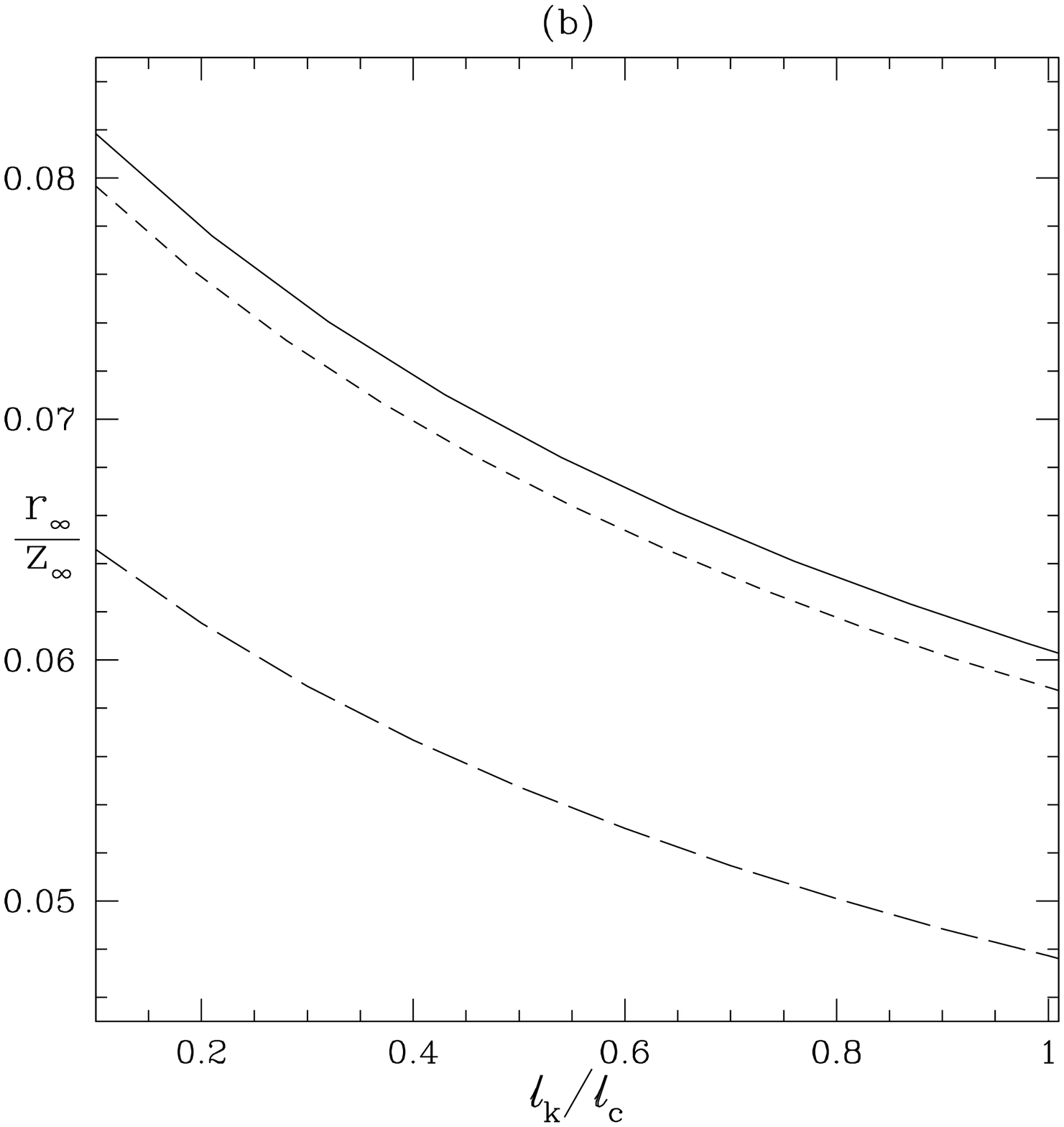}
\vskip 0.2cm
\includegraphics[scale=0.4]{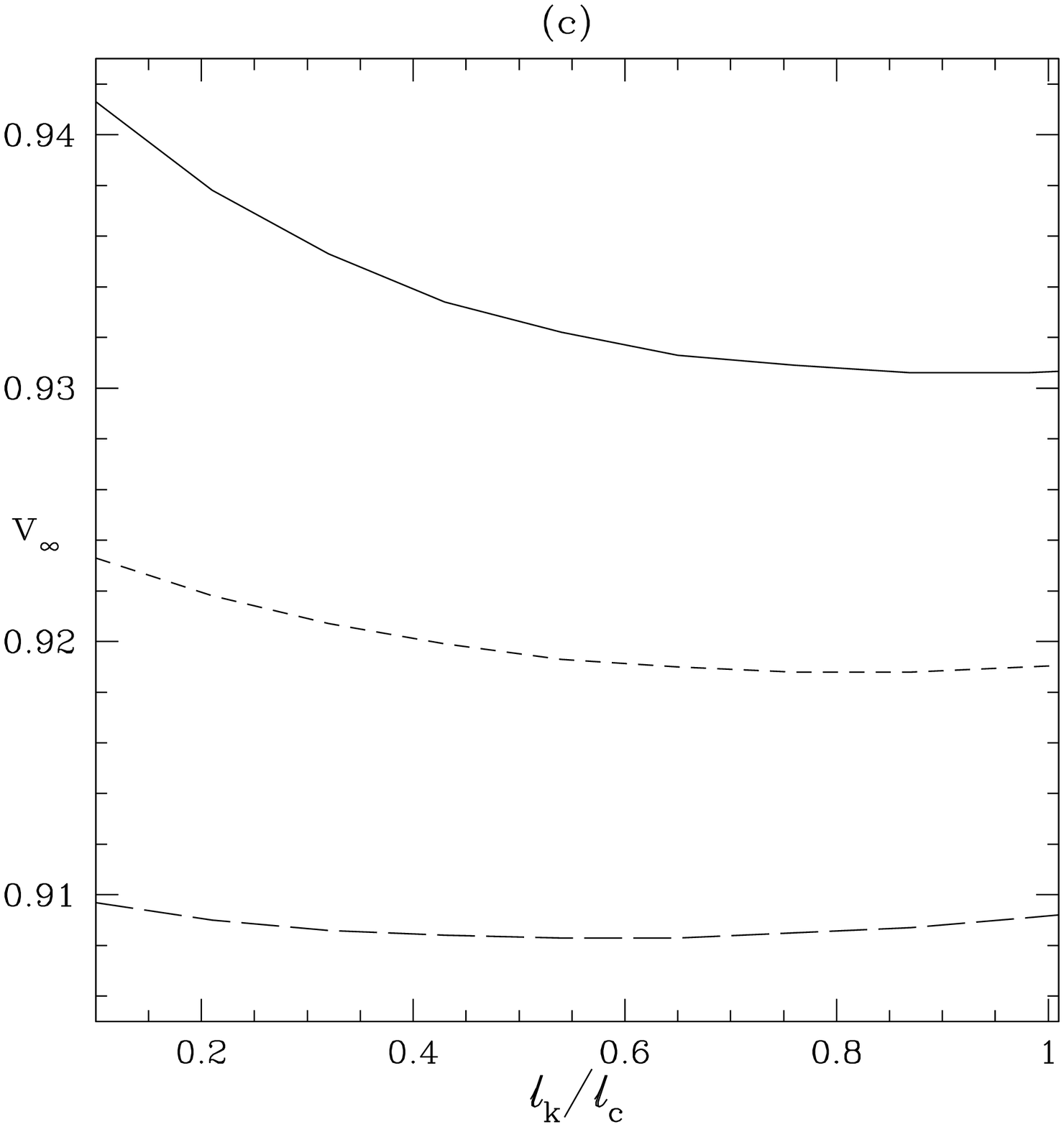}
\hskip 1.0cm
\includegraphics[scale=0.4]{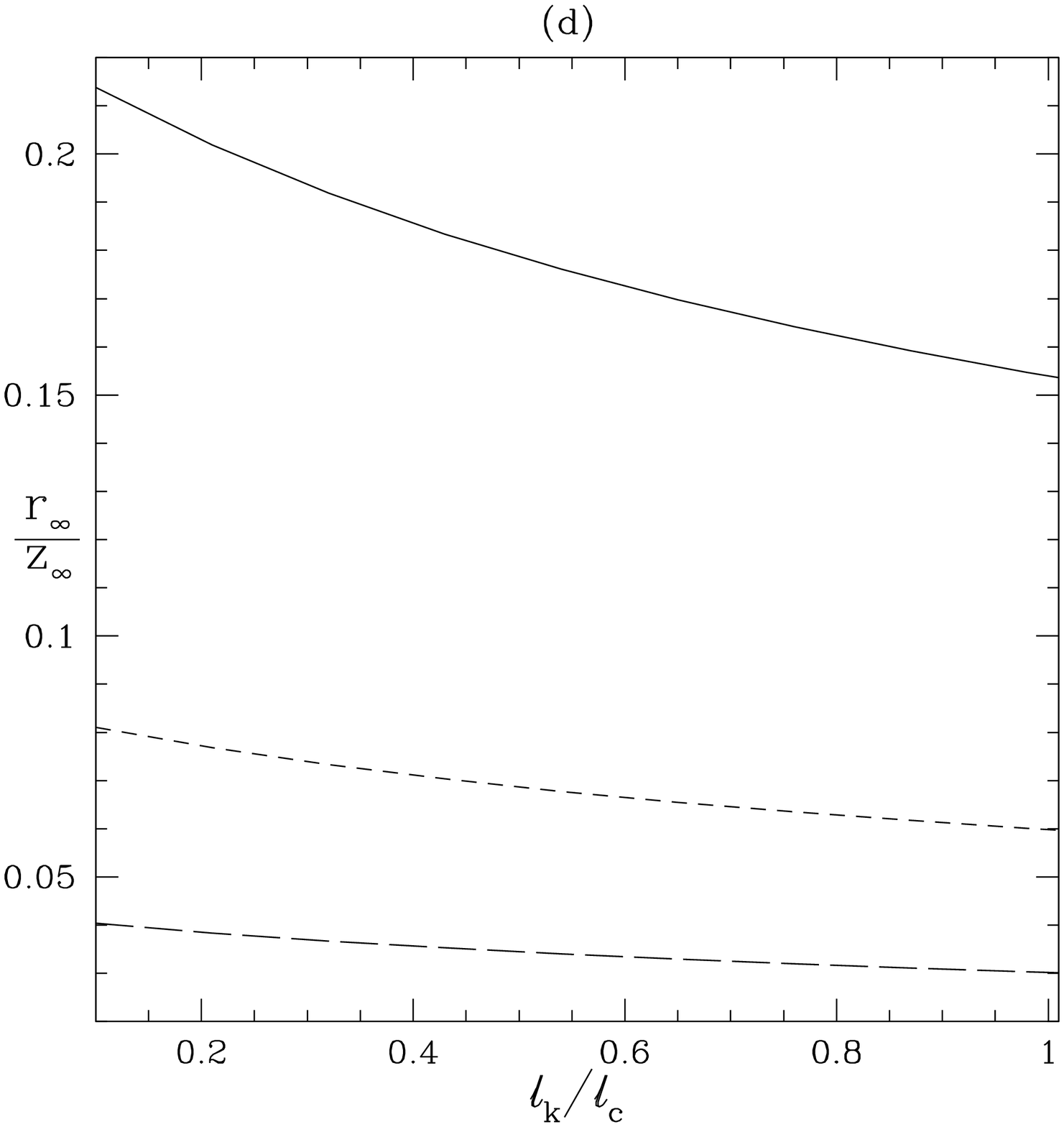}
\vskip 0.0cm
\hskip 0.0cm
\caption[]{Variation of $v_{\infty}$ (a) and $r_{\infty}/z_{\infty}$
(b), with ${\ell}_k/{\ell}_c$,
for ${\ell}_c=0.5$ (solid), ${\ell}_c=0.3$ (dashed)
and ${\ell}_c=0.1$ (long-dashed); $x_s=20r_g$. Variation of $v_{\infty}$ (c) and $r_{\infty}/z_{\infty}$
(d), with ${\ell}_k/{\ell}_c$, for $x_s=10r_g$ (solid), $x_s=20r_g$ (dashed) and
$x_s=30r_g$ (long-dashed); ${\ell}_c=0.4$. For all the figures $r_i=2r_g$.}
\end{figure}

In Fig. (12a), $v_{\infty}$ is plotted with ${\ell}_k/{\ell}_c$,
for ${\ell}_c=0.5$ (solid), ${\ell}_c=0.3$ (dashed)
and ${\ell}_c=0.1$ (long-dashed), and $x_s=20r_g$, $r_i=2r_g$
are kept fixed. As in Fig. (7e), we find that the increasing
${\ell}_k$ from $10\%$ to $100\%$ of ${\ell}_c$, has a marginal effect
on $v_{\infty}$. In Fig. (12b), corresponding
$r_{\infty}/z_{\infty}$ is plotted with ${\ell}_k/{\ell}_c$,
\ie for ${\ell}_c=0.5$ (solid), ${\ell}_c=0.3$ (dashed)
and ${\ell}_c=0.1$ (long-dashed), and $x_s=20r_g$, $r_i=2r_g$
are kept fixed. Though $v_{\infty}$ is marginally dependent on
${\ell}_k/{\ell}_c$, $r_{\infty}/z_{\infty}$ has a stronger dependence
on ${\ell}_k/{\ell}_c$. In other words, KD radiations are not
strong accelerators but definitely a better collimator.
As has been discussed in \S 4.1, the radial flux due to KD
is directed towards the axis in a larger domain, and as the
ratio ${\ell}_k/{\ell}_c{\gsim}0.5$ we find highly relativistic
and collimated jets.

We have seen in the preceding sub-section that we achieve better
collimation with larger shock location.
In Fig. (12c), we plot $v_{\infty}$ with ${\ell}_k/{\ell}_c$,
for $x_s=10r_g$ (solid), $x_s=20r_g$ (dashed) and
$x_s=30r_g$ (long-dashed), where ${\ell}_c=0.4$ and $r_i=2r_g$
are kept fixed. We find for $x_s=10r_g$ (solid), $v_{\infty}$
varies from above $0.94c$ to just above $0.93c$, on the other hand
for $x_s=30r_g$ (long-dashed), $v_{\infty}$ varies slightly
and is always a little less than $0.91c$. 
Interestingly for smaller values of $x_s$, increasing
${\ell}_k/{\ell}_c$, or in other words proportionally increasing
the KD radiation has relatively larger effect (although
much smaller than equivalent increase in ${\ell}_c$). For
larger $x_s$, the most luminous part of KD (${\sim}4r_g$) is missing
so its contribution to the components of radiative moments
are quite low, and as a result increasing ${\ell}_k$ has a negligible
effect on $v_{\infty}$.
In Fig. (12d),
corresponding $r_{\infty}/z_{\infty}$ is plotted with ${\ell}_k/
{\ell}_c$, for $x_s=10r_g$ (solid), $x_s=20r_g$ (dashed) and
$x_s=30r_g$ (long-dashed), where ${\ell}_c=0.4$ and $r_i=2r_g$
are kept fixed. 
It is also seen that, decrease of
$r_{\infty}/z_{\infty}$ with increasing ${\ell}_k/{\ell}_c$,
is more stronger for smaller values of $x_s$ (solid) than
the larger values (long-dashed).

Thus we see that, if ${\ell}_c>0.2$, $x_s>20r_g$, $1>{\ell}_k/{\ell}_c
<0.5$, for $r_i{\sim}2r_g$, we get relativistic ($v_{\infty}>0.9c$)
and collimated jets ($r_{\infty}/z_{\infty}<0.1$), which means
jets will be better collimated in intermediate hard states.

\begin{figure}

\includegraphics[scale=0.4]{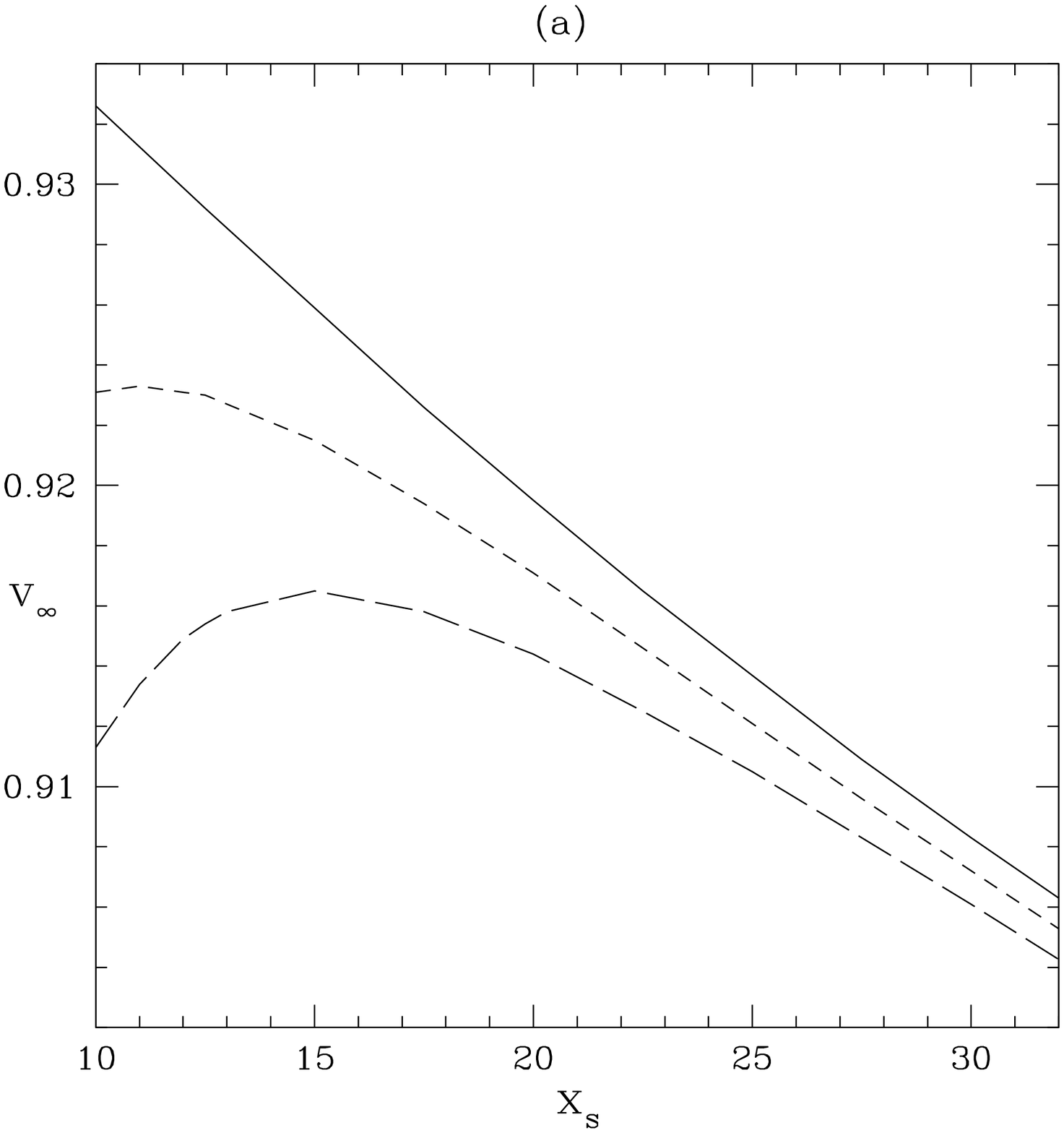}
\hskip 1.0cm
\includegraphics[scale=0.4]{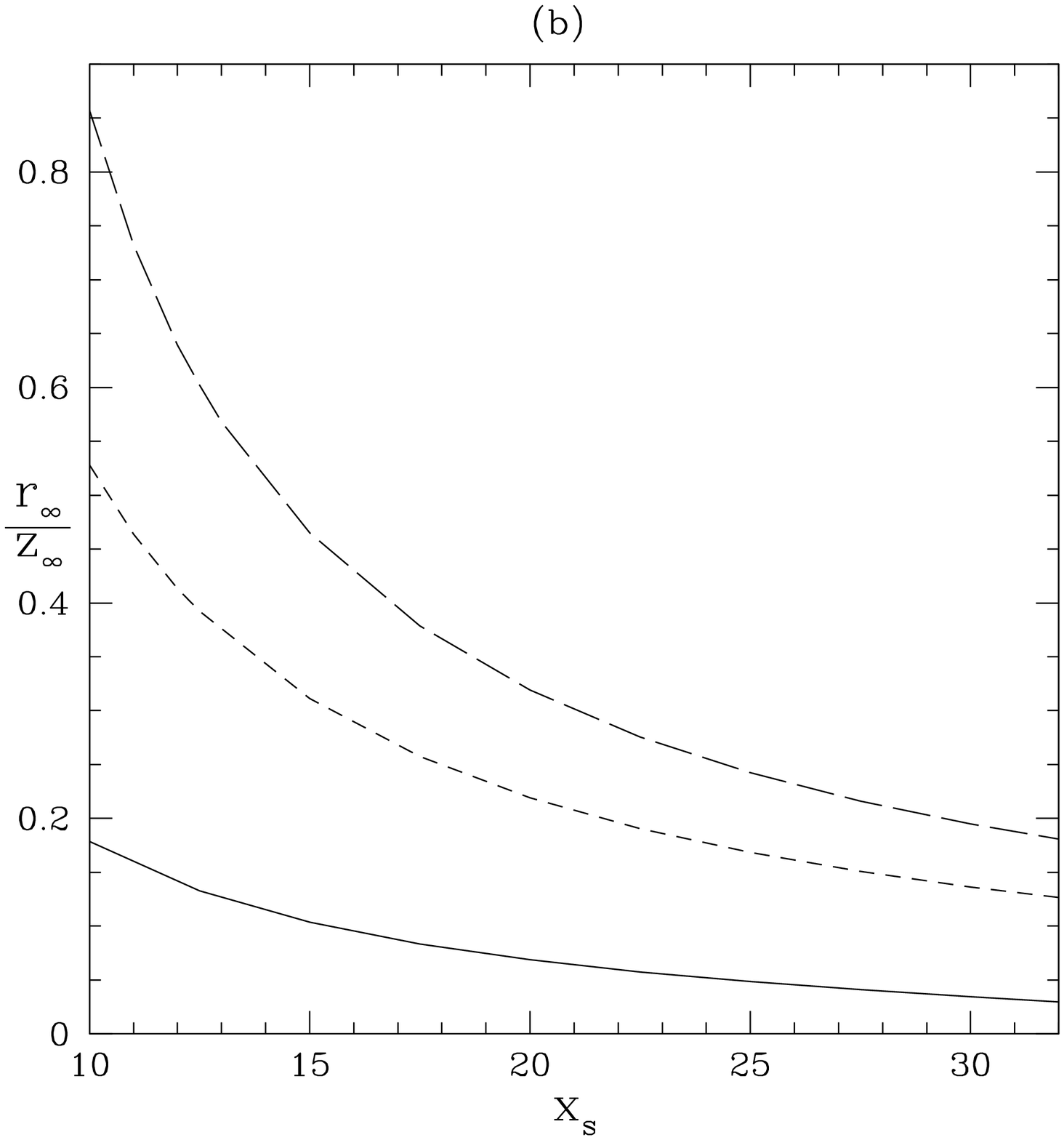}

\vskip 0.0cm
\hskip 0.0cm
\caption[]{Variation of $v_{\infty}$ (a) and $r_{\infty}/z_{\infty}$ (b)
with $x_s$, for $r_i=2r_g$ (solid), $r_i=4r_g$ (dashed) and $r_i=6r_g$
(long-dashed). For both the figures ${\ell}_c=0.4$ and ${\ell}_k=0.2$.}
\end{figure}

We already know that higher proportions of KD radiation, higher values
of $x_s$ etc are needed for collimation, while ${\ell}_c$ accelerates
the jets.
Still one should not forget about injection radius of the
jets, because we have seen that increase in $r_i$ can disturb
collimation as well as produce slower jets.
We now study the effect of $r_i$, on jet solution for higher
values of ${\ell}_k/{\ell}_c$ and increasing $x_s$.

In Fig. (13a), $v_{\infty}$ is plotted with $x_s$ for $r_i=2r_g$
(solid), $r_i=4r_g$ (dashed), $r_i=6r_g$ (long-dashed), where
${\ell}_c=0.4$ and ${\ell}_k=0.2$, \ie ${\ell}_k/{\ell}_c=0.5$.
The ratio of CENBOL and KD luminosity is assumed to be
around $0.5$, in order to aid collimation.
We now see a new feature, for lower $r_i$ (solid) $v_{\infty}$
decreases monotonically with $x_s$, but for higher $r_i$
(long-dashed) $v_{\infty}$ at first increases up to $x_s{\sim}15r_g$
and then decreases.
From Figs. (5a-j), we see that all the radiative moments peaks around
${\sim}2r_g$, just above the inner surface of the CENBOL, and also
have very strong gradients as one moves along the CENBOL surface
towards $x_s$. For larger $x_s$ the gradients are smoother,
but for smaller $x_s$ gradients becomes stronger.
So when $r_i(=6r_g)$ is larger for small $x_s$ (long-dashed),
then, the radiative moments received at $r_i$ is small,
because, the presence of strong gradients in all the moments, makes
all of them peak around
$2r_g$, and sharply decrease at $6r_g$.
As $x_s$ increases, for $r_i=6r_g$, the gradients are becoming smoother,
and $r_i=6r_g$ relatively becomes closer to the axis for larger $x_s$.
So, as $x_s$ increases from $10r_g$ to $15r_g$, the jet material
at $r_i=6r_g$, receives larger values of radiative moments, resulting
in increase of $v_{\infty}$. Increasing $x_s$ further,  decreases
$v_{\infty}$, because the CENBOL intensity decreases so much that
the driving force of radiation decreases anyway.

In Fig. (13b), $r_{\infty}/z_{\infty}$ is plotted with $x_s$
for $r_i=2r_g$
(solid), $r_i=4r_g$ (dashed), $r_i=6r_g$ (long-dashed), where
${\ell}_c=0.4$ and ${\ell}_k=0.2$, \ie ${\ell}_k/{\ell}_c=0.5$.
The relative spread decreases monotonically with $x_s$, as
the collimation depends on whether $f^r<0$ or $f^r>0$.
With increasing $x_s$, $f^r$ becomes negative (directed towards the axis)
in a larger domain, thus helping in collimation.
Thus we see that for $x_s>22r_g$, $r_i<4r_g$, ${\ell}_c{\sim}0.4$
and ${\ell}_k/{\ell}_c{\gsim}0.5$, we have jets with relativistic
terminal speed ${\gsim}0.9c$ and $r_{\infty}/z_{\infty}{\lsim}0.2$,
in other words we have collimated and relativistic jets.

\section{Discussion and Concluding Remarks}

In this paper, we have studied the interaction of radiation with
pair dominated jets from TCAF disc model.
We have ignored the details of the mechanism of production of pair
dominated jets. High energy photons
can produce particle-antiparticle pairs close to the
inner edge of a disk. It is well known,
if the photon energy $h{\nu} \gsim 2mc^2$, then an
electron-
positron pair may be created, where $h$ is the Planck's constant,
${\nu}$ is photon frequency and $m$ is the electron (or positron)
mass. If, on the other hand, electron and positron collide
it will annihilate each other to produce two Gamma-ray photons,
a process called pair annihilation.
Evidently, to produce pair dominated jets pair production has to
dominate pair annihilation process, as has been theoretically investigated by \citet{b30} and \citet{b41}.
%{Mishra \& Melia}{1993}]{b30},{Yamasaki \etal}{1999}]{b41}
Observationally, electron-positron jets were suggested in
galactic black hole candidate Nova Muscae 1991/GS 1124-684
\citep{b38}, GRS 1915+105 \citep{b29}, in quasar
%{Sunyaev \etal}{1992}]{b38},{Mirabel \& Rodriguez}{1998}]{b29}
3C279 \citep{b42}. Though there is little doubt about the existence of pair dominated 
%{Wardle \etal}{1998}]{b42}
jets,
radiative acceleration/collimation of such jets on the other hand is a different
issue altogether. If the pairs are produced to such an extent that
the jet medium is optically thick then the process discussed in this paper fails.
For optically thick medium, radiation drag terms are not there,
but disc intensity will fall off exponentially. We have not
considered such details, because considering such details involves
self-consistent treatment of the inflow-outflow solutions around
black holes.
The present effort confines itself to extend our earlier
work \citep{b14} to rotating jets.

The post-shock torus of the TCAF model,
produces normal plasma jets and high energy photons.
In this paper we have considered the electron
positron jets were produced within the funnel like
region of the post-shock torus.
In contrast to \citet{b14}, we have computed all
the different moments of radiation field for axial
and off-axial points. Furthermore, in \citet{b14},
we did not consider the Doppler shift of radiation
due to the disc motion. The rotational motion of
matter on the disc surface will generate an azimuthal
component of radiative flux. The Doppler effect
term also induces non-uniformity in frequency
integrated intensity of the CENBOL, while in \citet{b14}
the CENBOL intensity was uniform. This non-uniformity of
CENBOL intensity resulted
in strong gradients in the radiative moments around the
inner edge of the CENBOL.

The motion of the Keplerian disc is primarily rotation
dominated and its expression is known analytically
even in general relativity. While the motion
of matter in the post-shock torus has no simple
analytical expression. 
Since we were not considering inflow-outflow solutions
self-consistently, we needed to make a proper estimate
of the motion of post-shock matter.
Close to the black hole
the infall timescale is much smaller than the viscous
time scale so the angular momentum of in-falling matter
is almost constant near the black hole. Thus we assumed
wedge flow for the motion of in-falling matter along the inner surface
of the post shock torus, and solved geodesic equations.
The post-shock surface motion may be a little over estimated,
as pressure gradient terms are ignored.
One could have used Paczy\'nski-Wiita potential, but 
as this potential blows up at the horizon, the geodesic
equations are solved in general relativistic realm.
Another reason for estimating the inflow velocities
in general relativity is to ensure that the rotational velocities
of in-falling matter
should tend to become zero as one comes closer to the central object.

The Doppler effect on CENBOL intensity expression [Eq. (8e)],
is correct up to first order in ${\tilde u}/c$.
We have chopped off the CENBOL at $x_{in}=2r_g$,
and the velocity field of the CENBOL is such that
at $x_{in}<x<x_s$, ${\gamma}_{\tilde u}{\sim}1$.
At $x<x_{in}$, ${\gamma}_{\tilde u}$ sharply increases.
Thus taking first order correct Doppler term in Eq. (8e) is
consistant.

In this paper, we solved for three equations to find
three variables $v_s$, $\lambda$ and $r$, for which
three injection values corresponding to the the three variables were
supplied. All the other parameters like $x_{in}$, $x_s$, $x_o$,
${\ell}_c$, ${\dot m}_k$, and the disc velocity components ${\tilde u}$
and ${\tilde u}_K$ were supplied to compute various
independent components of radiation stress tensor or, in other words,
all the moments of radiation field.

It has been noticed, that the radiation from Keplerian disc
has marginal influence in determining $v_s$ and $\lambda$, while
has a greater role in determining $r$.
As has been explained above, collimation depends partly on reducing
angular momentum of the jet as well as pushing it towards
the axis by the radial component of radiative flux.
Within the funnel like region radiative flux from the Keplerian disc
is negative because of the geometry of the TCAF model.
Furthermore, the azimuthal component of the flux from it is also
weak compared to the other components, hence the angular
momentum gained by Keplerian radiation is small. In contrast, the radial flux
from the CENBOL is weakest amongst all the other components, although, because of
the special directiveness of the CENBOL, the radial flux is also
towards the axis, close to the black hole. However,
because of the small size of the CENBOL, at
$z{\sim}100r_g$, it approaches a point source and hence the radial flux
is positive and spreads the jet.
But drag terms along $r$ and also the fact that radial flux from
CENBOL is weakest amongst its all other components, makes the spreading
small.
The situation dramatically changes if the
injection radius is increased. As $r_i$ is increased, the radial flux increases and is directed more and more away from the axis.
On the top of that, the drag terms along $\phi$ becomes weaker
hence reduction of angular momentum decreases. The net effect is,
the jet spreads. On the top of that, with increasing $r_i$, the
CENBOL intensity goes down so the force driving the jets are less, 
resulting in slower jets. On the other hand, collimation is better
with increasing $x_s$. Increasing $x_s$, makes the radial flux from CENBOL
become directed towards the axis in a greater part of the
funnel like region, which pushes the jet and helps in collimation,
although as $x_s$ is increased the radiation from
Keplerian disc has lesser influence in determining $r$.

In the present paper we have restricted our analysis
to non-rotating black holes only. It will be interesting
to extend this study, to a Kerr black hole, because the higher efficiency
around a Kerr black hole will produce more intense radiation field,
and may produce terminal speeds higher than what we have observed in this
paper. As the spectrum from TCAF
disc around a Kerr black hole is yet to be computed, so the issue
of finding the spatial dependence of the disc intensity for a Kerr
black hole is an open problem as yet.

We conclude that, if radiative process is the main accelerating
and collimating process then,
\begin{enumerate}
\item Electron-positron jets are accelerated to highly relativistic
terminal speed, as well as, is collimated by the radiations from
two component accretion disc model.
\item The space dependent part of the radiative moments from
the post-shock region, dominates the corresponding moments from the
Keplerian disc.
\item The CENBOL radiation is the main accelerating agent, but
the Keplerian disc radiation has marginal influence in acceleration.
\item The drag terms in the azimuthal direction is greater than
the radiative flux term in the same direction, hence the radiation
removes angular momentum from the jet, near the jet base.
\item Collimation is partly brought about by removal of angular momentum,
and partly by the inward direction of the radial flux.
\item As the radial flux of the Keplerian disc is towards the axis, Keplerian radiation helps in collimation.
\item Collimation is better achieved for larger values of shock location
and lower values of injection radius.
\item As CENBOL radiation is the main accelerating agent and Keplerian disc
radiation is a good collimator, we conclude that, if this is the main process
for acceleration and collimation then, highly relativistic and collimated
jets should be observed in intermediate hard states (${\ell}_k/{\ell}_c{\lsim}1$),
and not in extreme hard states (${\ell}_k/{\ell}_c{\ll}1$). 
\item Drawing concrete conclusion, our study shows,
if shock in accretion is between $20r_g-30r_g$, injection radius $r_i<4r_g$, CENBOL luminosity ${\ell}_c{\gsim}0.2L_{\rm Edd}$ and  
${\ell}_k/{\ell}_c{\sim}0.5$, then the jets will have terminal speed greater
than $90\%$ the velocity of light, and the terminal relative spread will be less than $20\%$.

\end{enumerate}
\section*{Acknowledgements}

The author acknowledges a grant from ESA Prodex project, ESA contract No. 14815/00/NL/SFe.
The author also thanks Professor Sandip K. Chakrabarti of the S. N. Bose National Centre for Basic Sciences (India), for making valuable suggestions in preparing the
manuscript.

\appendix

\section[]{Estimation of Post-shock inflow velocity}

The CENBOL is assumed conical, \ie motion of post-shock
flow is assumed wedge flow. Let us solve the geodesic equation along
the radial coordinate $\overline{r}$, such that
$\overline{r}^2=x^2+y^2$ where $x$ \& $y$ defines
cylindrical radial and axial coordinated of CENBOL inner surface.
The geodesic equation for conical inflow around a non-rotating black-hole
is;
\begin{equation}
{\tilde u}^{\nu}\frac{{\partial {\tilde u}^{\overline{r}}}}{{\partial}x^{\nu}}
+{\Gamma}^{\overline{r}}_{{\mu}{\nu}}{\tilde u}^{\mu}{\tilde u}^{\nu}=0.
%\eqno{(A.1)}
\end{equation}

The radial velocity \& azimuthal velocity is defined as
\begin{equation}
{\tilde u}^2_{st}=-\frac{{\tilde u}_{\overline{r}}{\tilde u}^{\overline{r}}}
{{\tilde u}^0{\tilde u}_0}  
\hskip 0.6cm  {\mbox {and} }  \hskip 0.6cm
{\tilde u}^2_{\phi}=\frac{({\overline{r}}-1){\lambda}^2_{in}}{{\overline{r}}^3}
\end{equation}

Defining ${\tilde u}={\tilde u}_{st}/(1-{\tilde u}^2_{\phi})^{1/2}$,
reduces Eq. (A.1) to;

\begin{equation}
\frac{d{\tilde u}}{d{\overline{r}}}=-\frac{1/2
+({\tilde u}^2/2-\overline{r}+1){\tilde u}^2_{\phi}}
{\overline{r}(\overline{r}-1){\gamma}_{\tilde u}{\tilde u}},
\end{equation}

where, ${\gamma}^2_{\tilde u}=1/(1-{\tilde u}^2)$.

Equation (A.3) is solved by supplying the value of
${\tilde u}_{{st}{in}}=-0.01$ at ${\overline{r}}_{in}={\sqrt {x^2_s+h^2_s}}$.
As ${\tilde u}_{st}$ is the 3-velocity of inflow, so ${\tilde u}_{st}<0$,
i.e., at ${\overline{r}}=1r_g$, ${\tilde u}_{st}=-1$, ${\tilde u}_{\phi}=0$.

\label{lastpage}

\end{document}